\documentclass[10pt]{amsart}
\UseRawInputEncoding
\usepackage{amsmath,amsfonts,amssymb,color,a4,epsfig,graphics}
\usepackage[colorlinks=true,hyperindex=true]{hyperref}
\usepackage[english]{babel}
\usepackage{enumerate}
\usepackage{booktabs} 
\allowdisplaybreaks
\RequirePackage{ifthen}
\usepackage[utf8]{inputenc}
\usepackage[T1]{fontenc}
\usepackage{array,multirow,makecell}
\setcellgapes{1pt}
\makegapedcells
\usepackage[table]{xcolor}
\newcolumntype{R}[1]{>{\raggedleft\arraybackslash }b{#1}}
\newcolumntype{L}[1]{>{\raggedright\arraybackslash }b{#1}}
\newcolumntype{C}[1]{>{\centering\arraybackslash }b{#1}}
\usepackage{caption}
\usepackage{boldline}
\usepackage{booktabs}
\usepackage{tikz}
\def\build#1_#2^#3{\mathrel{\mathop{\kern 0pt#1}\limits_{#2}^{#3}}}
\usetikzlibrary{arrows}
\usepackage{array}
\usepackage{colortbl}
\usepackage{geometry}
\renewcommand{\arraystretch}{1.3}
\rowcolors{1}{}{white}
\newcommand\blue[1]{\textcolor{blue}{#1}}
\newcommand\red[1]{\textcolor{red}{#1}}

\newcommand{\R}{{\mathbb{R}}}

\newcommand{\Z}{{\mathbb{Z}}}
\newcommand{\N}{{\mathbb{N}}}

\newcommand{\Bc}{\mathcal{B}}
\newcommand{\Cc}{\mathcal{C}}
\newcommand{\Dc}{\mathcal{D}}
\newcommand{\Ec}{\mathcal{E}}
\newcommand{\Fc}{\mathcal{F}}
\newcommand{\Gc}{\mathcal{G}}
\newcommand{\Hc}{\mathcal{H}}
\newcommand{\Ic}{\mathcal{I}}

\newcommand{\Nc}{\mathcal{N}}
\newcommand{\Sc}{\mathcal{S}}
\newcommand{\Vc}{\mathcal{V}}

\def\rmi{{\rm i}}

\def\rmid{{\rm id}}
\newtheorem{theorem}{Theorem}[section]
\newtheorem{proposition}[theorem]{Proposition}
\newtheorem{lemma}[theorem]{Lemma}
\newtheorem{remark}[theorem]{Remark}
\newtheorem{corollary}[theorem]{Corollary}
\begin{document}
\title[\tiny Limiting absorption principle for long-range perturbation in a graphene...]{Limiting absorption principle for long-range perturbation in a graphene setting}
\author[\tiny Nassim Athmouni]{Nassim Athmouni$^{1}$}
\address{$^{1}$Universit\'e de Gafsa, Campus Universitaire, 2112, Tunisie}
\email{\tt nassim.athmouni@fsgf.u-gafsa.tn}
 \author[\tiny Marwa Ennaceur]{Marwa Ennaceur$^{2}$}
 \address{$^{2}$Department of Mathematics, College of Science, University of Hail, Ha’il 81451, \hspace*{0,5cm}Saudi Arabia}
\email{\tt mar.ennaceur@uoh.edu.sa}
\author[\tiny Sylvain Gol\'enia]{Sylvain Gol\'enia$^{3}$}
\address{$^{3}$Univ. Bordeaux,
Bordeaux INP, CNRS, IMB, UMR 5251, F-33400 Talence, France}
\email{\tt sylvain.golenia@math.u-bordeaux.fr}
\author[\tiny Amel Jadlaoui]{Amel Jadlaoui$^{4}$ }
 \address{$^{4}$Universit\'e de Sfax. Route de la Soukra km 4 - B.P. n° 802 - 3038 Sfax, Tunisie}
\email{\tt amel.jadlaoui@yahoo.com}

\begin{abstract}
In this paper, we examine the discrete Laplacian acting on a hexagonal lattice by introducing long-range modifications in both the metric and the potential. Our objective is to establish a Limiting Absorption Principle, excluding possible embedded eigenvalues. To this end, we employ the positive commutator technique as our method.
\end{abstract}
\maketitle
\tableofcontents
\section{Introduction and main result}
Over the last years, spectral graph theory has garnered research interest, particularly with regards to the study of various types of discrete Laplacians \cite{AD,AEG,AEGJ,BGJ,S,GG,Gk,AT,KM, KS} and their magnetic counterparts \cite{GT,BGKLM,GoMo,ABDE,PR1,HiSh}. One method for exploring the essential spectrum of these operators is using a positive commutator technique. With respect to previous research, \cite{S,BoSa,T1} investigate the case of
$\Z^d$, while \cite{AF} and \cite{GG} explore binary trees. On the other hand, \cite{MRT} analysed a general class of graphs. Moreover, \cite{AEG} examines a discrete version of cusps and funnels. Last, \cite{AEGJ} focuses on a triangular lattice graph.

In \cite{AEGJ}, our research is centered on the discrete Laplacian acting on the triangular lattice. We aim to obtain spectral results for a wider class of perturbations. Specifically, we introduce long-range perturbations in both the potential and the metric. To this end, we employ a Mourre estimate technique.
In the present work, we analyze the hexagonal lattice (see Figure 1), which is more complex due to its close resemblance to the analysis of a Dirac operator acting on a triangular lattice. It is highly similar to the structure of materials such as graphene, which exhibits unique electronic properties due to its hexagonal lattice. The mathematical framework developed here provides a foundation for understanding quantum phenomena behaviors, including particle wave interferences, which emerge in lattice systems. This structure allows for the observation
of phenomena like Dirac fermions, which play a key role in quantum mechanics. In this paper, we seek spectral results, such as resolvent estimates, which in turn yield propagation estimates. The model was already studied in \cite{PR, RT} through Floquet decomposition and analytic deformation. However, those approaches fail to treat the class of long-rang perturbations. With this in mind, the aim of our work is precisely to cover this long-range regime. As in \cite{AEGJ}, we implement long-range modifications of both the potential and the metric, relying on the Mourre estimate method.
Moreover, thanks to the specific structure of the model, we can explicitly determine the set of thresholds (Theorem \ref{t_LAP}) by means of the Fourier transform, see also \cite{AIM}. This distinguishes our approach from that of \cite{PR}, where the authors adopt a general Floquet-Bloch method. By means an abstract reasoning, they assert the existence of these points in the spectrum merely via a direct integral decomposition. We note that \cite{T2} analyses the short-range and long-range potentials with another choice of conjugate operator, but does not accommodate metric perturbations, see also Remark \ref{RX}. Concerning the perturbation of the metric, this is considered in \cite{PR} but due to the abstract context, they do not succeed in dealing with the long range perturbation of the metric. Here, our choice of conjugate operator is different from the two other approaches it helps us to attain this goal. However, our approach is heavier in term of computation. 

We begin by recalling definitions from some graph theory, which will form the foundation for our analysis. An infinite, connected
\textit{graph} $\Gc$ is a triple $(\Vc,\Ec,m)$, where
$\Vc$ is a countable set of \textit{vertices}, $m : \Vc \rightarrow (0, \infty)$ is a \textit{weight}, and
$\Ec:\Vc\times\Vc \rightarrow  [0,+\infty)$ (the edges) is symmetric. For two vertices
$n$ and $l$, we say they are \textit{neighbors} if
$\Ec(n,l)>0$.
We denote this relationship by
$n\sim l$ and define the set of neighbors of $n$ as
$\Nc_\Gc(n)$. We let $\Cc(\Vc):=\{f:\Vc\longrightarrow\mathbb{C}^d\}$ denote the space of complex-valued functions defined on the set of vertices $\Vc$. We define the Hilbert space \[\ell^2(\Vc,m;\mathbb{C}^d):=\left\{ f\in \Cc(\Vc), \sum_{n\in\Vc} m(n)|f(n)|^2_{\mathbb{C}^d} <\infty \right\},\]
equipped with the scalar product, $\langle f,g\rangle:= \sum_{n\in \Vc}\sum_{k=1}^d m(n)\overline{f_k(n)}g_k(n).$
In general, we define the Laplacian $\Delta_H$ on $\Gc$ by:
\begin{align}\label{delta}
\Delta_{H}f(x)&:=\frac{1}{m(x)}\sum_{y\in \Vc}\Ec(x,y)f(y).\end{align}
Here, we investigate a hexagonal lattice where:
\[\Vc_0:=\Big\{\sum_{j=1}^2k_j v_j\hbox{; } k:=(k_1,k_2)\in \mathbb{Z}^2\Big\}\hbox{, } v_1:=\left(\frac{3}{2},\frac{\sqrt{3}}{2}\right)\hbox{, }v_2:=(0,\sqrt{3}).\]
Let \begin{align*}p_1:=(\frac{1}{2},-\frac{\sqrt{3}}{2})\hbox{, }p_2:=(1,0),\end{align*}
and define the vertex set $\widetilde{\Vc}$ by:
\begin{align}\label{isom}\widetilde{\Vc}:=\Vc_1\cup\Vc_2,\ \Vc_i:=p_i+\Vc_0.\end{align}
Note that $\Vc_1\cap\Vc_2=\varnothing.$ Let $a_1\in \Vc_1$ and $a_2\in \Vc_2$ such that $a_1=p_1+(n_1,n_2)$ and $a_2=p_2+(n_1,n_2)$  we have:
\begin{align}\label{Na}\nonumber\Nc_{a_1}&=\left\{l\in\mathbb{R}^2\hbox{; }|l-a_1|_{\mathbb{R}^2}=1,\ a_1\in\Vc_1 \right\}\cap \Vc_2\\
&=\left\{a_1+\frac{v_{1}+v_2}{3}\hbox{, }a_1+\frac{v_{1}-2v_2}{3}\hbox{, }a_1-\frac{2v_{1}-v_2}{3} \right\},\end{align}
\begin{align}\label{Na1}\nonumber\Nc_{a_2}&=\left\{l\in\mathbb{R}^2\hbox{; }|l-a_2|_{\mathbb{R}^2}=1,\ a_2\in\Vc_2 \right\}\cap \Vc_1\\
&=\left\{a_2+\frac{2v_{1}-v_2}{3}\hbox{, }a_2-\frac{v_{1}-2v_2}{3}\hbox{, }a_2-\frac{v_{1}+v_2}{3} \right\}.\end{align}
Given $f\in \ell^2(\widetilde{\Vc},1;\mathbb{C}),\ f_i$ is its projection onto $\ell^2(\Vc_i,1;\mathbb{C})$. Namely, \begin{align}\label{XX}f_i(n_1,n_2):=f(p_i+(n_1,n_2))=:f(n_1,n_2,p_i).\end{align}
Using \eqref{isom} and the following isomorphisms:
 \begin{align}\label{isomorphisme1}
\ell^2(\widetilde{\Vc},1;\mathbb{C})&\to \ell^2(\Vc_1,1;\mathbb{C})\oplus \ell^2(\Vc_2,1;\mathbb{C})\to\ell^2(\Vc_0,1;\mathbb{C}^2)
\\
\nonumber f&\mapsto \left(f_1,f_2\right)\mapsto {^t}(f_1,f_2),
\end{align}
 we conclude that:
\begin{align}\label{isomorphisme11}\ell^2(\widetilde{\Vc},1;\mathbb{C})\simeq \ell^2(\Vc_1,1;\mathbb{C})\oplus \ell^2(\Vc_2,1;\mathbb{C})\simeq\ell^2(\Vc_0,1;\mathbb{C}^2) .\end{align}

In this sequel, we often identify the vertices of $\Vc_i$ with $\mathbb{Z}^2$, using the canonical map:
\begin{align}\label{map}
\nonumber G_i:\mathbb{Z}^2 &\to \Vc_i
\\
(k_1,k_2)&\mapsto k_1v_1+k_2v_2+p_i,
\end{align}
where $i \in \{1,2\}$.
\begin{figure}
\begin{tikzpicture}
  \draw (0,0) -- (1,0);
  \draw (1,0) -- (1.5,{sqrt(3)/2});
  \draw (1,0) -- (1.5,-{sqrt(3)/2});
  \draw (1.5,{sqrt(3)/2}) -- (1,{sqrt(3)});
  \draw (1,{sqrt(3)}) -- (0,{sqrt(3)});
  \draw (0,{sqrt(3)}) -- (-0.5,{sqrt(3)/2});
  \draw (-0.5,{sqrt(3)/2}) -- (0,0);
  \draw (0,0) -- (-0.5,-{sqrt(3)/2});
  \draw (1.5,{sqrt(3)/2}) -- (2.5,{sqrt(3)/2});
  \draw (2.5,{sqrt(3)/2}) -- (3,{sqrt(3)});
  \draw (3,{sqrt(3)}) -- (2.5,{sqrt(3)*3/2});
  \draw (2.5,{sqrt(3)*3/2}) -- (1.5,{sqrt(3)*3/2});
  \draw (1.5,{sqrt(3)*3/2}) -- (1,{sqrt(3)});
  \draw (2.5,-{sqrt(3)/2}) -- (3,0);
  \draw (3,0) -- (2.5,-{sqrt(3)/2});
  \draw (3,0) -- (2.5,{sqrt(3)/2});
  \draw (4.5,-{sqrt(3)/2}) -- (4,0);
\draw (4,0) -- (4.5,{sqrt(3)/2});
\draw (3,0) -- (4,0);
\draw (4.5,{sqrt(3)/2}) -- (4,{sqrt(3)});
\draw (4,{sqrt(3)}) -- (3,{sqrt(3)});
\draw (4,{sqrt(3)}) -- (4.5,{sqrt(3)*3/2});
\draw (4.5,{sqrt(3)/2}) -- (5.5,{sqrt(3)/2});
\draw (4.5,{sqrt(3)*3/2}) -- (4,{sqrt(3)*2});
\draw (4,{sqrt(3)*2}) -- (3,{sqrt(3)*2});
\draw (4,{sqrt(3)*2}) -- (4.5,{sqrt(3)*5/2});
\draw (3,{sqrt(3)*2}) -- (2.5,{sqrt(3)*3/2});
\draw (4.5,{sqrt(3)*3/2}) -- (5.5,{sqrt(3)*3/2});
\draw (3,{sqrt(3)*2}) -- (2.5,{sqrt(3)*5/2});
\draw (1.5,{sqrt(3)*3/2}) --(1,{sqrt(3)*2});
\draw (1,{sqrt(3)*2}) -- (1.5,{sqrt(3)*5/2});
\draw (1,{sqrt(3)*2}) -- (0,{sqrt(3)*2});
\draw (0,{sqrt(3)*2}) -- (-0.5,{sqrt(3)*5/2});
\draw (0,{sqrt(3)*2}) -- (-0.5,{sqrt(3)*3/2});
\draw (-0.5,{sqrt(3)*3/2}) -- (0,{sqrt(3)});
\draw (-0.5,{sqrt(3)*3/2}) -- (-1.5,{sqrt(3)*3/2});
\draw (-0.5,{sqrt(3)/2}) -- (-1.5,{sqrt(3)/2});
\node[draw=none, label={[font=\smaller\smaller\tiny]above right:($-2,-\sqrt{3}$)}] at (0-0.3, 0-0.1) {};
\node[draw=none, label={[font=\smaller\tiny] right:($-1,-\sqrt{3}$)}] at (1-0.2, 0) {};
\node[draw=none, label={[font=\smaller\tiny] below right:($1,-\sqrt{3}$)}] at (3-0.3, 0+0.1) {};
\node[draw=none, label={[font=\smaller\tiny] right:($2,-\sqrt{3}$)}] at (4-0.2, 0) {};
\node[draw=none, label={[font=\smaller\tiny] right:($-\frac{5}{2},-\frac{\sqrt{3}}{2}$)}] at ({-0.5-0.2}, {sqrt(3)/2}) {};
\node[draw=none, label={[font=\smaller\tiny] above right:($-\frac{1}{2},-\frac{\sqrt{3}}{2}$)}] at (1.2, 0.7) {};
\node[draw=none, label={[font=\smaller\tiny] right:($\frac{1}{2},-\frac{\sqrt{3}}{2}$)}] at (2.3, 0.7) {};
\node[draw=none, label={[font=\smaller\tiny] above right:($-\frac{5}{2},-\frac{\sqrt{3}}{2}$)}] at (4.2, 0.7) {};
\node[draw=none, label={[font=\smaller\tiny]above right:($-2,0$)}] at (0-0.3, 1.6) {};
\node[draw=none, label={[font=\smaller\tiny] above right:($-1,0$)}] at (0.9, 1.4) {};
\node[draw=none, label={[font=\smaller\tiny] above right:($1,0$)}] at (2.7, 1.6) {};
\node[draw=none, label={[font=\smaller\tiny] above right:($2,0$)}] at (3.8, 1.4) {};
\node[draw=none, label={[font=\smaller\tiny] right:($-\frac{5}{2},\frac{\sqrt{3}}{2}$)}] at ({-0.5-0.2}, 2.6) {};
\node[draw=none, label={[font=\smaller\tiny] above right:($-\frac{1}{2},\frac{\sqrt{3}}{2}$)}] at (1.2, 2.3) {};
\node[draw=none, label={[font=\smaller\tiny] right:($\frac{1}{2},\frac{\sqrt{3}}{2}$)}] at (2.3, 2.6) {};
\node[draw=none, label={[font=\smaller\tiny] above right:($-\frac{5}{2},\frac{\sqrt{3}}{2}$)}] at (4.2, 2.3) {};
\node[draw=none, label={[font=\smaller\tiny]above right:($-2,\sqrt{3}$)}] at (0-0.3, 3.3) {};
\node[draw=none, label={[font=\smaller\tiny] right:($-1,\sqrt{3}$)}] at (0.9, 3.45) {};
\node[draw=none, label={[font=\smaller\tiny] above right:($1,\sqrt{3}$)}] at (2.7, 3.3) {};
\node[draw=none, label={[font=\smaller\tiny] right:($2,\sqrt{3}$)}] at (3.8, 3.45) {};

 \draw[fill=red] (0-0.1,0) -- ++(0.1,0.1) -- ++(0.1,-0.1) -- ++(-0.1,-0.1) -- cycle;
\draw[fill=blue] (1,0) circle (0.1);
\draw[fill=red] (3-0.1,0) -- ++(0.1,0.1) -- ++(0.1,-0.1) -- ++(-0.1,-0.1) -- cycle;
\draw[fill=blue] (4,0) circle (0.1);
\draw[fill=blue] ({-0.5}, {sqrt(3)/2}) circle (0.1);
\draw[fill=red] (1.4, {sqrt(3)/2}) -- ++(0.1,0.1) -- ++(0.1,-0.1) -- ++(-0.1,-0.1) -- cycle;
\draw[fill=blue] (2.5, {sqrt(3)/2}) circle (0.1);
\draw[fill=red] (4.4, {sqrt(3)/2}) -- ++(0.1,0.1) -- ++(0.1,-0.1) -- ++(-0.1,-0.1) -- cycle;
\draw[fill=red] (0-0.1,{sqrt(3)}) -- ++(0.1,0.1) -- ++(0.1,-0.1) -- ++(-0.1,-0.1) -- cycle;
\draw[fill=blue] (1,{sqrt(3)}) circle (0.1);
\draw[fill=red] (3-0.1,{sqrt(3)}) -- ++(0.1,0.1) -- ++(0.1,-0.1) -- ++(-0.1,-0.1) -- cycle;
\draw[fill=blue] (4,{sqrt(3)}) circle (0.1);
\draw[fill=blue] ({-0.5}, {3*sqrt(3)/2}) circle (0.1);
\draw[fill=red] (1.4, {3*sqrt(3)/2}) -- ++(0.1,0.1) -- ++(0.1,-0.1) -- ++(-0.1,-0.1) -- cycle;
\draw[fill=blue] (2.5, {3*sqrt(3)/2}) circle (0.1);
\draw[fill=red] (4.4, {3*sqrt(3)/2}) -- ++(0.1,0.1) -- ++(0.1,-0.1) -- ++(-0.1,-0.1) -- cycle;
\draw[fill=red] (0-0.1,{2*sqrt(3)}) -- ++(0.1,0.1) -- ++(0.1,-0.1) -- ++(-0.1,-0.1) -- cycle;
\draw[fill=blue] (1,{2*sqrt(3)}) circle (0.1);
\draw[fill=red] (3-0.1,{2*sqrt(3)}) -- ++(0.1,0.1) -- ++(0.1,-0.1) -- ++(-0.1,-0.1) -- cycle;
\draw[fill=blue] (4,{2*sqrt(3)}) circle (0.1);
\end{tikzpicture}
 \caption{Hexagonal lattice\\ \blue{$\bullet$}: element of $\Vc_1;$ \red{\tiny{$\blacklozenge$}}: element of $\Vc_2.$ }
\end{figure}
In other words, we have the following identifications:
\begin{align}\label{identification}
\ell^2\left(\mathbb{Z}^2\times \{p_i\};\mathbb{C}\right)\simeq \ell^2(\Vc_i;\mathbb{C}),
\end{align}
where $i\in\{1,2\}.$
From \eqref{isomorphisme1} and \eqref{identification}, we have the following isomorphism.
\begin{align}\label{iden}
\ell^2\left(\mathbb{Z}^2\times \{p_1,p_2\};\mathbb{C}\right)\simeq \ell^2(\mathbb{Z}^2;\mathbb{C}^2)\simeq \ell^2(\widetilde{\Vc},1;\mathbb{C}).
\end{align}

For all $f\in\ell^2(\mathbb{Z}^2,1)$, we have:
\begin{align*}
&U_1f(n_1,n_2):=f(n_1-1,n_2),\ U_1^{*}f(n_1,n_2)=f(n_1+1,n_2),\\
&U_2f(n_1,n_2):=f(n_1,n_2-1),\ U_2^{*}f(n_1,n_2)=f(n_1,n_2+1).
\end{align*}
Recall \eqref{delta}. When $m=1$ and \begin{align*}
\Ec: \widetilde{\Vc}\times\widetilde{\Vc}&\to \{0,1\}
\\
(l,n)&\mapsto \Ec(l,n):=\left\{
                        \begin{array}{ll}
                          1, & \hbox{if } |l-n|_{\mathbb{R}^2}=1; \\
                          0,
                        \end{array}
                       \right.
\end{align*}
where $l=(l_1,l_2)$, $n=(n_1,n_2)$ and $|n|_{\mathbb{R}^2}:=\sqrt{n_1^2+n_2^2}$. Under the identification \eqref{iden} we rewrite $\Delta_H$ on the hexagonal lattice as follows:
 \begin{align*}\Delta_{H}&:=
\frac{1}{3}
\left(
  \begin{array}{cc}
    0 & \Delta_{2,H} \\
    \Delta_{1,H} & 0 \\
  \end{array}
\right)
=\frac{1}{3}
\left(
  \begin{array}{cc}
    0 & \rmid+U_1+U_2 \\
    \rmid+U^*_1+U^*_2 & 0 \\
  \end{array}
\right).
\end{align*}
Now, we perturb the metric of the previous case, by an amount that remains small at infinity.
Using \eqref{delta} and \eqref{iden} the Laplacian $\Delta_{m,\Ec}$ is given by
\begin{align}\label{dm}
\nonumber\Delta_{m,\Ec}: \ell^2(\widetilde{\Vc},m)&\to \ell^2(\widetilde{\Vc},m)
\\
f&\mapsto\Delta_{m,\Ec}f:=\frac{1}{3}\left(\begin{array}{cc}
                                         0 & \Delta_{2,m,\Ec} \\
                                       \Delta_{1,m,\Ec} & 0 \\
                                       \end{array}\right)
\left(
  \begin{array}{c}
    f_1 \\
    f_2 \\
  \end{array}
\right)
,
\end{align}
with 
\begin{align*}&\Delta_{1,m,\Ec}f_1(n_1,n_2):=\frac{1}{m(n_1,n_2,p_2)} \Bigg(\Bigg(\Ec((Q_1,Q_2,p_1),(Q_1,Q_2,p_2))
 +\Ec((Q_1,Q_2,p_1),(Q_1+1,Q_2,p_2))U^*_1\\
\nonumber &+\Ec((Q_1,Q_2,p_1),(Q_1,Q_2+1,p_2) U^*_2\Bigg)f_1\Bigg)(n_1,n_2),\end{align*}
\begin{align*}&\Delta_{2,m,\Ec}f_2(n_1,n_2):=\frac{1}{m(n_1,n_2,p_1)}\Bigg(\Bigg(\Ec((Q_1,Q_2,p_2),(Q_1,Q_2,p_1))
 +\Ec((Q_1,Q_2,p_2),(Q_1-1,Q_2,p_1))U_1\\
\nonumber &+\Ec((Q_1,Q_2,p_2),(Q_1,Q_2-1,p_1) U_2\Bigg)f_2\Bigg)(n_1,n_2),\end{align*}
where $m$ and $\Ec$ respectively satisfy the following hypotheses:
 \begin{align*}
 (H_0)\quad& m(n_1,n_2,p_i):=(1+\eta(n_1,n_2,p_i)),\
\mbox {where } \inf_{(n_1,n_2)\in\mathbb{Z}^2}\eta(n_1,n_2,p_i)>-1,\\
&\hspace*{4cm}\eta(n_1,n_2,p_i)\rightarrow0  \mbox{ if } |n|\rightarrow\infty,\end{align*}

 \begin{align*}
 (H_1)\quad& \Ec((n_1,n_2,,p_i),(n,p_j)):=(1+\varepsilon((n_1,n_2,p_i),(n_1,n_2,p_j))),\
\mbox {where } \\
&\inf_{(n_1,n_2),(k_1,k_2)\in\mathbb{Z}^2} \varepsilon((n_1,n_2,p_i),(k_1,k_2,p_j))>-1,\\
&\hspace*{1cm}\varepsilon((n_1,n_2,p_i),(k_1,k_2,p_j))\rightarrow0  \mbox{ if } |(n_1,n_2),(k_1,k_2)|\rightarrow\infty,\ \text{ where }i,j\in\{1,2\}.\end{align*}
Note that $\Ec$ has the following properties:
\begin{itemize}
\item  By symmetry of $\Ec$, we have $\Ec((n_1,n_2,p_i),(n'_1,n'_2,p_j))=\Ec((n'_1,n'_2,p_j),(n_1,n_2,p_i))$.
\item Note that $\Ec((n_1,n_2,p_i),(n'_1,n'_2,p_j))\neq 0$, then $p_i\neq p_j.$\end{itemize}
\hspace*{1.5cm}$\text{If } i=1, \text{ then } (n'_1,n'_2)=(n_1,n_2)\text{ or }(n_1+1,n_2)\text{ or }(n_1,n_2+1).$\\
\hspace*{1.5cm}$\text{If } i=2, \text{ then } (n'_1,n'_2)=(n_1,n_2)\text{ or }(n_1-1,n_2)\text{ or }(n_1,n_2-1).$

The operator 
 $\Delta_{m,\Ec}$ is self-adjoint and bounded on $\ell^2(\widetilde{\Vc},m)$.
Moreover, in the case where $\eta=0$ and $\varepsilon=0$, we have $\Delta_{m,\Ec}=\Delta_H$ and $\sigma(\Delta_{H})=\left[-1,1\right]=\sigma_{\rm ac}(\Delta_H)$, where $\sigma_{\rm ac}(\cdot)$ is the absolutely continuous spectrum, see \cite{AIM} and Lemma \ref{specter}.
Given a function $W: \widetilde{\Vc}\longrightarrow\mathbb{C}$, we denote by $W(Q_1,Q_2)$ the operator of multiplication by $W$. In particular $(W(Q_1,Q_2)f)(n_1,n_2,p_i):=W(n_1,n_2,p_i)f(n_1,n_2,p_i)$, for all $f\in \Dc(W(Q_1,Q_2))$. By \eqref{isomorphisme1}, \eqref{isomorphisme11} and Proposition \ref{send}, we have $W=W_1\oplus W_2$, with $W_i:=(\cdot,\cdot,p_i)$. It acts as follows:
 \begin{align*}
W: \ell^2(\widetilde{\Vc},m;\mathbb{C})&\to \ell^2(\Vc_1,m;\mathbb{C})\oplus \ell^2(\Vc_2,m;\mathbb{C})
\\
\nonumber (f_1,f_2)&\mapsto \left(W_1f_1,W_2f_2\right).
\end{align*}
Hence, $\Dc(W)=\Dc(W_1)\oplus \Dc(W_2)$ and
 $$\Dc(W_i(Q_1,Q_2)):=\left\{f\in \ell^2(\Vc_i,m);\ (n_1,n_2)\mapsto W_i(n_1,n_2)f(n_1,n_2,p_i)\in \ell^2(\Vc_i,m)\right\}.$$
Let $V: \widetilde{\Vc}\longrightarrow\mathbb{R}^2$ and $H:=\Delta_{m,\Ec}+V(Q)$ such that: \begin{align*}(H_2)\quad&  V(n_1,n_2,p_i)\to0 \mbox{ if } |(n_1,n_2)|\to\infty,\end{align*}
where $i\in\{1,2\}$. Note $V(Q_1,Q_2)$ is a compact operator, as uniform limit of finite rank operators given by $1_{\|\cdot\|_{\mathbb{R}^2}\leq R}V$, with $R\in\mathbb{N}$.
The operator $H_{m,\Ec}$ is self-adjoint on $\ell^2(\widetilde{\Vc},m)$. In fact, it can be regarded as a compact perturbation of $\Delta_{H}$, see Proposition \ref{com} for a precise statement. Moreover, we have $\sigma_{\rm ess}(H_{m,\Ec})=\sigma_{\rm ess}(\Delta_H)$, where $\sigma_{\rm ess}(\cdot)$ is the essential spectrum.

We now seek to refine the spectral property and ask for more decay. We set:
\begin{align}\label{Lambda}\Lambda_H(\cdot):=\Lambda(\cdot)\left(
            \begin{array}{cc}
             \rmid & 0 \\
             0 &  \rmid
              \\
            \end{array}
          \right),\end{align}
where $\Lambda(n_1,n_2)=\langle n_1\rangle+\langle n_2\rangle$ and $\langle \cdot\rangle:=\sqrt{\frac{1}{2}+|\cdot|^2}$. Note that $\Lambda(Q_1,Q_2)$ is an unbounded self-adjoint operator. We fix  $\gamma>0,$
we assume a long-range decay for $V,\ \eta$ and $\varepsilon$: Let $l_1,l_2\in \{0,1\}$. Let $k\in \mathbb{N}^*,\ h\in\{0,1,2\},\ b\in \mathbb{N}$.
We say that $G:(\mathbb{Z}^2\times \{p_1,p_2\})^k\longrightarrow\mathbb{C}$ satisfies the hypothesis $(H_{3,k})$ if the following condition hold
 \begin{align*}
&(H_{3,k})\quad \max_{h\in\{1,2\}}\max_{(l_1,l_2)}\sup_{(n_1,n_2)\in\mathbb{Z}^2}\Lambda^{\gamma}(n_1,n_2)\langle  n_1l_1-n_2l_2\rangle|G(J_{k,h,1}(n_1,n_2,p_1))-G(J_{k,h,0}(n_1,n_2,p_2))|<\infty,
 \end{align*}
which is equivalent to
 \begin{align*}
& \max_{h\in\{1,2\}}\max_{(l_1,l_2)}\sup_{(n_1,n_2)\in\mathbb{Z}^2}\Lambda^{\gamma}(n_1,n_2)\langle  n_1l_1-n_2l_2\rangle|G(J_{k,h,1}(n_1,n_2,p_2))
-G(J_{k,h,0}(n_1,n_2,p_1))|<\infty,
 \end{align*}
where  
 \begin{align*}
J_{k,h,b}: (\mathbb{Z}^2\times \{p_1,p_2\})&\to (\mathbb{Z}^2\times \{p_1,p_2\})^k
\\
\nonumber (n_1,n_2,p_i)&\mapsto \left((n_1-(-1)^ib\delta_{h,1},n_2-(-1)^ib\delta_{h,2},p_j),\underbrace{(n_1,n_2,p_i)}_{k-1\ times}\right)
\end{align*}
and $\delta_{h,r}$ is the Kronecker’s delta symbol. 

We denote by $\sigma_{\rm p}(\cdot)$ the set of pure point spectrum.
We now state our main theorem:
\begin{theorem}\label{t_LAP}
Suppose that $(H_0),\ (H_1)$ and $(H_2)$ are true. We assume that $\eta$ and $V$ satisfy $(H_{3,1})$ and $\varepsilon$ satisfy $(H_{3,2})$ and we take $s>\frac{1}{2}$. We obtain the following assertions:
\item[1.] $\sigma_{\rm ess}(H_{m,\Ec})=\sigma_{\rm ess}(\Delta_{H})$.
\item[2.] The eigenvalues of $H_{m,\Ec}$ distinct from $\kappa$ are of finite multiplicity and can accumulate
only toward $\kappa$, where $\kappa:=\left\{-1,\ -\frac{1}{3},\ 0,\ \frac{1}{3},\ 1\right\}$, the set of thresholds.

\item[3.] The singular continuous spectrum of $H_{m,\Ec}$ is empty.

\item[4.] The following limit exists and is finite:
$$\lim_{\rho\rightarrow 0^+}\sup_{\lambda\in[a,b]}\|\Lambda^{-s}(Q)(H_{m,\Ec}-\lambda-\rmi\rho)^{-1}\Lambda^{-s}(Q)\|<\infty,$$
where $[a,b]$ is included in $\mathbb{R}\setminus\left(\kappa\cup\sigma_{\rm p}(H_{m,\Ec})\right).$
Moreover, in the norm topology of bounded operators, the boundary values of the resolvent:
\[ [a,b] \ni\lambda\mapsto\lim_{\rho\to0^{\pm}}\Lambda^{-s}(Q)(H_{m,\Ec}-\lambda-\rmi\rho)^{-1}\Lambda^{-s}(Q) \mbox{ exists and is continuous}.\]

\item[5.] There exists $c>0$ such that for all $f\in\ell^2(\widetilde{\Vc})$, we have:
\[\int_{\R}\|\Lambda^{-s}(Q)e^{-\rmi tH_{m,\Ec}}E_{[a,b]}(H_{m,\Ec})f\|^2dt\leq c\|f\|^2,\]
where $E_{[a,b]}(H_{m,\Ec})$ is the spectral projection of $H_{m,\Ec}$ above $[a,b].$
\end{theorem}
We now explain the hypothesis $(H_{3,k})$ in more detail in our setting.
The hypotheses on $V$ correspond to the following cases: If $k=1$, 
we take $G\equiv V$.  
For $h=1$, the hypothesis $(H_{3,1})$ gives 
 \begin{align*}
& \max_{(l_1,l_2)}\sup_{(n_1,n_2)\in\mathbb{Z}^2}\Lambda^{\gamma}(n_1,n_2)\langle  n_1l_1-n_2l_2\rangle\left|V(J_{1,1,1}(n_1,n_2,p_1))-V(J_{1,1,0}(n_1,n_2,p_2))\right|,\\
&=\max_{(l_1,l_2)}\sup_{(n_1,n_2)\in\mathbb{Z}^2}\Lambda^{\gamma}(n_1,n_2)\langle  n_1l_1-n_2l_2\rangle\left|V(n_1+1,n_2,p_1)-V(n_1,n_2,p_2)\right|<\infty,
 \end{align*} 
 For $h=2$, the hypothesis $(H_{3,1})$ gives
  \begin{align*}
& \max_{(l_1,l_2)}\sup_{(n_1,n_2)\in\mathbb{Z}^2}\Lambda^{\gamma}(n_1,n_2)\langle  n_1l_1-n_2l_2\rangle\left|V(J_{1,2,1}(n_1,n_2,p_1))-V(J_{1,2,0}(n_1,n_2,p_2))\right|,\\
&=\max_{(l_1,l_2)}\sup_{(n_1,n_2)\in\mathbb{Z}^2}\Lambda^{\gamma}(n_1,n_2)\langle  n_1l_1-n_2l_2\rangle\left|V(n_1,n_2-1,p_1)-V(n_1,n_2,p_2)\right|<\infty,
 \end{align*} 
 Similarly for $G\equiv \eta$. 
The hypotheses on $\varepsilon$ correspond to the following cases:
If $k=2$,
we take $G\equiv \varepsilon$. For $h=1$, the hypothesis $(H_{3,2})$ gives 
 \begin{align*}
& \max_{(l_1,l_2)}\sup_{(n_1,n_2)\in\mathbb{Z}^2}\Lambda^{\gamma}(n_1,n_2)\langle  n_1l_1-n_2l_2\rangle\left|\varepsilon(J_{2,1,1}(n_1,n_2,p_1))-\varepsilon(J_{2,1,0}(n_1,n_2,p_2))\right|,\\
&=\max_{(l_1,l_2)}\sup_{(n_1,n_2)\in\mathbb{Z}^2}\Lambda^{\gamma}(n_1,n_2)\langle  n_1l_1-n_2l_2\rangle|\varepsilon((n_1+1,n_2,p_1),(n_1,n_2,p_2))\\
&\hspace*{4cm}-\varepsilon((n_1,n_2,p_2),(n_1,n_2,p_1))|<\infty,
 \end{align*}
 For $h=2$, the hypothesis $(H_{3,2})$ gives
   \begin{align*}
&\max_{(l_1,l_2)}\sup_{(n_1,n_2)\in\mathbb{Z}^2}\Lambda^{\gamma}(n_1,n_2)\langle  n_1l_1-n_2l_2\rangle\left|\varepsilon(J_{2,2,1}(n_1,n_2,p_1))-\varepsilon(J_{2,2,0}(n_1,n_2,p_2))\right|,\\
&=\max_{(l_1,l_2)}\sup_{(n_1,n_2)\in\mathbb{Z}^2}\Lambda^{\gamma}(n_1,n_2)\langle  n_1l_1-n_2l_2\rangle|\varepsilon((n_1,n_2-1,p_1),(n_1,n_2,p_2))\\
&\hspace*{4cm}-\varepsilon((n_1,n_2,p_1),(n_1,n_2,p_2))|<\infty,
 \end{align*}
In point $1.$ we only need hypotheses $(H_0),\ (H_1)$ and $(H_2)$. Points $2.$$-5.$ are typical results derived from Mourre's theory. Specifically, we establish a Mourre estimate and verify the regularity assumptions. This is the most technical part of our analysis. For historical context and an introduction to the topic, refer to Section \ref{s_mourre}. Point $4.$ is called a \emph{Limiting Absorption Principle}, which implies that the spectrum is purely absolutely continuous above $\mathbb{R}\setminus\left(\kappa\cup \sigma_{\rm p}(H_{m,\Ec})\right)$. Particularly, Riemann-Lebesgue Theorem ensures that the solution of the Schr{\"o}dinger equation dissipates at infinity.
That is, for $f$ belonging to the absolutely continuous subspace of $H_{m,\Ec}$ and $n\in\widetilde{\Vc},$ \begin{align}\label{1_}\lim_{|t|\rightarrow\infty}\left(e^{\rmi tH_{m,\Ec}}f\right)(n)=0.\end{align}
This can be interpreted as the particle moving off to infinity, while point $5.$ suggests that the particle focuses where $\Lambda_H^{\gamma}$ is large. Point $5.$ corresponds to the observation that $\Lambda_H^{\gamma}$ is locally $H$-regular over $[a,b]$, for example, \cite[Section VIII.C]{RS}.

Thanks to the concrete framework of this work, we can explicitly define the set of the critical points $\kappa$. It corresponds to the energy where, after Fourier transform, the symbol of $\Delta_H$ is zero, at this energies see Lemma \ref{Fourier}. Intuitively, there is no propagation, see Lemma \ref{c}. In \cite{PR}, the authors use a general and abstract Floquet Bloch approach. This ensures the existence of critical points by using the direct integral decomposition, see also \cite{GN} for a general theory. They do not give this set explicitly.

We now outline the structure of our paper. Section \ref{s_mourre} provides a brief overview of Mourre theory, and Subsection \ref{LTL} analyzes the model and establishs the Mourre estimate for the Laplacian operating on a hexagonal lattice. Moreover, \ref{section7} introduces a perturbation to the metric and incorporates the potential. Lastly, Subsection \ref{P.result}, we proves Theorem \ref{t_LAP}.
\section{The Mourre theory}\label{s_mourre}
In 1956, C.R. Putnam introduced a criterion ensuring that the spectrum of a self-adjoint operator $H$ is purely absolutely continuous, based on the existence of a bounded, self-adjoint operator $B$ satisfying $[H,\rmi B]>0$. However, the boundedness of
$B$ presents a significant limitation for practical applications. The Mourre Theory has attracted significant interest since its introduction in 1980 (cf., \cite{Mo81,Mo83}). Numerous works have proved the importance of Mourre's commutator theory for understanding the point and continuous spectra of a sufficiently large class of self-adjoint operators. Notable contributions include \cite{GGM1,GG,JMP,S}, as well as the book \cite{ABG}, the master courses \cite{G}, and more recent results such as \cite{GJ,Ge,GoMa}.

Now, we recall
Mourre's commutator theory. Let $H$ and $A$ be two self-adjoint operators acting on a complex Hilbert
space $\Hc$. Suppose also $H\in\Bc(\Hc)$. We denote by $\|\cdot\|$ the norm of bounded operators on $\Hc$. Thanks to the operator $A$, we
study several spectral properties of $H$.
Given $k\in \N$,  we say that $H\in \Cc^k(A)$ if for all $f\in \Hc$ the
map $\R\ni t\mapsto e^{\rm i t A}H e^{-\rm i t A}f$ has the usual $\Cc^k(\mathbb{R})$ regularity, i.e. $\R\ni t\mapsto e^{\rm i t A}H e^{-\rm i t A}\in \Bc(\Hc)$ has the usual $\Cc^k(\mathbb{R})$ regularity with  $\Bc(\Hc)$ endowed with the strong operator topology.
We say that $H\in \Cc^{k,u}(A)$ if the map $\R\ni t\mapsto e^{\rm i t A}H e^{-\rm i t A}\in\Bc(\Hc)$ has the usual $\Cc^k(\mathbb{R})$ regularity, with $\Bc(\Hc)$ endowed with the norm operator topology.
The form $[H,\rmi A]$ is defined on $\Dc(A)\times\Dc(A)$ by $\langle f,[H,\rmi A]g\rangle:=\rmi\left(\langle H^{*}f,Ag\rangle+\langle Af,Hg\rangle\right)$. By \cite[Lemma 6.2.9]{ABG} $H\in\Cc^1(A)$ if and only if the form $[H,\rmi A]$ extends to a bounded operator in which case we denote by $[H,\rmi A]_{\circ}$.
We say that $H\in\Cc^{0,1}(A)$ if
\[\int^1_0\| [H,e^{\rm i t A}]\| \frac{dt}{t}<\infty\]
and that $H\in\Cc^{1,1}(A)$ if
\[\int^1_0\| [[H,e^{\rm i t A}],e^{\rm i t A}]\| \frac{dt}{t^2}<\infty.\]
Thanks to \cite[p. 205]{ABG}, we have the following of vector spaces inclusions:
\begin{align}\label{classe}\Cc^2(A)\subset\Cc^{1,1}(A)\subset\Cc^{1,u}(A)\subset\Cc^1(A)\subset\Cc^{0,1}(A).\end{align}

The \emph{Mourre estimate} for $H$ on an open interval $\Ic$ of $\R$ holds if there exist $c>0$ and a compact operator $K$ such that:
\begin{align}\label{eqmourre}
E_\Ic (H)[H, \rm i A]_{\circ}E_\Ic(H)\geq E_\Ic(H)\, (c\, +\, K)\, E_\Ic(H),
\end{align}
where $E_{\Ic}(H)$ the spectral measure of $H$ above $\Ic$.
The aim of  Mourre's commutator theory is to
prove a \emph{Limiting Absorption Principle} (LAP), see \cite[Theorem 7.6.8]{ABG}.
\begin{theorem}Let $H$ be a self-adjoint operator, with $\sigma(H)\neq \R$. Assume that $H\in\Cc^1(A)$ and the Mourre estimate \eqref{eqmourre} true for $H$ in $\Ic$. Then:
\item[1.] If $K=0$, then $H$ has no eigenvalues in $\Ic$.
\item[2.] The number of eigenvalues of $H$ in $\Ic$ counted with multiplicity is finite.
\item[3.] If $H\in\Cc^{1,1}(A)$, $\gamma>\frac{1}{2}$ and $\Ic'$ a compact sub-interval of $\Ic$ containing no eigenvalue, then \[\sup_{\Re(z)\in\Ic',\Im(z)\neq0}\|\langle A\rangle^{-\gamma}(H-z)^{-1}\langle A\rangle^{-\gamma}\| \mbox{ is finite}.\]
\item[4.] In the norm topology of bounded operators, the boundary values of the resolvent:
\[\Ic'\ni\lambda\mapsto\lim_{\rho\to0^{\pm}}\langle A\rangle^{-\gamma}(H-\lambda-\rmi\mu)^{-1}\langle A\rangle^{-\gamma} \mbox{ exists and is continuous}.\]
\end{theorem}
For further details, we refer to \cite[Proposition 7.2.10, Corollary 7.2.11, Theorem 7.5.2]{ABG}.
\section{Proof of main result}\label{section3}
We now prove Theorem \ref{t_LAP}. Subsection \ref{LTL} studies the Laplacian on a hexagonal lattice and proves its Mourre estimate. In Subsection \ref{section7} the metrics are perturbed and the potential is added. Finally in Subsection \ref{P.result}, we prove Theorem \ref{t_LAP}.
\subsection{Laplacian on the hexagonal lattice}\label{LTL}
Let, \begin{align*}&\Sc:=\Bigg\{f:\mathbb{Z}^2\rightarrow \mathbb{C}^2\mbox{ such that, for all }N\in\mathbb{N}\ \sup_{(n_1,n_2)\in\mathbb{Z}^2}\left|(1+n_1^2+n_2^2)^Nf(n_1,n_2)\right|_{\mathbb{C}^2}\\
&\hspace*{6cm}<\infty \Bigg\},\end{align*} it is the discrete Schwartz space.
We denote by $\Cc^{\infty}_{2\pi}([-\pi,\pi]^2;\mathbb{C}^2)$ the set of functions defined on $[-\pi,\pi]^2$ of class $\Cc^{\infty}$ and $2\pi$-periodic.
Now, we define the Fourier transform $\Fc:\ell^2(\mathbb{Z}^2\times\{p_1,p_2\};\mathbb{C})\longrightarrow L^2([-\pi,\pi]^2\times\{p_1,p_2\};\mathbb{C})$ through
\[\Fc f(x,p_i):=\frac{1}{2\pi}\sum_n f(n,p_i)e^{-\rmi\langle n,x\rangle},\ \forall f\in\ell^2(\mathbb{Z}^2\times\{p_1,p_2\};\mathbb{C}). \]
The inverse Fourier transform $\Fc^{-1} :(L^2([-\pi,\pi]^2\times\{p_1,p_2\};\mathbb{C})\longrightarrow\ell^2(\mathbb{Z}^2\times\{p_1,p_2\};\mathbb{C})$ is
then given by
\[\Fc^{-1} f(n,p_i)=\frac{1}{2\pi}\int_{[-\pi,\pi]^2} f(x,p_i)e^{\rmi\langle n,x\rangle}dx,\ \forall f\in L^2([-\pi,\pi]^2\times\{p_1,p_2\};\mathbb{C}),\]
where $i\in\{1,2\}.$
By \eqref{iden}, we rewrite the Fourier transform $\Fc:\ell^2(\mathbb{Z}^2;\mathbb{C}^2)\longrightarrow L^2([-\pi,\pi]^2;\mathbb{C}^2)$ through
\[\Fc f(x)=\frac{1}{2\pi}\sum_n f(n)e^{-\rmi\langle n,x\rangle},\ \forall f\in\ell^2(\mathbb{Z}^2;\mathbb{C}^2). \]
The inverse Fourier transform $\Fc^{-1} :L^2([-\pi,\pi]^2;\mathbb{C}^2)\longrightarrow\ell^2(\mathbb{Z}^2;\mathbb{C}^2)$ is
then given by
\[\Fc^{-1} f(n)=\frac{1}{2\pi}\int_{[-\pi,\pi]^2} f(x)e^{\rmi\langle n,x\rangle}dx,\ \forall f\in L^2([-\pi,\pi]^2;\mathbb{C}^2).\]

\begin{remark}Note that $\Fc(\Sc)=\Cc_{2\pi}^{\infty}([-\pi,\pi]^2;\mathbb{C}^2)$.\end{remark}
First, we rewrite the Laplacian on a hexagonal lattice.
\begin{lemma}\label{Laplacian}
With the \eqref{iden}, we have:
\begin{align*}\Delta_{H}&=
\frac{1}{3}
\left(
  \begin{array}{cc}
    0 & \Delta_{2,H} \\
    \Delta_{1,H} & 0 \\
  \end{array}
\right)
=\frac{1}{3}
\left(
  \begin{array}{cc}
    0 & \rmid+U_1+ U_2 \\
    \rmid+U^*_1+ U^*_2 & 0 \\
  \end{array}
\right).
\end{align*}
\end{lemma}
\proof
Recalling \eqref{Na}, \eqref{Na1} and \eqref{map}. Let $f\in \ell^2(\mathbb{Z}^2;\mathbb{C}^2)$, we infer
\begin{align*}(\Delta_{H}f)(n)
&=\frac{1}{3}
\left(
  \begin{array}{cc}
     f_2(n-V_1)+f_2(n-V_2)+f_2(n-V_3) \\
    f_1(n-V_1)+f_1(n-V_2)+f_1(n-V_3) \\
  \end{array}
\right)\\
&=\frac{1}{3}
\left(
  \begin{array}{cc}
     f_2(n_1,n_2)+f_2(n_1-1,n_2)+f_2(n_1,n_2-1) \\
    f_1(n_1,n_2)+f_1(n_1+1,n_2)+f_1(n_1,n_2+1)  \\
  \end{array}
\right),\\
\end{align*}
where $V_1:=\frac{v_{1}+v_2}{3}\hbox{, }V_2:=\frac{v_{1}-2v_2}{3}\hbox{ and }V_3:=\frac{2v_{1}-v_2}{3}.$
This gives the result.
\qed

We turn to the Fourier transform.
\begin{lemma}\label{Fourier}
For all $f\in L^2([-\pi,\pi]^2;\mathbb{C}^2)$, we have:
\begin{align*}\left(\left(\Fc\Delta_{H} \Fc^{-1}\right)f\right)(x_1,x_2):=(F(Q_1,Q_2)f)(x_1,x_2)=F(x_1,x_2)f(x_1,x_2),\end{align*}
with
\begin{align*}
F(x_1,x_2):=\frac{1}{3}\left(
                   \begin{array}{cc}
                     0 & 1+e^{-\rmi x_1}+e^{-\rmi x_2} \\
                     1+e^{\rmi x_1}+e^{\rmi x_2} & 0 \\
                   \end{array}
                 \right).
\end{align*}
\end{lemma}
\proof
Let $f\in L^2([-\pi,\pi]^2;\mathbb{C}^2)$, we have:
\begin{align*}
\Fc(U_1\otimes1\ \Fc^{-1}f)(x)
=&\frac{1}{2\pi}\sum_{n}(U_1\Fc^{-1}f)(n)e^{-\rmi\langle x, n\rangle}\\
=&\frac{1}{2\pi}\sum_{n}(\Fc^{-1}f)(n_1-1,n_2)e^{-\rmi\langle x, n\rangle}\\
=&\frac{1}{2\pi}\sum_{n}(\Fc^{-1}f)(n_1,n_2)e^{-\rmi\langle x_1, n_1+1\rangle+\langle x_2,n_2\rangle}\\
=&\frac{1}{2\pi}\sum_{n}(\Fc^{-1}f)(n_1,n_2)e^{-\rmi x_1}e^{-\rmi\langle x, n\rangle}.
\end{align*}
Then,
$\Fc (U_1 \Fc^{-1}f)(x)=e^{-\rmi x_1}f(x)$. The other terms are treated in the same way. We obtain the result.
\qed

Now, we turn to the diagonalization of the matrix $F.$
\begin{lemma}\label{CP}
For all $(x_1,x_2)\in \Xi:= [-\pi,\pi]^2\backslash\left\{\left(-\frac{2\pi}{3},\frac{2\pi}{3}\right),\left(\frac{2\pi}{3},-\frac{2\pi}{3}\right)\right\},$ we have:\\
$F=PDP^{-1}$, with
$$D(x_1,x_2):=\left(
       \begin{array}{cc}
         \frac{\sqrt{\beta(x_1,x_2)}}{3} & 0 \\
         0 & -\frac{\sqrt{\beta(x_1,x_2)}}{3} \\
       \end{array}
     \right),$$
$$P(x_1,x_2):=\left(
    \begin{array}{cc}
      1 & 1 \\
      \frac{\sqrt{\beta(x_1,x_2)}}{1+e^{-\rmi x_1}+e^{-\rmi x_2}} & -\frac{1+e^{\rmi x_1}+e^{\rmi x_2}}{\sqrt{\beta(x_1,x_2)}} \\
    \end{array}
  \right),
$$
$$P^{-1}(x_1,x_2):=\left(
    \begin{array}{cc}
      \frac{1}{2} & \frac{1+e^{-\rmi x_1}+e^{-\rmi x_2}}{2\sqrt{\beta(x_1,x_2)}} \\
      \frac{1}{2} & -\frac{1+e^{-\rmi x_1}+e^{-\rmi x_2}}{2\sqrt{\beta(x_1,x_2)}} \\
    \end{array}
  \right)
$$
and $\beta(x):=|1+e^{\rmi x_1}+e^{\rmi x_2}|^2=3+2\left(\cos(x_1)+\cos(x_2)+\cos(x_1-x_2)\right).$\\
If $(x_1,x_2)\in\left\{\left(\frac{2\pi}{3},\frac{2\pi}{3}\right),\left(-\frac{2\pi}{3},-\frac{2\pi}{3}\right)\right\}$, then $$D(x_1,x_2)=F(x_1,x_2)=\left(
                                     \begin{array}{cc}
                                       0 & 0 \\
                                       0 & 0 \\
                                     \end{array}
                                   \right).$$
\proof
It suffices to verify that $\det F(x_1,x_2)\not=0$, thus $P$ is invertible and $\det(F-\lambda\rmid)=\lambda^2-\frac{\beta(x)}{9}.$
\qed
\end{lemma}
\begin{lemma}
On $\Xi$ the critical points of $\sqrt{\beta}$ is the set
\begin{align}\label{gamma}
\Gamma:=\left\{(-\pi,-\pi);\ (-\pi,\pi);\ (\pi,-\pi);\ (\pi,\pi);\ (-\pi,0);\ (\pi,0);\ (0,-\pi);\ (0,\pi);\ (0,0)\right\}
\end{align}
and $\sqrt{\beta\left(\Gamma\right)}=\left\{1,\ 3 \right \}$.
\end{lemma}
\proof  
We have $\sqrt{\beta}\in \Cc^{1}$ on $\Xi$. We obtain: 
\begin{equation}\label{1}\nabla \left(\sqrt{\beta}\right)(x_1,x_2)=\frac{\nabla\beta(x_1,x_2)}{2\sqrt{\beta}(x_1,x_2)}.\end{equation}
We introduce the functions:
\begin{equation*}
    X:=(x_1,x_2)\mapsto cos\left( \frac{x_1 + x_2}{2} \right), \quad Y:=(x_1,x_2)\mapsto \cos\left( \frac{x_1 - x_2}{2} \right).
\end{equation*}
Using trigonometric formulas, we obtain:
\begin{equation*}
    \beta =1+2\left( 2XY + 2Y^2\right).
\end{equation*}
Recall $\beta$ is a smooth on $\Xi$. We have:
\begin{equation*}
    \nabla \beta =4\left( Y \nabla X + (X + 2Y) \nabla Y\right),
\end{equation*}
where:
\begin{equation*}
    \nabla X(x_1,x_2) = -\frac{1}{2} \sin\left(\frac{x_1 + x_2}{2}\right) \left(
                                                                   \begin{array}{c}
                                                                     1 \\
                                                                     1 \\
                                                                   \end{array}
                                                                 \right)
    , \quad
    \nabla Y(x_1,x_2) = \frac{1}{2} \sin\left(\frac{x_1 - x_2}{2}\right)\left(
                                                                 \begin{array}{c}
                                                                   -1 \\
                                                                   1 \\
                                                                 \end{array}
                                                               \right)
     .
\end{equation*}
By \eqref{1}, $\nabla ( \sqrt{\beta})(x_1,x_2)=0$, then $\nabla \beta(x_1,x_2)=0$.
In particular, $\nabla X$ and $\nabla Y$ are orthogonal. Thus, they are independent vectors unless one of them is zero. Therefore:
\begin{align*}
   &\nabla \beta(x_1,x_2) = 0 \Longleftrightarrow 
        \left(\nabla X(x_1,x_2) = 0 = \nabla Y(x_1,x_2)\right) \\
        &\hspace*{3,2cm}\text{or } \left(\nabla X(x_1,x_2) = 0, \nabla Y(x_1,x_2) \neq 0, \text{ and } (X + 2Y)(x_1,x_2) = 0\right) \\
        &\hspace*{3,2cm}\text{or } \left(\nabla Y(x_1,x_2) = 0, \nabla X(x_1,x_2) \neq 0, \text{ and } Y (x_1,x_2)= 0\right) \\
        &\hspace*{3,2cm}\text{or } \left(\nabla X, \nabla Y(x_1,x_2) \text{ are independent, } Y(x_1,x_2) = 0, (X + 2Y)(x_1,x_2) = 0\right).
\end{align*}
We observe that we cannot simultaneously obtain that: $$Y(x_1,x_2) = 0\text{ and }\nabla Y (x_1,x_2)= 0.$$
In the case of independent $\nabla X(x_1,x_2)$ and $\nabla Y(x_1,x_2)$, we have $X(x_1,x_2) = Y(x_1,x_2) = 0$, hence $\sqrt{\beta(x_1,x_2)} = 1$. This actually occurs for $(x_1, x_2) \in\{(-\pi,0);\ (\pi,0);\ (0,-\pi);\ (0,\pi)\}$.\newline
In the case $\nabla X(x_1,x_2) = 0 = \nabla Y(x_1,x_2)$, we must have $|X(x_1,x_2)| = |Y(x_1,x_2)| = 1$. Then $\sqrt{\beta(x_1,x_2)} =3$ if $X(x_1,x_2)Y(x_1,x_2) > 0$ and $\sqrt{\beta(x_1,x_2)} = 1$ if $X(x_1,x_2)Y(x_1,x_2) < 0$. The first case occurs at $(x_1,x_2)\in\{(0,0);\ (-\pi,-\pi);\ (-\pi,\pi);\ (\pi,-\pi);\ (\pi,\pi)\}$.\newline
In the last case, we have $\nabla X(x_1,x_2) = 0$, $\nabla Y(x_1,x_2) \neq 0$, and $(X + 2Y)(x_1,x_2) = 0$. Since $\nabla Y(x_1,x_2) \neq 0$, $x_1 \neq x_2$ thus $-\pi < \frac{(x_1 + x_2)}{2} < \pi$, yielding $X(x_1,x_2) > -1$. Since $\nabla X(x_1,x_2) = 0$, $|X(x_1,x_2)| = 1$. Thus, $X(x_1,x_2) = 1$ and $Y(x_1,x_2) = -\frac{1}{2}$. Then $\sqrt{\beta(x_1,x_2)}= 0$
. This case occurs for $(x_1, x_2) = (-\frac{2\pi}{3}, \frac{2\pi}{3})$ and $(x_1, x_2) = (\frac{2\pi}{3}, -\frac{2\pi}{3})$, which does not belong $\Xi$.
We have shown that $\sqrt{\beta\left(\Gamma\right)}=\left\{ 1,\ 3 \right \}$.\qed\\

Next, we turn to the spectrum.
\begin{lemma} \label{specter}
$\sigma(F(Q_1,Q_2))=\left[-1,1\right].$
\end{lemma}
\proof
By \cite[Vol 1, p. 229]{RS} and Lemma \ref{CP}, we obtain $\sigma(F(Q_1,Q_2)):=\left\{\pm\frac{\sqrt{\beta(x)}}{3}, \forall x\in [-\pi,\pi]^2\right\}.$\newline
In the other hand, we have $\beta(x_1,x_2)=3+6\phi(x_1,x_2)$, where $$\phi(x_1,x_2):=\frac{1}{3}\left(\cos(x_1)+\cos(x_2)+\cos(x_1-x_2)\right).$$ By \cite[Lemma 3.3]{AEGJ}, we obtain that $\sigma(\phi(Q_1,Q_2))=\left[-\frac{1}{2},1\right]$. We conclude that $\sigma(F(Q_1,Q_2))=\left[-1,1\right].$

\qed

Let $g\in \Cc_{2\pi}^{\infty}\left([-\pi,\pi]^2;\mathbb{C}\right)$, we set:
\begin{align*}
\widehat{A}g:
&=\frac{\rmi}{2}\left(\nabla\beta^{\frac{5}{2}}\cdot\nabla+\nabla\cdot\nabla\beta^{\frac{5}{2}}\right)g\\
&=\frac{\rmi}{2}\left(\widehat{A_1}+\widehat{A_2}\right)g.
\end{align*}
\begin{remark}\label{RX}
The power $\frac{1}{2}$ should be the natural try but it does not seem to lead to a self-adjoint operator. Here we take power $\frac{5}{2}$ for two reasons: First it allows to obtain a self-adjoint operator and second, it can be explicitly set back to the original space which, in turn, permits to deal with the perturbation of the metric. Note that in \cite{T2} he chose to localize in energy away from $0$ to circumvent the difficulty. However, he does not deal with the metric perturbation. 
\end{remark}
\begin{lemma}\label{Operator}
For $g\in \Cc_{2\pi}^{\infty}([-\pi,\pi]^2;\mathbb{C})$, we have:
\begin{align*}
\widehat{A}g(x_1,x_2)
:=&5\rmi \beta^{\frac{3}{2}}(x_1,x_2)\Big(\left(-\sin(x_1)-\sin(x_1-x_2)\right)\frac{\partial g}{\partial x_1}(x_1,x_2)\\
&+
\left(-\sin(x_2)+\sin(x_1-x_2)\right)
\frac{\partial g}{\partial x_2}(x_1,x_2)
\Big)\\
&+15\rmi \sqrt{\beta(x_1,x_2)}\Big(\sin^2(x_1)+\sin^2(x_2)+2\sin^2(x_1-x_2)\\
&+2\sin(x_1-x_2)\left(\sin(x_1)-\sin(x_2)\right)\Big)f(x_1,x_2)\\
&-5\rmi \beta^{\frac{3}{2}}(x_1,x_2)\Big(\cos(x_1)+2\cos(x_1-x_2)+\cos(x_2)\Big)g(x_1,x_2).
\end{align*}
\end{lemma}
\proof
Let $g\in \Cc_{2\pi}^{\infty}\left([-\pi,\pi]^2;\mathbb{C}\right)$, we have:
\[\left(\begin{array}{cc}
                               \partial_{x_1}\beta^{\frac{5}{2}}\\
                               \partial_{x_2}\beta^{\frac{5}{2}} \\
                            \end{array}\right)(x_1,x_2)=5\left(\begin{array}{cc}
                              \beta^{\frac{3}{2}}(x_1,x_2)\left(-\sin(x_1)-\sin(x_1-x_2)\right) \\
                              \beta^{\frac{3}{2}}(x_1,x_2)\left(-\sin(x_2)+\sin(x_1-x_2)\right) \\
                            \end{array}\right)\]
and
\begin{align*}
\left\langle\left(\begin{array}{cc}
                               \partial_{x_1}\beta\\
                               \partial_{x_2}\beta \\
                            \end{array}\right),\left(
                     \begin{array}{c}
                       \partial_{x_1}g \\
                       \partial_{x_2}g \\
                     \end{array}
                  \right)\right\rangle(x_1,x_2)
=&5\beta^{\frac{3}{2}}(x_1,x_2)\Big(\left(-\sin(x_1)-\sin(x_1-x_2)\right)\frac{\partial g}{\partial x_1}(x_1,x_2)\\
&+\left(-\sin(x_2)+\sin(x_1-x_2)\right)
\frac{\partial g}{\partial x_2}(x_1,x_2)
\Big).\end{align*}
This concludes the Lemma.
\qed
\begin{lemma}\label{XXX}
We next denote $A_{\Fc}:=PA_DP^{-1}$ the conjugate operator associated to $F(Q)$, where
$$A_D:=\left(
         \begin{array}{cc}
           \widehat{A} & 0 \\
           0 & -\widehat{A} \\
         \end{array}
       \right).$$
       Let $f= {^t}(f_1,f_2)\in \Cc_{2\pi}^{\infty}\left([-\pi,\pi]^2;\mathbb{C}^2\right)$, we have
\item[1.]  If $(x_1,x_2)\in\Xi$, we obtain
\begin{align*}&A_{\Fc}f(x_1,x_2)
=\left(
         \begin{array}{cc}
           0 & \widehat{A_1}\frac{1+e^{-\rmi x_1}+e^{-\rmi x_2}}{\sqrt{\beta(x_1,x_2)}}+\widehat{A_2}\frac{1+e^{-\rmi x_1}+e^{-\rmi x_2}}{\sqrt{\beta(x_1,x_2)}} \\
          \frac{1+e^{\rmi x_1}+e^{\rmi x_2}}{\sqrt{\beta(x_1,x_2)}}\widehat{A_1}+\frac{1+e^{\rmi x_1}+e^{\rmi x_2}}{\sqrt{\beta(x_1,x_2)}}\widehat{A_2} & 0 \\
         \end{array}
       \right)\\
&\hspace*{3cm}\times \left(\begin{array}{cc}
          f_1 \\
          f_2 \\
        \end{array}\right)
(x_1,x_2)       .
\end{align*}
\item[2.] If $(x_1,x_2)\in \left\{(-\frac{2\pi}{3},\frac{2\pi}{3}),(\frac{2\pi}{3},-\frac{2\pi}{3})\right\}$, we have
$A_{\Fc}f$ is extendable by continuity in $(x_1,x_2)$
and
    $$A_{\Fc}f(x_1,x_2):=0.$$

\end{lemma}
\proof
\item[1.] After a straightforward calculation.
\item[2.] We have
\begin{align} \label{5}
\nonumber\frac{1+e^{\rmi x_1}+e^{\rmi x_2}}{\sqrt{\beta(x_1,x_2)}}\widehat{A_1}f(x_1,x_2)
\nonumber=&\frac{1+e^{\rmi x_1}+e^{\rmi x_2}}{\sqrt{\beta(x_1,x_2)}}\nabla\beta^{\frac{5}{2}}\cdot\nabla f(x_1,x_2)\\
=&5\left(1+e^{\rmi x_1}+e^{\rmi x_2}\right)^2\left(1+e^{-\rmi x_1}+e^{-\rmi x_2}\right)\\
\nonumber&\times\Bigg(-\left(\sin(x_1)+\sin(x_1-x_2)\right)\frac{\partial f}{\partial x_1}(x_1,x_2)\\
\nonumber&+\left(-\sin(x_2)+\sin(x_1-x_2)\right)\frac{\partial f}{\partial x_2}(x_1,x_2)\Bigg),
\end{align}
\begin{align} \label{55}
\nonumber\frac{1+e^{\rmi x_1}+e^{\rmi x_2}}{\sqrt{\beta(x_1,x_2)}}\widehat{A_2}f(x_1,x_2)
\nonumber=&\frac{1+e^{\rmi x_1}+e^{\rmi x_2}}{\sqrt{\beta(x_1,x_2)}}\nabla\cdot\nabla\beta^{\frac{5}{2}} f(x_1,x_2)\\
=&15\left(1+e^{\rmi x_1}+e^{\rmi x_2}\right)\Bigg(\sin^2(x_1)+\sin^2(x_2)+2\sin^2(x_1-x_2)\\
\nonumber&+2\sin(x_1-x_2)\left(\sin(x_1)-\sin(x_2)\right)\Bigg)f(x_1,x_2)\\
\nonumber&-5\left(1+e^{\rmi x_1}+e^{\rmi x_2}\right)^2\left(1+e^{-\rmi x_1}+e^{-\rmi x_2}\right)\\
\nonumber&\times\Bigg(\cos(x_1)+2\cos(x_1-x_2)
+\cos(x_2)\Bigg)f(x_1,x_2).
\end{align}
By \eqref{5} and \eqref{55}, we have
\begin{align*}&\lim_{(x_1,x_2)\rightarrow\pm(\frac{2\pi}{3},-\frac{2\pi}{3})}\frac{1+e^{\rmi x_1}+e^{\rmi x_2}}{\sqrt{\beta(x_1,x_2)}}\widehat{A_1}f(x_1,x_2)\\
&=\lim_{(x_1,x_2)\rightarrow\pm(\frac{2\pi}{3},-\frac{2\pi}{3})}\frac{1+e^{\rmi x_1}+e^{\rmi x_2}}{\sqrt{\beta(x_1,x_2)}}\widehat{A_2}f(x_1,x_2)=0,\end{align*} then
 $A_{\Fc}f$ is extendable by continuity.
\qed

We now define the conjugate operator $A_H$ on $\Sc.$
We denote by $A_H:=\Fc A_{\Fc}\Fc^{-1}$, where $$A_H:=\left(
         \begin{array}{cc}
          0 & A_{2,H} \\
          A_{1,H}  & 0 \\
         \end{array}
       \right)$$ is the conjugate operator associated to $\Delta_{H}.$
\begin{lemma}\label{XXXX}
On $\Sc $, we have
\begin{align*}
A_{1,H}
=&\frac{5\rmi}{2}Q_1(1+U_1^*+U_2^*)^2(1+U_1+U_2)\left(U_1^*-U_1+U_1^*U_2-U_1U_2^*\right)\\
&+\frac{5\rmi}{2}Q_2(1+U_1^*+U_2^*)^2(1+U_1+U_2)\left(U_2^*-U_2-(U_1^*U_2-U_1U_2^*)\right)\\
&-\frac{5\rmi}{8}\Bigg(3(1+U_1^*+U_2^*)\Big((U_1^*-U_1)^2+(U_2^*-U_2)^2+2(U_1^*U_2-U_1U_2^*)^2\\
&+2(U_1^*U_2-U_1U_2^*)\left(U_1^*-U_1-(U_2^*-U_2)\right)\Big)\\
&+2(1+U_1^*+U_2^*)^2(1+U_1+U_2)\left(U_1^*+U_1+U_2^*+U_2+2(U_1^*U_2+U_1U_2^*)\right)\Bigg)
.\end{align*}
Similarly, we obtain $A_{2,H}$ as the adjoint operator of $A_{1,H}.$
\end{lemma}
On $\Sc$, we can rewrite $A_{1,H}$ as follows:
\begin{align*}
 A_{1,H}=&R_{0,0}(U_1,U_2)+\sum_{l_1,l_2 \in \{0,1\}}(Q_1l_1-Q_2l_2)R_{l_1l_2}(U_1,U_2)\\
 &=\sum_{i,j\in \{-3,-2,-1,0,1,2,3\}}\alpha_{i,j}^{0,0}U_1^iU_2^j+\sum_{l_1,l_2 \in \{0,1\}}(Q_1l_1-Q_2l_2)\\
 &\quad\times\sum_{i,j\in \{-3,-2,-1,0,1,2,3\}}\alpha_{i,j}^{l_1,l_2}U_1^iU_2^j.
\end{align*}
where
\begin{align*}
R_{0,0}(U_1,U_2):=&-\frac{5\rmi}{8}\Bigg(3(1+U_1^*+U_2^*)\Big((U_1^*-U_1)^2+(U_2^*-U_2)^2+2(U_1^*U_2-U_1U_2^*)^2\\
&\quad\quad\quad+2(U_1^*U_2-U_1U_2^*)\left(U_1^*-U_1-(U_2^*-U_2)\right)\Big)
+2(1+U_1^*+U_2^*)^2\\
&\times(1+U_1+U_2)\left(U_1^*+U_1+U_2^*+U_2+2(U_1^*U_2+U_1U_2^*)\right)\Bigg)\\
=&-\frac{5\rmi}{8}\Big(19(U_1^*)^2+5U_1^2+19(U_2^*)^2+5U_2^2+5U_1^*+5U_2^*+5(U_1^*)^3\\
&+5(U_2^*)^3+16U_1+16U_2+15(U_1^*)^2U_2^2+15U_1^2(U_2^*)^2+4U_1U_2\\
&+12U_1^2U_2^*+25U_1(U_2^*)^2+9U_1^*(U_2^*)^2+18U_1^*U_2^*+8(U_1^*)^3U_2^2\\
&+7(U_1^*)^3U_2+9(U_1^*)^2U_2^*+8(U_1)^2(U_2^*)^3+7U_1(U_2^*)^3+25(U_1^*)^2U_2\\
&+24U_1U_2^*+24U_1^*U_2+2\Big)\\
=&\sum_{i,j\in \{-3,-2,-1,0,1,2,3\}}\alpha_{i,j}^{0,0}U_1^iU_2^j,
\end{align*}
where
\begin{align}\label{un}\nonumber&\alpha_{0,0}^{0,0}=2,\ \alpha_{-2,0}^{0,0}=\alpha_{0,-2}^{0,0}=19,\ \alpha_{2,0}^{0,0}=\alpha_{0,2}^{0,0}=5,\ \alpha_{-3,0}^{0,0}=\alpha_{0,-3}^{0,0}=5,\\
\nonumber& \alpha_{0,1}^{0,0}=\alpha_{1,0}^{0,0}=16,\ \alpha_{-2,-1}^{0,0}=\alpha_{-1,-2}^{0,0}=9,\ \alpha_{1,-2}^{0,0}=\alpha_{-2,1}^{0,0}=25,\\
&\alpha_{1,-1}^{0,0}=\alpha_{-1,1}^{0,0}=24,\ \alpha_{-1,2}^{0,0}=\alpha_{2,-1}^{0,0}=12,\ \alpha_{-3,1}^{0,0}=\alpha_{1,-3}^{0,0}=7,\\
\nonumber&\alpha_{-2,2}^{0,0}=\alpha_{2,-2}^{0,0}=15,\ \alpha_{2,-3}^{0,0}=\alpha_{-3,2}^{0,0}=8,\ \alpha_{-1,-1}^{0,0}=18,\ \alpha_{1,1}^{0,0}=4\\
\nonumber&\alpha_{-1,0}^{0,0}=\alpha_{0,-1}^{0,0}=5.
\end{align}

\begin{align*}
R_{1,0}(U_1,U_2):=&\frac{5\rmi}{2}\Bigg(-4+4U_1^*-5U_1-U_1^2-2U_2+(U_1^*)^3+5(U_1^*)^2+(U_2^*)^2\\
&+(U_1^*)^3U_2+2(U_1^*)^2U_2^*+2(U_1^*)^2U_2+5U_1^*U_2^*+U_1^*(U_2^*)^2\\
&-5U_1U_2^*-U_1U_2-U_1(U_2^*)^2-2U_1^2U_2^*-U_1^2(U_2^*)^2\Bigg),\\
=&\frac{5\rmi}{2}\sum_{i,j\in \{-3,-2,-1,0,1,2,3\}}\alpha_{i,j}^{1,0}U_1^iU_2^j,
\end{align*}
with
\begin{align}\label{2}\nonumber&\alpha_{0,0}^{1,0}=-4,\ \alpha_{-3,0}^{1,0}=\alpha_{0,-2}^{1,0}=\alpha_{-3,1}^{1,0}=\alpha_{-1,-2}^{1,0}=1,\ \alpha_{2,0}^{1,0}=\alpha_{1,1}^{1,0}=\alpha_{1,-2}^{1,0}=\alpha_{2,-2}^{1,0}=-1,\\
&\alpha_{-2,-1}^{1,0}=\alpha_{-2,1}^{1,0}=2,\ \alpha_{0,1}^{1,0}=\alpha_{2,-1}^{1,0}=-2,\ \alpha_{-2,0}^{1,0}=\alpha_{-1,-1}^{1,0}=5,\ \alpha_{1,-1}^{1,0}=\alpha_{1,0}^{1,0}=-5,\\
\nonumber& \alpha_{-1,0}^{1,0}=4.
\end{align}
\begin{align*}
R_{0,1}(U_1,U_2):=&\frac{5\rmi}{2}\Bigg(-4+4U_2^*-5U_2-U_2^2+U_2^3+5(U_2^*)^2+(U_1^*)^2+U_1(U_2^*)^3\\
&+2U_1^*(U_2^*)^2+2U_1(U_2^*)^2+5U_1^*U_2^*+(U_1^*)^2U_2^*-5U_1^*U_2\\
&-U_1U_2-(U_1^*)^2U_2-2U_1^*U_2^2-(U_1^*)^2U_2^2\Bigg)\\
=&\frac{5\rmi}{2}\sum_{i,j\in \{-3,-2,-1,0,1,2,3\}}\alpha_{i,j}^{0,1}U_1^iU_2^j,
\end{align*}
with
\begin{align}\label{3}\nonumber&\alpha_{0,0}^{0,1}=-4,\ \alpha_{0,-3}^{0,1}=\alpha_{-2,0}^{0,1}=\alpha_{1,-3}^{0,1}=\alpha_{-2,-1}^{0,1}=1,\ \alpha_{0,2}^{0,1}=\alpha_{1,1}^{0,1}=\alpha_{-2,1}^{0,1}=\alpha_{-2,2}^{0,1}=-1,\\
&\alpha_{-1,-2}^{0,1}=\alpha_{1,-2}^{0,1}=2,\ \alpha_{-1,2}^{0,1}=\alpha_{1,0}^{0,1}=-2,\ \alpha_{0,-2}^{0,1}=\alpha_{-1,-1}^{0,1}=5,\ \alpha_{-1,1}^{0,1}=\alpha_{0,1}^{0,1}=-5,\\
\nonumber &\alpha_{0,-1}^{0,1}=4.
\end{align}
\begin{align*}
R_{1,1}(U_1,U_2):=&\frac{5\rmi}{2}\Bigg(4U_1^*+6U_1^*+6U_2^*+U_1+U_2+2(U_1^*)^2+2(U_2^*)^2+2U_1^*U_2^*\\
&+5U_1^*U_2+5U_1U_2^*+5(U_1^*)^2U_2+5U_1(U_2^*)^2+U_1^*U_2^2\\
&+U_1^2U_2^*+2(U_1^*)^2U_2^2+2U_1^2(U_2^*)^2+(U_1^*)^3U_2\\
&+U_1(U_2^*)^3+(U_1^*)^3U_2^2+U_1^2(U_2^*)^3\Bigg)\\
=&\frac{5\rmi}{2}\sum_{i,j\in \{-3,-2,-1,0,1,2,3\}}\alpha_{i,j}^{1,1}U_1^iU_2^j,
\end{align*}
with
\begin{align}\label{4}
\nonumber&\alpha_{-1,0}^{1,1}=4,\ \alpha_{0,1}^{1,1}=\alpha_{-1,2}^{1,1}=\alpha_{-3,1}^{1,1}=\alpha_{-3,2}^{1,1}=1,\ \alpha_{2,-2}^{1,1}=\alpha_{0,-2}^{1,1}=-2,\\
&\alpha_{1,0}^{1,1}=\alpha_{2,-3}^{1,1}=\alpha_{1,-3}^{1,1}=\alpha_{2,-1}^{1,1}=-1,\ \alpha_{-2,0}^{1,1}=\alpha_{-2,2}^{1,1}=2,\\
\nonumber& \alpha_{-1,1}^{1,1}=\alpha_{-2,1}^{1,1}=5,\ \alpha_{1,-1}^{1,1}=\alpha_{1,-2}^{1,1}=-5,\ \alpha_{0,-1}^{1,1}=-4.
\end{align}
Similarly, we rewrite $A_{2,H}.$
\begin{lemma}\label{Alambda}
There exists $C> 0$ such that for all $f\in\Sc$, we have: \[\left\|A_Hf\right\|^2\leq C\left\|\Lambda_H(Q) f\right\|^2.\]
\end{lemma}
\proof
Let $f\in\Sc$. Here, all constants are denoted by $C$ and are independent of $f$. We have:
\begin{align*}
\|A_{H}f\|^2
=&\|A_{1,H}f_1\|^2+\|A_{2,H}f_2\|^2.
\end{align*}
So, we obtain:
\begin{align*}
\|A_{1,H}f_1\|^2
=&\sum_{(n_1,n_2)\in \mathbb{Z}^2}\left|\left(R_{0,0}(U_1,U_2)+\sum_{l_1,l_2\in\{0,1\}}(Q_1l_1-Q_2l_2)R_{l_1,l_2}(U_1,U_2)\right) f_1(n_1,n_2)\right|^2\\
\leq&\sum_{(n_1,n_2)\in \mathbb{Z}^2}2 \left| R_{0,0}(U_1,U_2)f_1(n_1,n_2)\right|^2\\
&+2\left|\sum_{l_1,l_2\in\{0,1\}}(Q_1l_1-Q_2l_2)R_{l_1,l_2}(U_1,U_2) f_1(n_1,n_2)\right|^2\\
\leq& C\left(\|(Q_1^2+\frac{1}{2})^{\frac{1}{2}}f_1\|^2+\|(Q_2^2+\frac{1}{2})^{\frac{1}{2}}f_1\|^2+\|f_1\|^2\right)
\leq C\|\Lambda(Q) f_1\|^2.
\end{align*}
Similarly, we show that:\[\left\|A_{2,H}f_2\right\|^2\leq C\left\|\Lambda(Q) f_2\right\|^2.\] This concludes the result.\qed
\begin{remark}\label{Lambdavarepsilon}
By applying Lemma \ref{Alambda} and since $\|\langle A_H\rangle^0f\|^2\leq \|\Lambda^0 (Q)f\|^2, \text{ for all } f\in\Sc$, by real interpolation, e.g. \cite[Theorem 4.1.2,p.88]{BJLJ}, for all $\gamma\in [0,1]$, we remark that there is $C_{\gamma}$ such that \[\left\|\langle A_H\rangle^{\gamma}f\right\|^2\leq C_{\gamma}\left\|\Lambda_H^{\gamma} f\right\|^2, \text{ for all } f\in\Sc.\]
\end{remark}
\begin{lemma}\label{Nelson}
$A_H$ is essentially self-adjoint on $\Sc$. We write $A_H$ for its closure in the sequel.
\end{lemma}
\proof
First, we recall that 
$A_H$ is symmetric operator on $\Sc$. By Lemma \ref{Alambda}, there exists $C$ such that for all $f\in\Sc$, we have: \[\left\|A_Hf\right\|^2\leq C\left\|\Lambda(Q) f\right\|^2.\] By the Nelson's Lemma, e.g. \cite[Theorem X.37]{RS}, it suffices to prove 
 \[\exists C>0,\ \forall f\in\Sc,\ \Big|\langle f,[\Lambda(Q),A_H]f\rangle\Big|\leq C\big\|\Lambda^\frac{1}{2}(Q)f\big\|.\] to ensure to that $A_H$, defined on $\Sc$, extends to a self-adjoint operator.  
Let $f\in\Sc$. We denote all constants by $C$, we have:
\begin{align*}
\left|\langle f,[\Lambda_H(Q),A_H]f\rangle\right|
\leq& \left|\langle f_2,[\Lambda(Q),A_{1,H}]f_1\rangle\right|+\left|\langle f_1,[\Lambda(Q),A_{2,H}]f_2\rangle\right|\\
=&2\left|\langle f_2,[\Lambda(Q),A_{1,H}]f_1\rangle\right|.
\end{align*}
Then,
\begin{align*}
&\left|\langle f_2,[\Lambda(Q),A_{1,H}]f_1\rangle\right|\\
=&\Bigg|\sum_{(n_1,n_2)\in \mathbb{Z}^2}\overline{f_2(n_1,n_2)}\Bigg(\Lambda(n_1,n_2)\left(R_{0,0}(U_1,U_2)+\sum_{(l_1,l_2)\in\{0,1\}^2, (l_1,l_2)\neq(0,0)}(n_1l_1-n_2l_2)R_{l_1,l_2}(U_1,U_2)\right)\\
&-\left(R_{0,0}(U_1,U_2)+\sum_{(l_1,l_2)\in\{0,1\}^2, (l_1,l_2)\neq(0,0)}(n_1l_1-n_2l_2)R_{l_1,l_2}(U_1,U_2)\right)\Lambda(n_1,n_2)\Bigg)f_1(n_1,n_2)\Bigg|\\
=&\Bigg|\sum_{(n_1,n_2)\in \mathbb{Z}^2}\overline{f_2(n_1,n_2)}\Bigg(\sum_{i,j\in\{-3,-2,-1,0,1,2,3\}}\alpha_{i,j}^{0,0}\left(\Lambda(n_1,n_2)-\Lambda(n_1+i,n_2+j)\right)\\
&+\sum_{(l_1,l_2)\in\{0,1\}^2, (l_1,l_2)\neq(0,0)}(n_1l_1-n_2l_2)\\
&\times\sum_{i,j\in\{-3,-2,-1,0,1,2,3\}}\alpha_{i,j}^{l_1,l_2}\left(\Lambda(n_1,n_2)-\Lambda(n_1+i,n_2+j)\right)\Bigg)f_1(n_1+i,n_2+j)\Bigg|\\
=&\Bigg|\sum_{(n_1,n_2)\in \mathbb{Z}^2}\overline{f_2(n_1,n_2)}\Bigg(\sum_{i,j\in\{-3,-2,-1,0,1,2,3\}}\alpha_{i,j}^{0,0}\left(\frac{\Lambda^2(n_1,n_2)-\Lambda^2(n_1+i,n_2+j)}
{\Lambda(n_1,n_2)+\Lambda(n_1+i,n_2+j)}\right)\\
&+\sum_{(l_1,l_2)\in\{0,1\}^2, (l_1,l_2)\neq(0,0)}(n_1l_1-n_2l_2)\\
&\times\sum_{i,j\in\{-3,-2,-1,0,1,2,3\}}\alpha_{i,j}^{l_1,l_2}\left(\frac{\Lambda^2(n_1,n_2)-\Lambda^2(n_1+i,n_2+j)}{\Lambda(n_1,n_2)+\Lambda(n_1+i,n_2+j)}\right)\Bigg) f_1(n_1+i,n_2+j)\Bigg|\\
=&\Bigg|\sum_{(n_1,n_2)\in \mathbb{Z}^2}\overline{f_2(n_1,n_2)}\Bigg(\sum_{i,j\in\{-3,-2,-1,0,1,2,3\}}\alpha_{i,j}^{0,0}\left(\frac{-2i n_1-2jn_2+i^2+j^2}
{\Lambda(n_1,n_2)+\Lambda(n_1+i,n_2+j)}\right)\\
&+\sum_{(l_1,l_2)\in\{0,1\}^2, (l_1,l_2)\neq(0,0)}(n_1l_1-n_2l_2)\\
&\times\sum_{i,j\in\{-3,-2,-1,0,1,2,3\}}\alpha_{i,j}^{l_1,l_2}\left(\frac{-2i n_1-2jn_2+i^2+j^2}{\Lambda(n_1,n_2)+\Lambda(n_1+i,n_2+j)}\right)\Bigg) f_1(n_1+i,n_2+j)\Bigg|\\
\leq&\sum_{(n_1,n_2)\in \mathbb{Z}^2}|f_2(n_1,n_2)|\Bigg(\Bigg|\sum_{i,j\in\{-3,-2,-1,0,1,2,3\}}\alpha_{i,j}^{0,0}\left(\frac{-2i n_1-2jn_2+i^2+j^2}
{\Lambda(n_1,n_2)+\Lambda(n_1+i,n_2+j)}\right)f_1(n_1+i,n_2+j)\Bigg|\\
&+\Bigg|\sum_{(l_1,l_2)\in\{0,1\}^2, (l_1,l_2)\neq(0,0)}(n_1l_1-n_2l_2)\sum_{i,j\in\{-3,-2,-1,0,1,2,3\}}\alpha_{i,j}^{l_1,l_2}\left(\frac{-2i n_1-2jn_2+i^2+j^2}{\Lambda(n_1,n_2)+\Lambda(n_1+i,n_2+j)}\right)\\
&\quad\times f_1(n_1+i,n_2+j)\Bigg|\Bigg)\\
\leq& C \left(\|f_2(\cdot)\|\|f_1(\cdot)\|+\|\Lambda^\frac{1}{2}(Q)f_2(\cdot) \| \|\Lambda^\frac{1}{2}(Q)f_1(\cdot) \|\right), \text{ due to }\frac{-2i n_1-2jn_2+i^2+j^2}{\Lambda(n_1,n_2)+\Lambda(n_1+i,n_2+j)} \text{ is bounded. } \\
\leq& C \left(\|f_2(\cdot)\|^2+\|f_1(\cdot)\|^2+\|\Lambda^\frac{1}{2}(Q)f_2(\cdot)\|^2+ \|\Lambda^\frac{1}{2}(Q)f_1 (\cdot)\|^2\right)\\
\leq&C\|\Lambda_H^\frac{1}{2}(Q)f\|^2.
\end{align*}  
As $\Sc$ is a core for $\Lambda_H(Q)$, applying \cite[Theorem X.37]{RS}, we prove that: 
$$\left|\langle f,[\Lambda_H(Q),A_H]f\rangle\right|\leq C\left\|\Lambda_H^\frac{1}{2}(Q)f\right\|^2.$$
\qed
\begin{lemma}\label{FA}
On $\Cc^{\infty}_{2\pi}\left([-\pi,\pi]^2;\mathbb{C}^2\right)$,
we have:
\begin{align}\label{fa1H}A_{\Fc}:=\Fc A_{H}\Fc^{-1}.\end{align}
\end{lemma}
\proof
We have:
$$ \left( \Fc U_1^*\Fc^{-1} f\right)(x)=e^{\rmi x_1}f(x),\quad \left( \Fc U_2^*\Fc^{-1} f\right)(x)=e^{\rmi x_2}f(x),$$
\begin{align*}
\frac{1}{2\rmi}\left( \Fc U_1^*-U_1\Fc^{-1} f\right)(x)
=&\frac{1}{2\rmi}\left(e^{\rmi x_1}-e^{-\rmi x_1}\right)f(x)=\sin(x_1)f(x),
\end{align*}
\[\frac{1}{2\rmi}\left(\Fc\left(U_1 U_2^*-U_1^* U_2\right)\Fc^{-1}f\right)(x)=\sin(x_1-x_2)f(x)\]
and
\begin{align*}
\left(\Fc(-\rmi Q_i)\Fc^{-1} f\right)(x)
=\frac{\partial f}{\partial x_i}(x),\end{align*}
where $i
\in\{1,2\}.$
Similarly, we treat the other terms. So, we obtain the result.
\begin{remark}
Since $\Fc(\Sc)=\Cc_{2\pi}^{\infty}([-\pi,\pi]^2;\mathbb{C}^2)$, we now recall Lemma \ref{FA} by density we have, $A_{\Fc}$ is essentially self-adjoint on $\Cc_{2\pi}^{\infty}\left([-\pi,\pi]^2;\mathbb{C}^2\right)$ and we denote by  $A_{\Fc}$ its closure. Note that \eqref{fa1H} extends to the closure and $\Dc(A_{H})=\Fc^{-1}\Dc(A_{\Fc}).$
\end{remark}
\begin{lemma}\label{c}
We have, $\sqrt{\beta(Q)}\in \Cc^1(\widehat{A})$ and therefore $\Delta_{H}\in\Cc^1(A_H)$. Moreover,
\begin{align}\label{lab}
\left[\sqrt{\beta(Q)},\rmi \widehat{A}\right]_{\circ}
=\frac{5}{4}\Big\|\sqrt{\beta}\nabla \beta\Big\|^2(Q).
\end{align}
\end{lemma}
 \proof
For $f\in \Cc_{2\pi}^{\infty}([-\pi,\pi]^2;\mathbb{C})$, we have:
\begin{align*}
\left[\sqrt{\beta(Q)},\rmi \widehat{A}\right]f(x_1,x_2)
=&-\left[\sqrt{\beta(Q)},\nabla\cdot\nabla\beta^{\frac{5}{2}}(Q)\right]f(x_1,x_2)\\
=&-\left[\sqrt{\beta(Q)},\sum_{j=1}^{2}\partial_{x_j}\left(\partial_{x_j}\beta^{\frac{5}{2}}\right)(Q)\right]f(x_1,x_2)\\
=&-\frac{5}{2}\left[\sqrt{\beta(Q)},\sum_{j=1}^{2}\partial_{x_j}\left(\beta^{\frac{3}{2}}\partial_{x_j}\beta\right)(Q)\right]f(x_1,x_2)\\
=&-\frac{5}{2}\sum_{j=1}^{2}\sqrt{\beta(x_1,x_2)}\partial_{x_j}\left(\beta^{\frac{3}{2}}\partial_{x_j}\beta f\right)(x_1,x_2)\\
&+\partial_{x_j}\left(\beta^{2}\partial_{x_j}\beta f\right)(x_1,x_2)\\
=&\frac{5}{4}\beta(x_1,x_2)(\partial_{x_j}\beta)^2(x_1,x_2)f(x_1,x_2).
\end{align*}
As $\beta \nabla \beta \in L^{\infty}([-\pi,\pi]^2;\mathbb{C}) \hbox{ then there exists }c> 0 \hbox{ such that } \left\|\left[\sqrt{\beta(Q)},\rmi \widehat{A}\right]f\right\|\leq c\|f\|$. By density and thanks to \cite[Lemma 6.2.9]{ABG}, we obtain $\sqrt{\beta(Q)}\in \Cc^1(\widehat{A})$. Recalling $\Fc(\Sc)=\Cc_{2\pi}^{\infty}([-\pi,\pi]^2;\mathbb{C}^2)$, as the Fourier transform is unitary, we obtain $\left\|\left[\Delta_H,\rmi A_H\right]g\right\|\leq c\|g\|$ for all $g\in\Sc$ which ensures that $\Delta_H\in\Cc^1(A_H).$
\qed\\

We now focus on the Mourre estimate for the unperturbed Laplacian.
\begin{proposition}\label{Mourre}
We have $\sqrt{\beta(Q)}\in \Cc^1(\widehat{A})$. Moreover, let $\Ic$ be an open interval such that its closure is included in $[-1,1]\backslash\kappa$, where $\kappa:=\pm\frac{\sqrt{\beta(\Gamma)}}{3}\cup\{0\}$, with $\Gamma$ is given in \eqref{gamma},
there exists $c>0$ such that:
\begin{align}\label{eq_mourre}
E_\Ic (\sqrt{\beta(Q)})[\sqrt{\beta(Q)}, \rm i \widehat{A}]_{\circ}E_\Ic(\sqrt{\beta(Q)})\geq c E_\Ic(\sqrt{\beta(Q)}).
\end{align}\\
Equivalently, we have:
 \begin{align}\label{eq_mour}
E_\Ic (\Delta_{H})[\Delta_{H}, \rm i A_{H}]_{\circ}E_\Ic(\Delta_{H})\geq c E_\Ic(\Delta_{H}).
\end{align}
The set $\kappa$ is called the set of \emph{thresholds}.
\end{proposition}
\proof
We work in $L^2([-\pi,\pi]^2;\mathbb{C}).$ The $\Cc^1$ property is given in Lemma \ref{c}.
Let $\Ic$ be an open interval such that its closure is included in $[-1,1]\backslash\kappa$. Since $\Ic$ is bounded then by Bolzano-Weierstrass Theorem, its closure is compact. There exists $c>0$, such that for all $(x_1,x_2)\in \sqrt{\beta}^{-1}(\Ic)$, we have:
$\frac{5}{4}\Big\|\sqrt{\beta(Q)}\nabla \beta\Big\|^2(x_1,x_2)\geq c.$ Recalling $\left[\sqrt{\beta(Q)},\rmi \widehat{A}\right]_{\circ}
=\frac{5}{4}\Big\|\sqrt{\beta(Q)}\nabla \beta\Big\|^2(Q)$, by functional calculus, we have:
$$E_\Ic (\sqrt{\beta(Q)})[\sqrt{\beta(Q)}, \rm i \widehat{A}]_{\circ}E_\Ic(\sqrt{\beta(Q)})\geq c E_\Ic(\sqrt{\beta(Q)}).$$
Using \eqref{isomorphisme11}, \eqref{iden}, from \eqref{eq_mourre} and thanks to Lemma \ref{XXX} and Lemma \ref{XXXX}, we obtain \eqref{eq_mour} by going back to $\ell^2(\widetilde{V};\mathbb{C}).$ \qed
\begin{proposition}\label{C2}
We have $\sqrt{\beta(Q)}\in\Cc^2(\widehat{A})$ and therefore $\Delta_{H}\in\Cc^2(A_H).$
\end{proposition}
\proof
We denote the constant by $c$. By Lemma \ref{c} and for $f\in \Cc^{\infty}_{2\pi}([-\pi,\pi]^2)$, we obtain:
\begin{align*}
\left[\left[\sqrt{\beta(Q)},\rmi\widehat{A}\right]_{\circ},\rmi \widehat{A}\right]f
=\left[\frac{5}{4}\left\|\sqrt{\beta}\nabla \beta\right\|^2(Q),\rmi\widehat{A}\right]f(x_1,x_2).
\end{align*}
Hence,
\begin{align*}
&\left[\frac{5}{4}\left\|\sqrt{\beta}\nabla \beta\right\|^2(Q), \rmi 5\rmi \beta^{\frac{3}{2}}(Q)\left(-\sin(Q_1)-\sin(Q_1-Q_2)\right)\frac{\partial }{\partial x_1}\right]f(x_1,x_2)\\
=&\frac{25}{4} \beta^{\frac{3}{2}}(Q)\left(-\sin(x_1)-
\sin(x_1-x_2)\right)\frac{\partial\left\|\sqrt{\beta}\nabla \beta\right\|^2}{\partial x_1}(x_1,x_2)f(x_1,x_2).
\end{align*}
As $\frac{5\partial\left\|\sqrt{\beta}\nabla \beta\right\|^2}{4\partial x_1}\in L^{\infty}([-\pi,\pi]^2),$
we deduce:
\begin{align*}
\left\|\left[\frac{5}{4}\left\|\sqrt{\beta}\nabla \beta\right\|^2(Q),5\beta^{\frac{3}{2}}(Q)\left(-\sin(Q_1)-\sin(Q_1-Q_2)\right)\frac{\partial }{\partial x_1}\right]f\right\|^2\leq c\left\|f\right\|^2.
\end{align*}
Similarly, we treat the other terms. Hence, by density and \cite[Proposition 5.2.2]{ABG}. As in Lemma \ref{c}, we also obtain that $\Delta_H\in\Cc^2(A_H)$, this proves the lemma.
\qed
\subsection{The perturbed model}\label{section7}
In this subsection, we perturb the earlier case by modifying the metric and introducing a potential. We require both the modified metric and potential to be in some sense, small at infinity. We address some technical details and begin with the properties of $\Lambda_H.$
\begin{proposition}\label{p__hypothesis}
$\Lambda_H$ satisfies the following assertions:
\item[1.]$\Dc(\Lambda_H(Q))\subset \Dc(A_H)$.
\item[2.]There is $c > 0$ such that for all $r > 0,\ -\rmi r$ belongs to the resolvent set of
$\Lambda_H(Q)$ and $r\|(\Lambda_H (Q)+ \rmi r)^{-1}\|_{\Bc(\ell^2(\widetilde{\Vc},1))}\leq c.$
\item[3.]$t\rightarrow e^{\rmi t\Lambda_H(Q)}$ has a polynomial growth in $\ell^2(\widetilde{\Vc},1)$.
\item[4.] Given $\xi\in\Cc^{\infty}(\mathbb{R};\mathbb{R})$ such that $\xi(x)=0$ near $0$ and $1$ near infinity and
$T\in\Bc(\ell^2(\widetilde{\Vc},1))$ symmetric, if
\begin{align}\label{integral}
 \int_{1}^{\infty}\left\|\xi\left(\frac{\Lambda_H(Q)}{r}\right)T \right\|_{\Bc(\ell^2(\widetilde{\Vc},1))}dr<\infty,
\end{align}
then $T\in \Cc^{0,1}(A_H).$
\end{proposition}
\proof
See \cite[Proposition 3.14]{AEGJ} .\qed
\begin{corollary}\label{p_hypothesis}
With the notation of Proposition \ref{p__hypothesis}, let $\gamma\in (0, 1)$ and $T\in\Bc(\ell^2(\widetilde{\Vc},1))$ symmetric. Assume that
$$\langle\Lambda_H\rangle^{\gamma}T\in\Bc(\ell^2(\widetilde{\Vc},1)),$$
then $T\in \Cc^{0,1}(A_H).$ Moreover, if $T\in\Cc^1(A_H),\ T\in \Cc^{1,1}(A_H).$
\end{corollary}
\subsubsection{Unitary transformation}
The perturbation of the metric introduces a second Hilbert space. To address this difficulty, we apply the following transformation:
\begin{proposition}\label{send}Set the following map  \begin{align*} T_{1\rightarrow m}: \ell^2(\widetilde{\Vc},1;\mathbb{C})&\to\ell^2(\widetilde{\Vc},m;\mathbb{C})
 \\
f&\mapsto T_{1\rightarrow m}f(n):=\frac{1}{\sqrt{m(n)}}f(n).\end{align*} Then, the transformation $T_{1\rightarrow m}$ is unitary.
\end{proposition}
\proof See \cite[Proposition 3.17]{AEGJ}.
\qed

Recalling \eqref{dm} and the hypothesis $(H_0)$ and $(H_1)$.
We can now send $\Delta_{m,\Ec}$ in $\ell^2(\widetilde{\Vc},1)$ with the help of this unitary transformation. Namely,
let $\widetilde{\Delta}:= T^{-1}_{1\rightarrow m}\Delta_{m,\Ec}T_{1\rightarrow m}$. We have the following result:
\begin{proposition}\label{calcul}We have:
\begin{align*}\widetilde{\Delta}=\left(
                     \begin{array}{cc}
                       0 & \widetilde{\Delta_2} \\
                       \widetilde{\Delta_1} & 0 \\
                     \end{array}
                   \right),\end{align*}
where
 \begin{align}\label{2_transform}
\widetilde{\Delta_1}
=&\frac{1}{3\sqrt{m_2(Q_1,Q_2)}}
 \Bigg(\frac{\Ec((Q_1,Q_2,p_1),(Q_1,Q_2,p_2))}{\sqrt{m_1(Q_1,Q_2)}}
 +\frac{\Ec((Q_1,Q_2,p_1),(Q_1+1,Q_2,p_2))}{\sqrt{m_1(Q_1+1,Q_2)}}U^*_1\\
\nonumber &+\frac{\Ec((Q_1,Q_2,p_1),(Q_1,Q_2+1,p_2))}{\sqrt{m_1(Q_1,Q_2+1)}} U^*_2\Bigg).\end{align}
 and
 \begin{align}\label{1_transform}
\widetilde{\Delta_2}
=&\frac{1}{3\sqrt{m_1(Q_1,Q_2)}}
 \Bigg(\frac{\Ec((Q_1,Q_2,p_2),(Q_1,Q_2,p_1))}{\sqrt{m_2(Q_1,Q_2)}}
 +\frac{\Ec((Q_1,Q_2,p_2),(Q_1-1,Q_2,p_1))}{\sqrt{m_2(Q_1-1,Q_2)}}U_1\\
\nonumber& +\frac{\Ec((Q_1,Q_2,p_2),(Q_1,Q_2-1,p_1))}{\sqrt{m_2(Q_1,Q_2-1)}}U_2
 \Bigg)\end{align}
\end{proposition}
We establish the following expression for the perturbation:
\begin{proposition}\label{compact}
\item[1.]For all $i,j\in\{1,2\},\ i\neq j$, we have:
 \begin{align}\label{D} D_i:=\Delta_{i,H}-\widetilde{\Delta_i}: \ell^2(\Vc_i,1)&\to\ell^2(\Vc_j,1)
 \\
\nonumber f_1&\mapsto \frac{1}{3}\left(L_{i,H}+T_{i,H}+S_{i,H}\right)f_1(n),\end{align}
where 
$$\left(L_{i,H}f_i\right)(n):=\left(1-\frac{\Ec((Q_1,Q_2,p_i),(Q_1,Q_2,p_j))}{\sqrt{m_i(Q_1,Q_2)}\sqrt{m_j(Q_1,Q_2)}}\right)f_i(n),$$
$$\left(T_{i,H}f_1\right(n):=\left(1-\frac{\Ec((Q_1,Q_2,p_i),(Q_1-i+j,Q_2,p_j))}{\sqrt{m_i(Q_1-i+j,Q_2)}\sqrt{m_j(Q_1,Q_2)}}\right)U_1^{i-j}f_i(n)$$
and
$$\left(S_{i,H}f_i\right)(n):=\left(1-\frac{\Ec((Q_1,Q_2,p_i),(Q_1,Q_2-i+j,p_j))}{\sqrt{m_i(Q_1,Q_2-i+j)}\sqrt{m_j(Q_1,Q_2)}}\right)U_2^{i-j}f_i(n).$$
\item [2.]We have $D_i$ is a compact operator in $\ell^2(\Vc_i,1).$
\end{proposition}
\proof
\item[1.] This is by Proposition \ref{calcul} and straightforward calculation.
\item[2.]
We have 
\begin{align*}&\left(\Delta_{i,H}-\widetilde{\Delta_i}\right)f_i(n_1,n_2)\\
&:=\sum_{\sigma\in\{0,j-i\}} \sum_{(h,k)\in \{(0,j-1)\}^2\text{, }h+k=\sigma}\left(\left(1-\frac{\Ec((Q_1,Q_2,p_i),(Q_1+h,Q_2+k,p_j))}{\sqrt{m_i(Q_1+h,Q_2+k)}\sqrt{m_j(Q_1,Q_2)}}\right)U_1^{-h}U_2^{-k}f\right)(n_1,n_2)\\
&=\sum_{\sigma\in\{0,j-i\}}\sum_{(h,k)\in \{(0,j-1)\}^2} \Bigg(\Bigg(m_j(Q_1,Q_2)\left(\eta_i(Q_1+h,Q_2+k)-\varepsilon((Q_1,Q_2,p_i),(Q_1+h,Q_2+k,p_j))\right)\\
&+\Ec((Q_1,Q_2,p_i),(Q_1+h,Q_2+k,p_j))\left(\eta_j(Q_1,Q_2)
-\varepsilon((Q_1,Q_2,p_i),(Q_1+h,Q_2+k,p_j))\right)\\
&\times \frac{1}{\sqrt{m_i(Q_1+h,Q_2+k)}\sqrt{m_j(Q_1,Q_2)}}\\
&\times\frac{1}{\left(\sqrt{m_1(Q_1+h,Q_2+k)}\sqrt{m_2(Q_1,Q_2)}
+\Ec((Q_1,Q_2,p_i),(Q_1+h,Q_2+k,p_j))\right)}\Bigg) U_1^{-h}U_2^{-k}f_i\Bigg)(n_1,n_2).\end{align*}
By using the hypotheses $(H_0)$ and $(H_1)$, we have $\eta_1((n_1+h,n_2+k))-\varepsilon((n_1,n_2,p_i),(n_1+h,n_k+j,p_j))\longrightarrow 0$ if $(n_1,n_2)\longrightarrow\infty$ and $\eta_2(n_1,n_2)
-\varepsilon((n_1,n_2,p_i),(n_1+h,n_2+k,p_j))\longrightarrow 0$ if $(n_1,n_2)\longrightarrow\infty$ and $U_1^{-h}U_2^{-k}$ is bounded.
So have that the operator $\Delta_{i,H}-\widetilde{\Delta_i}$ is a finite sum of compact operators. This concludes the result. 
\qed

We have a first spectral and general result. 
\begin{proposition}\label{com}Let $\varepsilon$, $\eta$ and $V$ be three real-valued bounded functions satisfying respectively $(H_0),\ (H_1)$ and $(H_2)$. We have:
\item[1.]$H_{m,\Ec}$ is self-adjoint and bounded.
\item[2.]$\sigma_{\rm ess}(H_{m,\Ec})=\sigma_{\rm ess}(\Delta_{H})$.
\end{proposition}
\proof
\item[1.] By hypotheses $(H_2)$, we deduce that $V(Q)$ is compact. By using the fact that $\Delta_{H}$ is self-adjoint and thanks to \cite[Theorem XIII.14]{RS}, we conclude that $H_{m,\Ec}$ is self-adjoint.
\item[2.] Thanks to Proposition \ref{compact} and using the fact that $V(Q)$ is compact, we infer that $\sigma_{\rm ess}(\widetilde{\Delta}+V(Q))=\sigma_{\rm ess}(\Delta_{H})$.
The unitary transformation assures that $\sigma_{\rm ess}(H_{m,\Ec})=\sigma_{\rm ess}(\Delta_{H})$.\qed
\subsubsection{Perturbed potential}
We start focus on the potential $V$. The perturbation of the metric is more complicated. It will be treated in the following subsection. We start with same auxiliary and general results.  
\begin{remark}\label{new}
1. For all $k\in\mathbb{N}^*,\ J_{k,0,0}\equiv J_{k,1,0}\equiv J_{k,2,0}.$

2. If $h=0$ and $\forall k\in\mathbb{N},\ \forall b\in\mathbb{N}$, we have $J_{k,0,b}\equiv J_{k,0,0}.$ 
\end{remark}
\begin{lemma}\label{small2}
Let $G:(\mathbb{Z}^2\times\{p_1,p_2\})^k\longrightarrow\mathbb{C}$ satisfying the hypothesis $(H_{3,k})$, then we have
\begin{align*}
&(H_{4,k})\quad\max_{h\in\{1,2\}}\max_{(l_1,l_2)}\sup_{(n_1,n_2)\in\mathbb{Z}^2}\Lambda^{\gamma}(n_1,n_2)\langle  n_1l_1-n_2l_2\rangle|G(J_{k,h,0}(n_1,n_2,p_i))\\
&\hspace*{6cm}-G(J_{k,h,0}(n_1,n_2,p_j))|<\infty,
\end{align*}
where $k\in\mathbb{N}^*,\ \{i,j\}=\{1,2\}.$
\end{lemma}
\proof
Let $h\in\{1,2\},\ h'\in\{1,2\}\backslash\{h\}$. 
For all $l_1,l_2\in\{0,1\}$, we have:
\begin{align*}
&\Lambda^{\gamma}(n_1,n_2)\langle n_1l_1-n_2l_2\rangle\left|G(J_{k,h,0}(n_1,n_2,p_i))-G(J_{k,h,1}(n_1,n_2,p_i))\right|\\
&\leq \Lambda^{\gamma}(n_1,n_2)\langle n_1l_1-n_2l_2\rangle\Big( \left|G(J_{k,h,0}(n_1,n_2,p_i))-G(J_{k,h,1}(n_1,n_2,p_j))\right|\\
&+\left|G(J_{k,h,1}(n_1,n_2,p_j))-G(J_{k,h',1}(n_1,n_2,p_i))\right|+\left|G(J_{k,h',1}(n_1,n_2,p_i))-G(J_{k,h',0}(n_1,n_2,p_j))\right|\Big).
\end{align*}
Thanks to the hypothesis $(H_{3,k})$, we deduce the result.
\qed 
\begin{lemma}\label{small1}
Let $G:(\mathbb{Z}^2\times\{p_1,p_2\})^k\longrightarrow\mathbb{C}$ satisfying the hypothesis $(H_{3,k})$ and Lemma \ref{small2}, then we have
\begin{align*}
&(H_{5,k})\hspace*{0,1cm}\max_{i\in\{1,2\}}\max_{j\in\{1,2\}, j\neq i}\max_{h\in\{1,2\}}\max_{h'\in\{1,2\},h'\neq h}\max_{(l_1,l_2)}\sup_{(n_1,n_2)\in\mathbb{Z}^2}\Lambda^{\gamma}(n_1,n_2)\langle n_1l_1-n_2l_2\rangle\\
&\hspace*{2cm}\left(\left|G(J_{k,h,0}(n_1,n_2,p_i))-G(J_{k,h,1}(n_1-(-1)^j\delta_{h',1},n_2-(-1)^j\delta_{h',2},p_j))\right|\right)<\infty,
\end{align*}
where $k\in\mathbb{N}^*$ and $\{i,j\}\in\{1,2\}.$
\end{lemma}
\proof
Let $h\in\{1,2\}\text{ and } h'\in\{1,2\}\backslash\{h\}$. 
For all $l_1,l_2\in\{0,1\}$, we have:
\begin{align*}
&\Lambda^{\gamma}(n_1,n_2)\langle n_1l_1-n_2l_2\rangle\left(\left|G(J_{k,h,0}(n_1,n_2,p_i))-G(J_{k,h,1}(n_1-(-1)^j\delta_{h',1},n_2-(-1)^j\delta_{h',2},p_j))\right|\right)\\
&\hspace*{-0,5cm}\leq \Lambda^{\gamma}(n_1,n_2)\langle n_1l_1-n_2l_2\rangle\Big( \left|G(J_{k,h,0}(n_1,n_2,p_i))-G(J_{k,h,1}(n_1,n_2,p_j))\right|\\
&+\left|G(J_{k,h,1}(n_1,n_2,p_j))-G(J_{k,h,0}(n_1,n_2,p_i))\right|\\
&+\left|G(J_{k,h',0}(n_1,n_2,p_i))-G(J_{k,h',1}(n_1,n_2,p_j))\right|,\quad\text{ by Remark \ref{new} } \\
&+\left|G(J_{k,h,0}(n_1+(-1)^j\delta_{h',1},n_2-(-1)^j\delta_{h',2},p_j))-G(J_{k,h,0}(n_1+(-1)^i\delta_{h',1},n_2-(-1)^i\delta_{h',2},p_i))\right|\\
&+|G(J_{k,h,0}(n_1+(-1)^i\delta_{h',1},n_2-(-1)^j\delta_{h',2},p_i))
-G(J_{k,h,1}(n_1+(-1)^j\delta_{h',1},n_2-(-1)^j\delta_{h',2},p_j))|\Big).
\end{align*}
Thanks to the hypothesis $(H_{3,k})$ and Lemma \ref{small2}, we deduce the result.
\qed

We turn to the regularity of $V$ with respect to $A_H.$
\begin{proposition}\label{H'}
Let $V:\widetilde{\Vc}\rightarrow \mathbb{R}^2$ be a function. We assume that $V$ satisfying $(H_2)$ and $(H_{3,1})$ holds true, then $V(Q)\in \Cc^{1}(A_H)$ and $[V(Q),A_H]_{\circ}\in \Cc^{0,1}(A_H)$. In particular, we obtain $V(Q)\in \Cc^{1,1}(A_H).$
\end{proposition}
\proof
We show this proposition in two steps. First, we prove that $V(Q)\in \Cc^{1}(A_H)$. We establish there is $c>0$ such that
$$\left\|\left[V(Q),\rmi A_H\right]\right\|^2\leq c\|f\|^2,\ \forall f\in \Sc.$$
Second, we take $\gamma\in[0,1)$ such that $\gamma'<\gamma$. We prove that $[V(Q),\rmi A_H]_{\circ}\in\Cc^{0,1}(A_H)$. We show there exists $c_{\gamma'}>0$ such that
$$\left\|\Lambda^{\gamma}(Q)\left[V(Q),\rmi A_H\right]\right\|^2\leq c_{\gamma'}\|f\|^2,\ \forall f\in \Sc.$$
We have \begin{align*}\left[V(Q),\rmi A_{H}\right]&=\left(
                                        \begin{array}{cc}
                                          V_1 & 0 \\
                                          0 & V_2 \\
                                        \end{array}
                                      \right)
                                      \left(
                                        \begin{array}{cc}
                                          0 & \rmi A_{2,H} \\
                                          \rmi A_{1,H} & 0 \\
                                        \end{array}
                                      \right)
                                      -\left(
                                         \begin{array}{cc}
                                           0 & \rmi A_{2,H} \\
                                           \rmi A_{2,H} & 0 \\
                                         \end{array}
                                       \right)
                                       \left(
                                         \begin{array}{cc}
                                           V_1 & 0 \\
                                           0 & V_2 \\
                                         \end{array}
                                       \right)\\
                                       &=\left(
                                           \begin{array}{cc}
                                             0 & \rmi \left(V_1A_{2,H}-A_{2,H}V_2\right) \\
                                             \rmi \left(V_2A_{1,H}-A_{1,H}V_1\right) & 0 \\
                                           \end{array}
                                         \right).
                                       \end{align*}

We are going to treat the term in $V_2A_{1,H}-A_{1,H}V_1$.
We work on $\Sc$. We infer:
 \begin{align*}
&\left(V_2A_{1,H}-A_{1,H}V_1\right)f_1(n_1,n_2)\\
=&\Bigg(V_2\left(R_{0,0}(U_1,U_2)+\sum_{l_1,l_2\in \{0,1\}}(Q_1l_1-Q_2l_2)R_{l_1,l_2}(U_1,U_2)\right)\\
&-\left(R_{0,0}(U_1,U_2)+\sum_{l_1,l_2\in \{0,1\}}(Q_1l_1-Q_2l_2)R_{l_1,l_2}(U_1,U_2)\right)V_1\Bigg)f_1(n_1,n_2)\\
=&\sum_{i,j\in\{-3,-2,-1,0,1,2,3\}}\alpha_{i,j}^{0,0}\left(V_2(n_1,n_2)-V_1(n_1+i,n_2+j)\right)f_1(n_1+i,n_2+j)\\
&+\sum_{l_1,l_2\in\{0,1\}^2, (l_1,l_2)\neq(0,0)}(Q_1l_1-Q_2l_2)\\
&\times\sum_{i,j\in\{-3,-2,-1,0,1,2,3\}}\alpha_{i,j}^{l_1,l_2}\left(V_2(n_1,n_2)-V_1(n_1+i,n_2+j)\right) f_1(n_1+i,n_2+j)\\
=&\sum_{i,j\in\{-3,-2,-1,0,1,2,3\}}\alpha_{i,j}^{0,0}\left(V_2(n_1,n_2)-V_1(n_1,n_2)\right)f_1(n_1+i,n_2+j)\\
&+\sum_{i,j\in\{-3,-2,-1,0,1,2,3\}}\alpha_{i,j}^{0,0}\left(V_1(n_1,n_2)-V_1(n_1+i,n_2+j)\right)f_1(n_1+i,n_2+j)\\
&+\sum_{l_1,l_2\in\{0,1\}^2, (l_1,l_2)\neq(0,0)}(Q_1l_1-Q_2l_2)\\
&\times\sum_{i,j\in\{-3,-2,-1,0,1,2,3\}}\alpha_{i,j}^{l_1,l_2}\left(V_2(n_1,n_2)-V_1(n_1,n_2)\right) f_1(n_1+i,n_2+j)\\
&+\sum_{l_1,l_2\in\{0,1\}^2, (l_1,l_2)\neq(0,0)}(Q_1l_1-Q_2l_2)\\
&\times\sum_{i,j\in\{-3,-2,-1,0,1,2,3\}}\alpha_{i,j}^{l_1,l_2}\left(V_1(n_1,n_2)-V_1(n_1+i,n_2+j)\right)f_1(n_1+i,n_2+j)\\
=&\sum_{i,j\in\{-3,-2,-1,0,1,2,3\}}\alpha_{i,j}^{0,0}\left(V_2(n_1,n_2)-V_1(n_1,n_2)\right)f_1(n_1+i,n_2+j)\\
&+ \sum_{i,j\in\{-3,-2,-1,0,1,2,3\}}\alpha_{i,j}^{0,0}\Bigg(\sum_{h=0}^{|j|-1}\Big(V_2(n_1,n_2+sgn(j)h)-V_1(n_1,n_2+sgn(j)(h+1))\Big)\\
&+\sum_{h=0}^{|i|-1}V_2(n_1+sgn(i)h,n_2+j)-V_1(n_1+sgn(i)(h+1),n_2+j)\Bigg)f_1(n_1+i,n_2+j)\\
&+\sum_{l_1,l_2\in\{0,1\}^2, (l_1,l_2)\neq(0,0)}(Q_1l_1-Q_2l_2)\\
&\times\sum_{i,j\in\{-3,-2,-1,0,1,2,3\}}\alpha_{i,j}^{l_1,l_2}\left(V_2(n_1,n_2)-V_1(n_1,n_2)\right) f_1(n_1+i,n_2+j)\\
&+\sum_{l_1,l_2\in\{0,1\}^2, (l_1,l_2)\neq(0,0)}\sum_{i,j\in\{-3,-2,-1,0,1,2,3\}}\alpha_{i,j}^{l_1,l_2}(Q_1l_1-Q_2l_2)\\
&\quad\times\Bigg(\sum_{h=0}^{|i|-1}V_2(n_1+sgn(i)h,n_2)-V_1(n_1+sgn(i)(h+1),n_2)\\
&+\sum_{h=0}^{|j|-1}V_2(n_1+i,n_2+sgn(j)h)-V_1(n_1+i,n_2+sgn(j)(h+1))\Bigg)f_1(n_1+i,n_2+j).
 \end{align*}
We assume that $(H_2),\ (H_{3,1})$ and $(H_{4,1})$ are true. Let $f_1\in\Sc$, we obtain:
\begin{align}\label{part0}
&\nonumber\left\|\Lambda^{\gamma'}(Q_1,Q_2)\left(V_2A_{1,H}-A_{1,H}V_1\right)f_1\right\|\\
\nonumber\leq&\Bigg\|\Lambda^{\gamma'}(Q_1,Q_2) \sum_{i,j}\alpha_{i,j}^{0,0}\Big(V_2(Q_1,Q_2)-V_1(Q_1,Q_2)\Big)f_1\Bigg\|\\
\nonumber&+\Bigg\|\Lambda^{\gamma'}(Q_1,Q_2) \sum_{i,j}\alpha_{i,j}^{0,0}\sum_{h=0}^{|j|-1}\Big(V_2(Q_1,Q_2+sgn(j)h)-V_1(Q_1,Q_2+sgn(j)(h+1))\Big)f_1\Bigg\|\\
&+\Bigg\|\sum_{l_1,l_2\in\{0,1\}^2, (l_1,l_2)\neq(0,0)}\sum_{i,j}\alpha_{i,j}^{l_1,l_2}(Q_1l_1-Q_2l_2)\Bigg(V_2(Q_1,Q_2)-V_1(Q_1,Q_2)\Bigg)f_1\Bigg\|\\
\nonumber&+\Bigg\|\sum_{l_1,l_2\in\{0,1\}^2, (l_1,l_2)\neq(0,0)}\sum_{i,j}\alpha_{i,j}^{l_1,l_2}(Q_1l_1-Q_2l_2)\\
\nonumber&\quad\times\Bigg(\sum_{h=0}^{|i|-1}V_2(Q_1+sgn(i)h,Q_2)-V_1(Q_1+sgn(i)(h+1),Q_2)\Bigg)f_1\Bigg\|\\
\nonumber\leq& c_{\gamma'} \|f_1\|,
\end{align}
where $sgn(i):=1,  \hbox{ if } i\geq0, \text{ and } :=-1,  \hbox{ otherwise.}$
The other terms are treated similarly.
Then, by \cite[Lemma 6.2.9]{ABG}, we obtain $V(Q)\in \Cc^{1}(A_H)$. Now, to prove that $[V(Q),A_H]_{\circ}\in \Cc^{0,1}(A_H)$ it is enough to use \eqref{part0}, there is $c$ such that
$\left\|\Lambda^{\gamma'}(Q)\left[V(Q),\rmi A_H\right]_{\circ}f\right\|\leq c_{\gamma'}\|f\|,\ \forall f\in\Sc.$
Finally, thanks to Corollary \ref{p_hypothesis},
we apply \cite[Proposition 7.5.7]{ABG} and the result
follows.\qed
\subsubsection{Perturbed metric.}
We now move to the most technical section. We start with technical lemmata.
\begin{lemma}
Let $G:(\mathbb{Z}^2\times\{p_1,p_2\})^k\longrightarrow\mathbb{C}$ satisfying the hypothesis $(H_{3,k})$ and Lemma \ref{small2}, then for $k\in\mathbb{N}^*$, we have
\begin{align*} (H_{6,k})\quad\max_{h\in\{1,2\}}\max_{i\in\{1,2\}}\sup_{(n_1,n_2)\in\mathbb{Z}^2}\Lambda^{\gamma}(n_1,n_2)\langle n_1l_1-n_2l_2\rangle \left|G(J_{k,h,0}(n_1,n_2,p_i))-G(J_{k,h,1}(n_1,n_2,p_i))\right|<\infty.\end{align*}
\end{lemma}
\proof
Let $\{i,j\}\in\{1,2\},\ h\in\{1,2\}$. For all $l_1,l_2\in\{0,1\}$, we have:
\begin{align*}
&\Lambda^{\gamma}(n_1,n_2)\langle n_1l_1-n_2l_2\rangle\left|G(J_{k,h,0}(n_1,n_2,p_i))-G(J_{k,h,1}(n_1,n_2,p_i))\right|\\
&\leq \Lambda^{\gamma}(n_1,n_2)\langle n_1l_1-n_2l_2\rangle\Big( \left|G(J_{k,h,0}(n_1,n_2,p_i))-G(J_{k,h,0}(n_1,n_2,p_j))\right|\\
&+\left|G(J_{k,h,0}(n_1,n_2,p_j))-G(J_{k,h,1}(n_1,n_2,p_i))\right|\Big).
\end{align*}
Thanks to $(H_{3,k})$ and Lemma \ref{small2}, we obtain the result.
\qed
\begin{lemma}\label{small}
Let $k\in\mathbb{N}^*$. Let $G:(\mathbb{Z}^2\times\{p_1,p_2\})^k\longrightarrow\mathbb{C}$ satisfying the hypothesis $(H_{3,k})$ and  
\begin{align*}(H_{7,k})\quad \max_{h\in\{1,2\}}\max_{j\in\{1,2\}}G(J_{k,h,b}(n_1,n_2,p_j))\longrightarrow 0\text{ if }b\longrightarrow\infty,\end{align*}for all $ (n_1,n_2)\in\mathbb{Z}^2$. Then we have:
\begin{align*} (H_{8,k})\quad\max_{i\in\{1,2\}}\sup_{(n_1,n_2)\in\mathbb{Z}^2}\Lambda^{\gamma}(n_1,n_2) \left|G(J_{k,h,0}(n_1,n_2,p_i))\right|<\infty.\end{align*}
\end{lemma}
\proof
 We fix $i\in\{1,2\}$. We suppose that $(H_{7,k})$ hold true, then for all $j\in\{1,2\}$, we have: 
\begin{align*}&\Lambda^{\gamma}(n_1,n_2)\left|G(J_{k,h,0}(n_1,n_2,p_i))\right|\\
&=\lim_{b\rightarrow +\infty}\Lambda^{\gamma}(n_1,n_2)\left|G(J_{k,h,0}(n_1,n_2,p_i))-G(J_{k,h,b}(n_1,n_2,p_j))\right|
\end{align*}
First of all, by using $(H_{3,k})$ and $(H_{6,k})$, for $i,j \in \{1,2\}$ and $\sigma\in\{0,1\}$, we have 
\begin{align*}
&\mathcal{M}_{k,h,\sigma,i,j} := \max_{(l_1,l_2)}\sup_{(n_1,n_2)\in\mathbb{Z}^2} \Lambda^\gamma(n_1,n_2) \langle n_1l_1-n_2l_2 \rangle 
\Big|G(J_{k,h,0}(n_1,n_2,p_i))\\
&\hspace*{4cm}-G(J_{k,h,\sigma}(n_1,n_2,p_j))\Big| < \infty.
\end{align*}
To get $(H_{8,k})$, it suffices to find an upper bound for the function $$(n_1,n_2) \mapsto \Lambda^\varepsilon(n_1,n_2)\left|G(J_{k,h,0}(n_1,n_2,p_i))\right|$$ on the set $\mathbb{Z}^2 \setminus \{(0,0)\}$.

\paragraph{First case:} $|n_1| \geq 1$ and $|n_1| \geq |n_2|$. Let $j \in \{1,2\}$ such that $(-1)^{j+1} n_1 = |n_1|$. For $b \in \mathbb{N}$, we have
\begin{align*}
f_b(n_1,n_2,p_i) &:= \Lambda^\gamma(n_1,n_2) \left| G(J_{k,1,b+1}(n_1,n_2,p_j))-G(J_{k,1,0}(n_1,n_2,p_i)) \right|
\\
&\leq \Lambda^\gamma(n_1,n_2) \Big(\left| G(J_{k,1,b+1}(n_1,n_2,p_j))-G(J_{k,1,0}(n_1,n_2,p_j)) \right|\\
&\quad+\left|G(J_{k,1,0}(n_1,n_2,p_j))-G(J_{k,1,0}(n_1,n_2,p_i)) \right|\Big)\\
&
\leq \Lambda^\gamma(n_1,n_2)\Big(\sum_{t=0}^b\left| G(J_{k,1,t+1}(n_1,n_2,p_j))-G(J_{k,1,t}(n_1,n_2,p_j)) \right|\\
&\quad+\left| G(J_{k,1,0}(n_1,n_2,p_j))-G(J_{k,1,0}(n_1,n_2,p_i)) \right|
 \Big)
\\
&= \Lambda^\gamma(n_1,n_2)\Big(\sum_{t=0}^b \left|G(n_1+(-1)^{j+1}(t+1),n_2,p_i) - G(n_1+(-1)^{j+1}t,n_2,p_i) \right|\\
&\quad+\left| G(J_{k,1,0}(n_1,n_2,p_j))-G(J_{k,1,0}(n_1,n_2,p_i)) \right|
 \Big)
\\
&
\leq \Lambda^\gamma(n_1,n_2)\Big(\sum_{t=0}^b \left|G(J_{k,1,1}(n_1+(-1)^{j+1}t,n_2,p_j)) - G(J_{k,1,0}(n_1+(-1)^{j+1}t,n_2,p_j)) \right|\\
&\quad+\mathcal{M}_{k,h,0,i,j} \Lambda^{-\gamma}(n_1 , n_2) \langle n_1  \rangle^{-1}
\Big),\ \text{by } (H_{4,k}), \text{ see Lemma \ref{small2}}.
\end{align*}

Changing the index in the last sum and using $|n_2| \leq |n_1|$, by $(H_{6,k})$,  we get:
\begin{align*}
f_b(n_1,n_2,p_i) &\leq \mathcal{M}_{k,h,0,i,i}\langle n_1  \rangle^{-1}+\mathcal{M}_{k,h,1,j,j} \Lambda^\gamma(n_1,n_2) \sum_{\ell = n_1}^{n_1+(b+1)(-1)^{j+1}} \Lambda^{-\gamma}(\ell,n_2) \langle \ell \rangle^{-1}
\\
&
\leq \mathcal{M} \left( 1 + 2^\gamma \langle n_1 \rangle^\gamma \sum_{\ell = n_1+(-1)^{j+1}}^{n_1+(b+1)(-1)^{j+1}} \langle \ell \rangle^{-1-\gamma}+\langle n_1  \rangle^{-1} \right)
\\
&
\leq \mathcal{M} \left( 3 + 2^\gamma \langle n_1 \rangle^\gamma \int_{|n_1|}^\infty \langle s\rangle^{-1-\gamma} ds \right)\leq
\mathcal{M} \left( 3 + 2^\gamma \langle n_1 \rangle^\gamma \int_{|n_1|}^\infty  s^{-1-\gamma} ds \right)\\
&\leq \mathcal{M} \left( 3 + \frac{2^\gamma}{\gamma} \cdot \frac{\langle n_1 \rangle^\gamma}{|n_1|^\gamma} \right) \leq  \mathcal{M} \left( 3 + \frac{ (2\sqrt{2})^{\gamma}}{\gamma
} \right),
\end{align*}
with $\mathcal{M}:=\max (\mathcal{M}_{k,h,0,i,j},\mathcal{M}_{k,h,1,j,j}).$
Since $G$ tends to 0 at infinity, we may take the limit $b \to \infty$
\[
\Lambda^\gamma(n_1,n_2) |G(J_{k,h,0}(n_1,n_2,p_i))| = \lim_{b \to \infty} f_b(n_1,n_2,p_i) \leq \mathcal{M} \left( 3 + \frac{ (2\sqrt{2})^{\gamma}}{\gamma} \right).
\]

\paragraph{Second case:} $|n_2| \geq 1$ and $|n_2| \geq |n_1|$. Take $j \in \{1,2\}$ such that $(-1)^{j+1} n_2 = |n_2|$.

\qed

Since explicitly writing out the expression for the commutator is extremely laborious (requiring thousands of elementary operations), we adopt a more systematic approach to simplify the process. In particular, we introduce an important lemma that serves as a computational algorithm in the following proof. The proof itself is provided in the appendix.
\begin{lemma}\label{remark}
We set: $$I_{l_1,l_2}^0:=\left\{(i,j)\in \mathbb{Z}^2,\text{ such that } \alpha_{i,j}^{l_1,l_2}\neq 0\right\},$$
$$I_{l_1,l_2}^1:=\left\{(i,j)\in I_{l_1,l_2}^0, \text{ such that } (i+1,j)\notin I_{l_1,l_2}^0\right\},$$
$$I_{l_1,l_2}^2:=\left\{(i,j)\in I_{l_1,l_2}^1, \text{ such that } (i,j+1)\notin I_{l_1,l_2}^0 \right\},$$
$$I_{l_1,l_2}^3:=\left\{(i,j)\in I_{l_1,l_2}^0, \text{ such that } (-i,-j)\notin I_{l_1,l_2}^0\right\},$$
$$I_{l_1,l_2}^4:=\left\{(i,j)\in I_{l_1,l_2}^0, \text{ such that } (-i-1,-j)\notin I_{l_1,l_2}^0\right\},$$
$$I_{l_1,l_2}^5:=\left\{(i,j)\in I_{l_1,l_2}^0, \text{ such that } (-i,-j-1)\notin I_{l_1,l_2}^0\right\},$$
$$I_{l_1,l_2}^6:=\left\{(i,j)\in I_{l_1,l_2}^2, \text{ such that } (i-1,j+1)\notin I_{l_1,l_2}^0\right\},$$
$$I_{l_1,l_2}^7:=\left\{(i,j)\in I_{l_1,l_2}^3, \text{ such that } (-i-1,-j)\notin I_{l_1,l_2}^0\text{ and } (-i,-j-1)\notin I_{l_1,l_2}^0\right\},$$
\begin{align*}&I_{l_1,l_2}^8:=\Big\{(i,j)\in I_{l_1,l_2}^4, \text{ such that } (-i-2,-j)\notin I_{l_1,l_2}^0\text{ and } (-i-1,-j-1)\notin I_{l_1,l_2}^0\Big\},\end{align*}
\begin{align*}&I_{l_1,l_2}^9:=\Big\{(i,j)\in I_{l_1,l_2}^5, \text{ such that } (-i-1,-j-1)\notin I_{l_1,l_2}^0, (-i,-j-2)\notin I_{l_1,l_2}^0, \\
&\hspace*{3cm} (i,j+1)\notin I_{l_1,l_2}^0\text{ and }(i+1,j+1)\notin I_{l_1,l_2}^0\Big\}.\end{align*}
  We denote
$$I_{l_1,l_2}^{3+\min(l_1,l_2)}:=\left\{(i,j)\in I_{l_1,l_2}^0, \text{ such that }
 (-i-\min(l_1,l_2),-j)\notin I_{l_1,l_2}^0\right\},$$  where
$I_{l_1,l_2}^{3+\min(l_1,l_2)}=I_{l_1,l_2}^3$ if $l_1,l_2\in \{(0,1),(1,0),(0,0)\}$ and $I_{l_1,l_2}^{3+\min(l_1,l_2)}=I_{l_1,l_2}^4$, if $l_1,l_2\in \{(1,1)\}.$

 \item[1.]Let $(i,j)\in I_{l_1,l_2}^{4-\min(l_1,l_2)}$, then at least one of the following conditions is satisfied :
 \begin{align}\label{Q4}
    &i.\ (-i-2+\min(l_1,l_2),-j)\in I_{l_1,l_2}^1,\quad ii.\ (-i-1+\min(l_1,l_2),-j-1)\in I_{l_1,l_2}^2,\\
      &\nonumber\hspace*{2cm} iii.\ (-i-\min(l_1,l_2)+1,j)\in I_{l_1,l_2}^{3+\min(l_1,l_2)}.\end{align}
      \item[2.]Let $(i,j)\in I_{l_1,l_2}^5$, then at least one of the following conditions is satisfied: \begin{align}\label{Q5} i.\ (-i-1,-j-1)\in I^1_{l_1,l_2}\quad ii.\ (-i,-j-2)\in I_{l_1,l_2}^2\quad iii.\ (-i-1,-j+1)\in I_{l_1,l_2}^4.\end{align}
We denote \begin{align*}&I_{l_1,l_2}^{7+\min(l_1,l_2)}:=\Big\{(i,j)\in I_{l_1,l_2}^{3+\min(l_1,l_2)}, \text{ such that } (i-1-\min(l_1,l_2),-j)\notin I_{l_1,l_2}^1\\
&\hspace*{4cm}\text{ and }(i-\min(l_1,l_2),-j-1)\notin I_{l_1,l_2}^2\Big\},\end{align*}
where
 $I_{l_1,l_2}^{7+\min(l_1,l_2)}=I_{l_1,l_2}^7$, if $(l_1,l_2)\in\left\{(0,1),(1,0),(0,0)\right\}$
and
  $I_{l_1,l_2}^{7+\min(l_1,l_2)}=I_{l_1,l_2}^8$, if $(l_1,l_2)\in\left\{(1,1)\right\}.$
\item[3.]We have $I^9_{l_1,l_2}=\emptyset$ and $I_{l_1,l_2}^{7+\min(l_1,l_2)}\subset I_{l_1,l_2}^1\cup I_{l_1,l_2}^2\cup I_{l_1,l_2}^{8-\min(l_1,l_2)}.$ \\
\item[4.] For all $(i,j)\in I_{l_1,l_2}^0$ we have: \begin{align}\label{Q1}\alpha _{i,j}^{l_1,l_2}+\alpha _{i-1,j}^{l_1,l_2}+\alpha _{i,j-1}^{l_1,l_2}=(-1)^{\max(l_1,l_2)}\left(\alpha _{-i,-j}^{l_1,l_2}+\alpha _{-i-1,-j}^{l_1,l_2}+\alpha _{-i,-j-1}^{l_1,l_2}\right).\end{align}
\item[5.]For all $(i,j)\in I_{l_1,l_2}^1$ we obtain: \begin{align}\label{Q2}\alpha _{i,j}^{l_1,l_2}+\alpha _{i+1,j}^{l_1,l_2}+\alpha _{i+1,j-1}^{l_1,l_2}=(-1)^{\max(l_1,l_2)}\left(\alpha _{-i-1,-j}^{l_1,l_2}
    +\alpha _{-i-2,-j}^{l_1,l_2}+\alpha _{-i-1,-j-1}^{l_1,l_2}\right).\end{align}
\item[6.]For all $(i,j)\in I_{l_1,l_2}^2$ we get: \begin{align}\label{Q3}\alpha _{i,j}^{l_1,l_2}+\alpha _{i,j+1}^{l_1,l_2}+\alpha _{i-1,j+1}^{l_1,l_2}=(-1)^{\max(l_1,l_2)}\left(\alpha _{-i,-j-1}^{l_1,l_2}+\alpha _{-i-1,-j-1}^{l_1,l_2}+\alpha _{-i,-j-2}^{l_1,l_2}\right).\end{align}
  \item[7.]For all $(i,j)\in I_{l_1,l_2}^4$ we get: \begin{align}\label{Q9}\alpha _{i,j}^{l_1,l_2}+\alpha _{i+1,j}^{l_1,l_2}+\alpha _{i+1,j-1}^{l_1,l_2}=(-1)^{\max(l_1,l_2)}\left(\alpha _{-i-1,-j}^{l_1,l_2}+\alpha _{-i-2,-j}^{l_1,l_2}+\alpha _{-i-1,-j-1}^{l_1,l_2}\right).\end{align}
\end{lemma}
We turn to the regularity of metric with respect of $A_H.$
\begin{proposition}\label{H}
We assume that $(H_0)$ and $(H_1)$ hold true, $\eta$ satisfy $(H_{3,1})$ and $\varepsilon$ satisfy $(H_{3,2})$, we have $\widetilde{\Delta}\in \Cc^{1}(A_H)$. Moreover $[\widetilde{\Delta},\rmi A_H]_{\circ}\in\Cc^{0,1}(A_H)$. In particular, $\widetilde{\Delta}\in \Cc^{1,1}(A_H).$
\end{proposition}
\proof
 We prove this proposition in two steps. In the first step, we will establish that $\widetilde{\Delta}\in \Cc^{1}(A_H)$. We show there is $c$, such that:
\begin{equation}\label{eq1}
\left\|\left[\Delta_H-\widetilde{\Delta}, \rmi A_{H}\right]f\right\|^2\leq c \|f\|^2,\ \forall f\in \Sc.
\end{equation}
Hence, by density and thanks to Proposition \ref{C2} and \cite[Proposition 6.2.9]{ABG}, we obtain that $\widetilde{\Delta}\in \Cc^{1}(A)$.
In the second step, we aim to show that $[\widetilde{\Delta},\rmi A_{H}]_{\circ}\in\Cc^{0,1}(A_H)$. Using $(H_0)$, $(H_1)$, $(H_{3,k})$, $(H_{4,k})$, $(H_{5,k})$, $(H_{6,k})$ and thanks to Lemma \ref{small}, we derive, for all $\gamma'<\gamma$ such that $\gamma'\in]0,1[$, there exists an integer $c_{\gamma'}$, such that:
\begin{equation}\label{eq2}
\left\|\Lambda^{\gamma'}(Q_1,Q_2)\left[\Delta_H-\widetilde{\Delta},\rmi A_H\right]f\right\|^2\leq c_{\gamma'}\|f\|^2,\ \forall f\in \Sc.
\end{equation}
So, by density, we obtain $[\Delta_H-\widetilde{\Delta},\rmi A_H]_{\circ}\in \Cc^{0,1}(A_H)
$. Therefore, thanks to
Corollary \ref{p_hypothesis} and by Proposition \ref{C2}, we deduce that $[\widetilde{\Delta},\rmi A_H]_{\circ}\in \Cc^{0,1}(A_H)$. In particular, by Proposition \ref{C2}, we get $\widetilde{\Delta}\in \Cc^{1,1}(A_H).$

Now, we proceed to prove \eqref{eq2}, from which \eqref{eq1} follows as a consequence. We recall \eqref{D}. We have:
\begin{align*}
[\Delta_H-\widetilde{\Delta},\rmi A_H]
                &=\rmi\left(\begin{array}{cc}
                   D_2A_{1,H}-A_{2,H}D_1 & 0 \\
                    0 & D_1A_{2,H}- A_{1,H}D_2\\
                  \end{array}\right).
\end{align*}
We denote $Diag(\mathbb{Z}^4):=\{(n,n),\ n\in \mathbb{Z}^2\}$. In the sequel, to simplify matters, for $((n_1,n_2),(n_1,n_2))\in Diag(\mathbb{Z}^4)$, we write 
$$\Ec|_{Diag(\mathbb{Z}^4)\times \{p_1,p_2\} }((n_1,n_2,p_1),(n_1,n_2,p_2))=\Ec (n_1,n_2,p_1,p_2)$$
and we write $$m(n_1,n_2,p_i):=m_i(n_1,n_2), \text{ where } i, j\in\{1,2\}.$$
We next treat the term $\rmi\left( D_1A_{2,H}-A_{1,H}D_2\right)$.
 It gives three types of terms, the ones in: $f_2(n_1+i,n_2+j),\ f_2(n_1+i+1,n_2+j)$ and $f_2(n_1+i,n_2+j+1)$.
We have:
\begin{align*}
&\rmi\left(D_1A_{2,H}-A_{1,H}D_2\right)f_2(n)\\
&=\rmi\left(\left(\Delta_{1,H}-\widetilde{\Delta_1}\right)A_{2,H}-A_{1,H}\left(\Delta_{2,H}-\widetilde{\Delta_2}\right)\right)f_2(n)\\
&=\frac{\rmi}{3}\left(\left(L_{1,H}+T_{1,H}+S_{1,H}\right)A_{2,H}-A_{1,H}\left(L_{2,H}+T_{2,H}+S_{2,H}\right)\right)f_2(n)\\
&=\frac{1}{3}\Bigg(\left(L_{1,H}+T_{1,H}+S_{1,H}\right)\left(R_{0,0}^*(U_1,U_2)+\sum_{l_1,l_2\in \{ 0,1\}}R_{l_1,l_2}^*(U_1,U_2)(Q_1l_1-Q_2l_2)\right)\\
&+\left(R_{0,0}(U_1,U_2)+\sum_{l_1,l_2\in \{ 0,1\}}(Q_1l_1-Q_2l_2)R_{l_1,l_2}(U_1,U_2)\right)\left(L_{2,H}+T_{2,H}+S_{2,H}\right)\Bigg)f_2(n).
\end{align*}
As the first step, we analyse the terms in $R_{0,0}$, we have
\begin{align*}
&\Bigg(\left(L_{1,H}+T_{1,H}+S_{1,H}\right)R_{0,0}^*(U_1,U_2)+R_{0,0}(U_1,U_2)\left(L_{2,H}+T_{2,H}+S_{2,H}\right)\Bigg)f_2(n)\\
=&\left(1-\frac{\Ec(n_1,n_2,p_1,p_2)}{\sqrt{m_1(n_1,n_2)}\sqrt{m_2(n_1,n_2)}}\right)\sum_{(i,j)\in I^0_{0,0}}\alpha_{i,j}^{0,0}f_2(n_1+i,n_2+j)\\
&+\sum_{(i,j)\in I^0_{0,0}}\left(1-\frac{\Ec(n_1-i,n_2-j,p_2,p_1)}{\sqrt{m_1(n_1-i,n_2-j)}\sqrt{m_2(n_1-i,n_2-j)}}\right)
\alpha_{i,j}^{0,0}f_2(n_1-i,n_2-j)\\
&+\left(1-\frac{\Ec((n_1,n_2,p_1),(n_1+1,n_2,p_2))}{\sqrt{m_1(n_1+1,n_2)}\sqrt{m_2(n_1,n_2)}}\right)\sum_{(i,j)\in I^0_{0,0}}
\alpha_{i,j}^{0,0}f_2(n_1+i+1,n_2+j)\\
&+\sum_{(i,j)\in I^0_{0,0}}\left(1-\frac{\Ec((n_1-i-1,n_2-j,p_2),(n_1-i,n_2-j,p_1))}{\sqrt{m_1(n_1-i,n_2-j)}\sqrt{m_2(n_1-i-1,n_2-j)}}\right)
\alpha_{i,j}^{0,0}f_2(n_1-i-1,n_2-j)\\
&+\left(1-\frac{\Ec((n_1,n_2,p_1),(n_1,n_2+1,p_2))}{\sqrt{m_1(n_1,n_2+1)}\sqrt{m_2(n_1,n_2)}}\right)\sum_{(i,j)\in I^0_{0,0}}
\alpha_{i,j}^{0,0}f_2(n_1+i,n_2+j+1)\\
&+\sum_{(i,j)\in I^0_{0,0}}\left(1-\frac{\Ec((n_1-i,n_2-j-1,p_2),(n_1-i,n_2-j,p_1))}{\sqrt{m_1(n_1-i,n_2-j)}\sqrt{m_2(n_1-i,n_2-j-1)}}\right)
\alpha_{i,j}^{0,0}f_2(n_1-i,n_2-j-1)\\
=&\sum_{(i,j)\in I^0_{0,0}}\Bigg(\left(1-\frac{\Ec(n_1,n_2,p_1,p_2)}{\sqrt{m_1(n_1,n_2)}\sqrt{m_2(n_1,n_2)}}\right)\alpha_{i,j}^{0,0}
+\left(1-\frac{\Ec((n_1,n_2,p_1),(n_1+1,n_2,p_2))}{\sqrt{m_1(n_1+1,n_2)}\sqrt{m_2(n_1,n_2)}}\right)\alpha_{i-1,j}^{0,0}\\
&+\left(1-\frac{\Ec((n_1,n_2,p_1),(n_1,n_2+1,p_2))}{\sqrt{m_1(n_1,n_2+1)}\sqrt{m_2(n_1,n_2)}}\right)\alpha_{i,j-1}^{0,0}\\
&+\left(1-\frac{\Ec(n_1+i,n_2+j,p_2,p_1)}{\sqrt{m_1(n_1+i,n_2+j)}\sqrt{m_2(n_1+i,n_2+j)}}\right)\alpha_{-i,-j}^{0,0}\\
&+\left(1-\frac{\Ec((n_1+i,n_2+j,p_2),(n_1+i+1,n_2+j,p_1))}{\sqrt{m_1(n_1+i+1,n_2+j)}\sqrt{m_2(n_1+i,n_2+j)}}\right)\alpha_{-i-1,-j}^{0,0}\\
&+\left(1-\frac{\Ec((n_1+i,n_2+j,p_2),(n_1+i,n_2+j+1,p_1))}{\sqrt{m_1(n_1+i,n_2+j+1)}\sqrt{m_2(n_1+i,n_2+j)}}\right)\alpha_{-i,-j-1}^{0,0}\Bigg)f_2(n_1+i,n_2+j).
\end{align*}
We turn to the norm, by Lemma \ref{small}, we infer
\begin{align*}
&\left\|\Lambda^{\gamma'}(n_1,n_2)\Bigg(\left(L_{1,H}+T_{1,H}+S_{1,H}\right)R_{0,0}^*(U_1,U_2)+R_{0,0}(U_1,U_2)\left(L_{2,H}+T_{2,H}+S_{2,H}\right)\Bigg)f_2(n)\right\|^2\\
\leq& \sum_n\Bigg|\Lambda^{\gamma'}(n_1,n_2)\sum_{(i,j)\in I^0_{l_2,l_2}}\Bigg(m_1(n_1,n_2)\left(\eta_2(n_1,n_2)-\varepsilon(n_1,n_2,p_1,p_2)\right)\\
&+ \Ec(n_1,n_2,p_1,p_2)\left(\eta_1(n_1,n_2)-\varepsilon(n_1,n_2,p_1,p_2)\right)\Bigg) \frac{1}{\sqrt{m_1(n_1,n_2)}\sqrt{m_2(n_1,n_2)}}\\
&\quad\times \frac{1}{\sqrt{m_1(n_1,n_2)}\sqrt{m_2(n_1,n_2)}+\Ec(n_1,n_2,p_1,p_2)}
\alpha_{i,j}^{0,0}f_2(n_1+i,n_2+j)\Bigg|^2\\
&+\sum_n\Bigg|\Lambda^{\gamma'}(n_1,n_2)\sum_{(i,j)\in I^0_{l_2,l_2}}\Bigg(m_1(n_1+1,n_2)\left(\eta_2(n_1,n_2)-\varepsilon((n_1,n_2,p_1),(n_1+1,n_2,p_2))\right)\\
&+ \Ec((n_1,n_2,p_1),(n_1+1,n_2,p_2))\left(\eta_1(n_1+1,n_2)-\varepsilon((n_1,n_2,p_1),(n_1+1,n_2,p_2))\right)\Bigg)\\
&\quad\times \frac{1}{\sqrt{m_1(n_1+1,n_2)}\sqrt{m_2(n_1,n_2)}}\\
&\quad\times \frac{1}{\sqrt{m_1(n_1+1,n_2)}\sqrt{m_2(n_1,n_2)}+\Ec((n_1,n_2,p_1),(n_1+1,n_2,p_2))}
\alpha_{i-1,j}^{0,0}f_2(n_1+i,n_2+j)\Bigg|^2\\
&+\sum_n\Bigg|\Lambda^{\gamma'}(n_1,n_2)\sum_{(i,j)\in I^0_{l_2,l_2}}\Bigg(m_1(n_1,n_2+1)\left(\eta_2(n_1,n_2)-\varepsilon((n_1,n_2,p_1),(n_1,n_2+1,p_2))\right)\\
&+ \Ec((n_1,n_2,p_1),(n_1,n_2+1,p_2))\left(\eta_1(n_1,n_2+1)-\varepsilon((n_1,n_2,p_1),(n_1,n_2+1,p_2))\right)\Bigg)\\
&\quad\times \frac{1}{\sqrt{m_1(n_1,n_2+1)}\sqrt{m_2(n_1,n_2)}}\\
&\quad\times \frac{1}{\sqrt{m_1(n_1,n_2+1)}\sqrt{m_2(n_1,n_2)}+\Ec((n_1,n_2,p_1),(n_1,n_2+1,p_2))}
\alpha_{i,j-1}^{0,0}f_2(n_1+i,n_2+j)\Bigg|^2\\
&+\sum_n\Bigg|\Lambda^{\gamma'}(n_1,n_2)\sum_{(i,j)\in I^0_{l_2,l_2}}\Bigg(m_1(n_1+i,n_2+j)\left(\eta_2(n_1+i,n_2+j)-\varepsilon(n_1+i,n_2+j,p_1,p_2)\right)\\
&+ \Ec(n_1+i,n_2+j,p_1,p_2)\left(\eta_1(n_1+i,n_2+j)-\varepsilon(n_1+i,n_2+j,p_1,p_2)\right)\Bigg)\\
&\quad\times \frac{1}{\sqrt{m_1(n_1+i,n_2+j)}\sqrt{m_2(n_1+i,n_2+j)}} \Big(\sqrt{m_1(n_1+i,n_2+j)}\sqrt{m_2(n_1+i,n_2+j)}\\
&\quad+\Ec(n_1+i,n_2+j,p_1,p_2)\Big)^{(-1)}\alpha_{-i,-j}^{0,0}f_2(n_1+i,n_2+j)\Bigg|^2\\
&+\sum_n\Bigg|\Lambda^{\gamma'}(n_1,n_2)\sum_{(i,j)\in I^0_{l_2,l_2}}\Bigg(m_1(n_1+i+1,n_2+j)\left(\eta_2(n_1+i,n_2+j)-\varepsilon(n_1+i,n_2+j,p_1,p_2)\right)\\
&+ \Ec((n_1+i+1,n_2+j,p_1),(n_1+i,n_2+j,p_2))\\
&\quad\times\left(m_1(n_1+i+1,n_2+j)-\Ec((n_1+i+1,n_2+j,p_1),(n_1+i,n_2+j,p_2))\right)\Bigg)\\
&\quad\times \frac{1}{\sqrt{m_1(n_1+i+1,n_2+j)}\sqrt{m_2(n_1+i,n_2+j)}} \Big(\sqrt{m_1(n_1+i+1,n_2+j)}\sqrt{m_2(n_1+i,n_2+j)}\\
&\quad+\Ec((n_1+i+1,n_2+j,p_1),(n_1+i,n_2+j,p_2))\Big)^{(-1)}\alpha_{-i-1,-j}^{0,0}f_2(n_1+i,n_2+j)\Bigg|^2\\
&+\sum_n\Bigg|\Lambda^{\gamma'}(n_1,n_2)\sum_{(i,j)\in I^0_{l_2,l_2}}\Bigg(m_1(n_1+i,n_2+j+1)\\
&\quad\times\left(\eta_2(n_1+i,n_2+j)-\varepsilon((n_1+i,n_2+j,p_1),(n_1+i,n_2+j+1,p_2))\right)\\
&+ \Ec((n_1+i,n_2+j,p_1),(n_1+i,n_2+j+1,p_2))\\
&\quad\times\left(\eta_1(n_1+i,n_2+j+1)-\varepsilon((n_1+i,n_2+j,p_1),(n_1+i,n_2+j+1,p_2))\right)\Bigg)\\
&\quad\times \frac{1}{\sqrt{m_1(n_1+i,n_2+j+1)}\sqrt{m_2(n_1+i,n_2+j)}} \Big(\sqrt{m_1(n_1+i,n_2+j+1)}\sqrt{m_2(n_1+i,n_2+j)}\\
&\quad+\Ec((n_1+i,n_2+j,p_1),(n_1+i,n_2+j+1,p_2))\Big)^{(-1)}\alpha_{-i,-j-1}^{0,0}f_2(n_1+i,n_2+j)\Bigg|^2\\
\leq c\|f_2\|^2.
\end{align*}
Let $S:=\{(1,0), (0,1), (1,1)\}$. Now, we treat the terms in $R_{l_1,l_2}$, we have:
\begin{align*}
&\Bigg(\left(L_{1,H}+T_{1,H}+S_{1,H}\right)\sum_{(l_1,l_2)\in S}R_{l_1,l_2}^*(U_1,U_2)(Q_1l_1-Q_2l_2)\\
&+\sum_{(l_1,l_2)\in S}(Q_1l_1-Q_2l_2)R_{l_1,l_2}(U_1,U_2)\left(L_{2,H}+T_{2,H}+S_{2,H}\right)\Bigg)f_2(n)\\
=&\left(1-\frac{\Ec(n_1,n_2,p_1,p_2)}{\sqrt{m_1(n_1,n_2)}\sqrt{m_2(n_1,n_2)}}\right)\sum_{(l_1,l_2)\in S}\sum_{(i,j)\in I^0_{l_2,l_2}}
\left((n_1+i)l_1-(n_2+j)l_2\right)\alpha_{i,j}^{l_1,l_2}f_2(n_1+i,n_2+j)\\
&+\sum_{(l_1,l_2)\in S}\sum_{(i,j)\in I^0_{l_2,l_2}}\left(n_1l_1-n_2l_2\right)\left(1-\frac{\Ec(n_1-i,n_2-j,p_2,p_1)}{\sqrt{m_1(n_1-i,n_2-j)}\sqrt{m_2(n_1-i,n_2-j)}}\right)\\ &\quad\times\alpha_{i,j}^{l_1,l_2}f_2(n_1-i,n_2-j)\\
&+\left(1-\frac{\Ec((n_1,n_2,p_1),(n_1+1,n_2,p_2))}{\sqrt{m_1(n_1+1,n_2)}\sqrt{m_2(n_1,n_2)}}\right)\sum_{(l_1,l_2)\in S}\sum_{(i,j)\in I^0_{l_2,l_2}}
\left((n_1+i+1)l_1-(n_2+j)l_2\right)\\
&\quad\times\alpha_{i,j}^{l_1,l_2}f_2(n_1+i+1,n_2+j)\\
&+\sum_{(l_1,l_2)\in S}\sum_{(i,j)\in I^0_{l_2,l_2}}\left(n_1l_1-n_2l_2\right)\left(1-\frac{\Ec((n_1-i-1,n_2-j,p_2),(n_1-i,n_2-j,p_1))}{\sqrt{m_1(n_1-i,n_2-j)}\sqrt{m_2(n_1-i-1,n_2-j)}}\right)\\
&\quad\times\alpha_{i,j}^{l_1,l_2}f_2(n_1-i-1,n_2-j)\\
&+\left(1-\frac{\Ec((n_1,n_2,p_1),(n_1,n_2+1,p_2))}{\sqrt{m_1(n_1,n_2+1)}\sqrt{m_2(n_1,n_2)}}\right)\sum_{(l_1,l_2)\in S}\sum_{(i,j)\in I^0_{l_2,l_2}}
\left((n_1+i)l_1-(n_2+j+1)l_2\right)\\
&\quad\times\alpha_{i,j}^{l_1,l_2}f_2(n_1+i,n_2+j+1)\\
&+\sum_{(l_1,l_2)\in S}\sum_{(i,j)\in I^0_{l_2,l_2}}\left(n_1l_1-n_2l_2\right)\left(1-\frac{\Ec((n_1-i,n_2-j-1,p_2),(n_1-i,n_2-j,p_1))}{\sqrt{m_1(n_1-i,n_2-j)}\sqrt{m_2(n_1-i,n_2-j-1)}}\right)\\
&\quad\times\alpha_{i,j}^{l_1,l_2}f_2(n_1-i,n_2-j-1).
\end{align*} 
Using $1.,\ 2.$ and $3.$ in Lemma \ref{remark}, and a change of variable, we derive:
\begin{align*}
&\Bigg(\left(L_{1,H}+T_{1,H}+S_{1,H}\right)\sum_{l_1,l_2\in \{ 0,1\}}R_{l_1,l_2}^*(U_1,U_2)(Q_1l_1-Q_2l_2)\\
&+\sum_{l_1,l_2\in \{ 0,1\}}(Q_1l_1-Q_2l_2)R_{l_1,l_2}(U_1,U_2)\left(L_{2,H}+T_{2,H}+S_{2,H}\right)\Bigg)f_2(n)\\
=&\left(\left(M_{\{1,0,1,0\}}+M_{\{1,1,1,0\}}+M_{\{1,0,1,1\}}+M_{\{-1,-\min(l_1,l_2),-1,0\}}\right)f_2\right)(n_1,n_2),
\end{align*}
where
\begin{align*}
&\left(M_{\{1,0,1,0\}}f_2\right)(n_1,n_2)\\
:=&\sum_{(l_1,l_2)\in S}\sum_{(i,j)\in I^0_{l_1,l_2}}\Bigg(\left((n_1+i)l_1-(n_2+j)l_2\right)\Bigg(\left(1-\frac{\Ec(n_1,n_2,p_1,p_2)}{\sqrt{m_1(n_1,n_2)}\sqrt{m_2(n_1,n_2)}}\right)\alpha_{i,j}^{l_1,l_2}\\
&+\left(1-\frac{\Ec((n_1,n_2,p_1),(n_1+1,n_2,p_2))}{\sqrt{m_1(n_1+1,n_2)}\sqrt{m_2(n_1,n_2)}}\right)\alpha_{i-1,j}^{l_1,l_2}
+\left(1-\frac{\Ec((n_1,n_2,p_1),(n_1,n_2+1,p_2))}{\sqrt{m_1(n_1,n_2+1)}\sqrt{m_2(n_1,n_2)}}\right)\alpha_{i,j-1}^{l_1,l_2}\Bigg)\\
&+\left(n_1l_1-n_2l_2\right)\Bigg(\left(1-\frac{\Ec(n_1+i,n_2+j,p_2,p_1)}{\sqrt{m_1(n_1+i,n_2+j)}\sqrt{m_2(n_1+i,n_2+j)}}\right)\alpha_{-i,-j}^{l_1,l_2}\\
&+\left(1-\frac{\Ec((n_1+i,n_2+j,p_2),(n_1+i+1,n_2+j,p_1))}{\sqrt{m_1(n_1+i+1,n_2+j)}\sqrt{m_2(n_1+i,n_2+j)}}\right)\alpha_{-i-1,-j}^{l_1,l_2}\\
&+\left(1-\frac{\Ec((n_1+i,n_2+j,p_2),(n_1+i,n_2+j+1,p_1))}{\sqrt{m_1(n_1+i,n_2+j+1)}\sqrt{m_2(n_1+i,n_2+j)}}\right)\alpha_{-i,-j-1}^{l_1,l_2}\Bigg)\Bigg)f_2(n_1+i,n_2+j),
\end{align*}
\begin{align*}
&\left(M_{\{1,1,1,0\}}f_2\right)(n_1,n_2)\\
:=&\sum_{(l_1,l_2)\in S}\sum_{(i,j)\in I^1_{l_1,l_2}}\Bigg(\left((n_1+i+1)l_1-(n_2+j)l_2\right)\Bigg(\left(1-\frac{\Ec((n_1,n_2,p_1),(n_1+1,n_2,p_2))}{\sqrt{m_1(n_1+1,n_2)}\sqrt{m_2(n_1,n_2)}}\right)
\alpha_{i,j}^{l_1,l_2}\\
&+\left(1-\frac{\Ec((n_1,n_2,p_1),(n_1,n_2+1,p_2))}{\sqrt{m_1(n_1,n_2+1)}\sqrt{m_2(n_1,n_2)}}\right)\alpha_{i+1,j-1}^{l_1,l_2}\Bigg)\\
&+\left(n_1l_1-n_2l_2\right)\Bigg(\left(1-\frac{\Ec(n_1+i+1,n_2+j,p_2,p_1)}{\sqrt{m_1(n_1+i+1,n_2+j)}\sqrt{m_2(n_1+i+1,n_2+j)}}\right)\alpha_{-i-1,-j}^{l_1,l_2}\\
&+\left(1-\frac{\Ec((n_1+i+1,n_2+j,p_2),(n_1+i+2,n_2+j,p_1))}{\sqrt{m_1(n_1+i-2,n_2+j)}\sqrt{m_2(n_1+i+1,n_2+j)}}\right)\alpha_{-i-2,-j}^{l_1,l_2}\\
&+\left(1-\frac{\Ec((n_1+i+1,n_2+j,p_2),(n_1+i+1,n_2+j+1,p_1))}{\sqrt{m_1(n_1+i+1,n_2+j+1)}\sqrt{m_2(n_1+i+1,n_2+j)}}\right)\alpha_{-i-1,-j-1}^{l_1,l_2}\Bigg)\Bigg)
f_2(n_1+i+1,n_2+j),
\end{align*}
\begin{align*}
&\left(M_{\{1,0,1,1\}}f_2\right)(n_1,n_2)\\
:=&\sum_{(l_1,l_2)\in S}\sum_{(i,j)\in I^6_{l_1,l_2}}\Bigg(\left((n_1+i)l_1-(n_2+j+1)l_2\right)\Bigg(
\left(1-\frac{\Ec((n_1,n_2,p_1),(n_1,n_2+1,p_2))}{\sqrt{m_1(n_1,n_2+1)}\sqrt{m_2(n_1,n_2)}}\right)\alpha_{i,j}^{l_1,l_2}\Bigg)\\
&+\left(n_1l_1-n_2l_2\right)\Bigg(\left(1-\frac{\Ec(n_1+i,n_2+j+1,p_2,p_1)}{\sqrt{m_1(n_1+i,n_2+j+1)}\sqrt{m_2(n_1+i,n_2+j+1)}}\right)\alpha_{-i,-j-1}^{l_1,l_2}\\
&+\left(1-\frac{\Ec((n_1+i,n_2+j+1,p_2),(n_1+i+1,n_2+j+1,p_1))}{\sqrt{m_1(n_1+i+1,n_2+j+1)}\sqrt{m_2(n_1+i,n_2+j+1)}}\right)\alpha_{-i-1,-j-1}^{l_1,l_2}\\
&+\left(1-\frac{\Ec((n_1+i,n_2+j+1,p_2),(n_1+i,n_2+j+2,p_1))}{\sqrt{m_1(n_1+i,n_2+j+2)}\sqrt{m_2(n_1+i,n_2+j+1)}}\right)\alpha_{-i,-j-2}^{l_1,l_2}\Bigg)\Bigg)f_2(n_1+i,n_2+j+1),
\end{align*}
\begin{align*}
&\left(M_{\{-1,-\min(l_1,l_2),-1,0\}}f_2\right)(n_1,n_2)\\
:=&\sum_{(l_1,l_2)\in S}\sum_{(i,j)\in I^{7+\min(l_1,l_2)}_{l_1,l_2}}\left(n_1l_1-n_2l_2\right)\\
&\Bigg(\left(1-\frac{\Ec((n_1-i,n_2-j-\min(l_1,l_2),p_2),(n_1-i,n_2-j,p_1))}{\sqrt{m_1(n_1-i,n_2-j)}\sqrt{m_2(n_1-i,n_2-j-\min(l_1,l_2))}}\right)
\alpha_{-i,-j-\min(l_1,l_2)}^{l_1,l_2}\\
&+\left(1-\frac{\Ec((n_1-i-\min(l_1,l_2),n_2-j,p_2),(n_1-i+(-1)^{\min(l_1,l_2)},n_2-j,p_1))}{\sqrt{m_1(n_1-i+(-1)^{\min(l_1,l_2)},n_2-j)}
\sqrt{m_2(n_1-i-\min(l_1,l_2),n_2-j)}}\right)\alpha_{i-(-1)^{\min(l_1,l_2)},j}^{l_1,l_2}\\
&+\left(1-\frac{\Ec((n_1-i-\min(l_1,l_2),n_2-j,p_2),(n_1-i-\min(l_1,l_2),n_2-j+1,p_1))}{\sqrt{m_1(n_1-i-\min(l_1,l_2),n_2-j+1)}\sqrt{m_2(n_1-i-\min(l_1,l_2),n_2-j)}}\right)
\alpha_{i+\min(l_1,l_2),j-1}^{l_1,l_2}\Bigg)\Bigg)\\
&\quad\times f_2(n_1-i-\min(l_1,l_2),n_2-j).
\end{align*}
We estimate on the norms. Firstly, we treat the terms on $f_2(n_1+i,n_2+j)$, the sum is over $I^0_{l_1,l_2}$, we obtain:
\begin{align}
&\nonumber\left\|\Lambda^{\gamma'}(n_1,n_2)\left(M_{\{1,0,1,0\}}f_2\right)(n_1,n_2)\right\|^2\\
\nonumber\leq& \sum_n\Bigg|\Lambda^{\gamma'}(n_1,n_2)\sum_{(l_1,l_2)\in S}\sum_{(i,j)\in I^{0}_{l_1,l_2}}\left(n_1l_1-n_2l_2\right)\Bigg(\sqrt{m_1(n_1+i,n_2+j)}\sqrt{m_2(n_1+i,n_2+j)}\\
\label{metric1}&\quad\times\left(\varepsilon(n_1+i,n_2+j,p_2,p_1)-
\varepsilon((n_1+i,n_2+j,p_2),(n_1+i+1,n_2+j,p_1))\right)\\
\nonumber&\quad\times\frac{1}{\sqrt{m_1(n_1+i,n_2+j)}\sqrt{m_2(n_1+i,n_2+j)}\sqrt{m_1(n_1+i+1,n_2+j)}\sqrt{m_2(n_1+i,n_2+j)}}\\ 
\label{metric2}&+\Ec(n_1+i,n_2+j,p_2,p_1)\Big(m_1(n_1+i+1,n_2+j)\left(\eta_2(n_1+i,n_2+j)-\eta_1(n_1+i,n_2+j)\right)\\
\label{metric3}&+m_1(n_1+i,n_2+j)\left(\eta_1(n_1+i+1,n_2+j)-\eta_2(n_1+i,n_2+j)\right)\Big)\\
\nonumber&\quad\times\frac{1}{\sqrt{m_1(n_1+i,n_2+j)}\sqrt{m_2(n_1+i,n_2+j)}\sqrt{m_1(n_1+i+1,n_2+j)}\sqrt{m_2(n_1+i,n_2+j)}}\\ 
\nonumber&\quad\times \frac{1}{m_1(n_1+i,n_2+j)m_2(n_1+i,n_2+j)+m_1(n_1+i+1,n_2+j)m_2(n_1+i,n_2+j)}\Bigg)\\
\nonumber&\quad\times\alpha_{-i-1,-j}^{l_1,l_2}f_2(n_1+i,n_2+j)\Bigg|^2\\
\nonumber&+ \Bigg|\Lambda^{\gamma'}(n_1,n_2)\left(n_1l_1-n_2l_2\right)\Bigg(\sqrt{m_1(n_1+i,n_2+j)}\sqrt{m_2(n_1+i,n_2+j)}\\
&\label{metric4}\times\left(\varepsilon(n_1+i,n_2+j,p_2,p_1)-
\varepsilon((n_1+i,n_2+j,p_2),(n_1+i,n_2+j+1,p_1))\right)\\
\nonumber&\quad\times\frac{1}{\sqrt{m_1(n_1+i,n_2+j)}\sqrt{m_2(n_1+i,n_2+j)}\sqrt{m_1(n_1+i,n_2+j+1)}\sqrt{m_2(n_1+i,n_2+j)}}\\ 
\label{metric5}&+\Ec(n_1+i,n_2+j,p_2,p_1)\Big(m_1(n_1+i,n_2+j+1)\left(\eta_2(n_1+i,n_2+j)-\eta_1(n_1+i,n_2+j)\right)\\
\label{metric6}&+m_1(n_1+i,n_2+j)\left(\eta_1(n_1+i,n_2+j+1)-\eta_2(n_1+i,n_2+j)\right)\Big)\\
\nonumber&\quad\times\frac{1}{\sqrt{m_1(n_1+i,n_2+j)}\sqrt{m_2(n_1+i,n_2+j)}\sqrt{m_1(n_1+i,n_2+j+1)}\sqrt{m_2(n_1+i,n_2+j)}}\\ 
\nonumber&\quad\times \frac{1}{m_1(n_1+i,n_2+j)m_2(n_1+i,n_2+j)+m_1(n_1+i,n_2+j+1)m_2(n_1+i,n_2+j)}\Bigg)\\
\nonumber&\quad \times \alpha_{-i,-j-1}^{l_1,l_2}f_2(n_1+i,n_2+j)\Bigg|^2\\
\nonumber&+\sum_n\Bigg|\Lambda^{\gamma'}(n_1,n_2)\sum_{(l_1,l_2)\in S}\sum_{(i,j)\in I^{0}_{l_1,l_2}}\left(il_1-j l_2\right)\Bigg(\Bigg(\Ec(n_1+i,n_2+j,p_2,p_1)\\
\label{metric7}&\quad\times\left(\eta_1(n_1+i,n_2+j)-\varepsilon(n_1+i,n_2+j,p_2,p_1)\right)\\
\label{metric8}&+m_1(n_1+i,n_2+j)\left(\eta_2(n_1+i,n_2+j)-\varepsilon(n_1+i,n_2+j,p_2,p_1)\right)\Bigg)\\
\nonumber&\quad\times \frac{1}{\sqrt{m_1(n_1+i,n_2+j)}\sqrt{m_2(n_1+i,n_2+j)}}\\
\nonumber&\quad\times \frac{1}{\sqrt{m_1(n_1+i,n_2+j)}\sqrt{m_2(n_1+i,n_2+j)}+\Ec((n_1+i,n_2+j,p_2),(n_1+i,n_2+j,p_1))}\Bigg)\\
\nonumber&\left(\alpha_{i,j}^{l_1,l_2}+\alpha_{i-1,j}^{l_1,l_2}+\alpha_{i,j-1}^{l_1,l_2}\right)f_2(n_1+i,n_2+j)\Bigg|^2\\
\nonumber&+\sum_n\Bigg|\Lambda^{\gamma'}(n_1,n_2)\sum_{(l_1,l_2)\in S}\sum_{(i,j)\in I^{0}_{l_1,l_2}}\left(il_1-jl_2\right)\Bigg(\sqrt{m_1(n_1+i,n_2+j)}\sqrt{m_2(n_1+i,n_2+j)}\\
\nonumber&\quad\times\left(\varepsilon(n_1+i,n_2+j,p_1,p_2)-
\varepsilon(n_1,n_2,p_1,p_2)\right)\\
\nonumber&\quad\times\frac{1}{\sqrt{m_1(n_1+i,n_2+j)}\sqrt{m_2(n_1+i,n_2+j)}\sqrt{m_1(n_1,n_2)}\sqrt{m_2(n_1,n_2)}}\\ 
\label{metric10}&+\Ec(n_1+i,n_2+j,p_1,p_2)\Big(m_1(n_1,n_2)\left(\eta_2(n_1,n_2)-\eta_1(n_1+i,n_2+j)\right)\\
\nonumber&+m_1(n_1+i,n_2+j)\left(\eta_1(n_1,n_2)-\eta_2(n_1+i,n_2+j)\right)\Big)\\
\nonumber&\quad\times\frac{1}{\sqrt{m_1(n_1+i,n_2+j)}\sqrt{m_2(n_1+i,n_2+j)}\sqrt{m_1(n_1,n_2)}\sqrt{m_2(n_1,n_2)}}\\ 
\nonumber&\quad\times \frac{1}{m_1(n_1,n_2)m_2(n_1,n_2)+m_1(n_1+i,n_2+j)m_2(n_1+i,n_2+j)}\Bigg)\\
\nonumber&\quad\times\alpha_{i,j}^{l_1,l_2}f_2(n_1+i,n_2+j)\Bigg|^2\\
\nonumber&+\sum_n\Bigg|\Lambda^{\gamma'}(n_1,n_2)\sum_{(l_1,l_2)\in S}\sum_{(i,j)\in I^{0}_{l_1,l_2}}\left(il_1-jl_2\right)\Bigg(\sqrt{m_1(n_1+i,n_2+j)}\sqrt{m_2(n_1+i,n_2+j)}\\
\nonumber&\quad\times\left(\varepsilon(n_1+i,n_2+j,p_1,p_2)-
\varepsilon((n_1,n_2,p_1),(n_1+1,n_2,p_2))\right)\\
\nonumber&\quad\times\frac{1}{\sqrt{m_1(n_1+i,n_2+j)}\sqrt{m_2(n_1+i,n_2+j)}\sqrt{m_1(n_1+1,n_2)}\sqrt{m_2(n_1,n_2)}}\\ 
\nonumber&+\Ec(n_1+i,n_2+j,p_1,p_2)\Big(m_1(n_1+1,n_2)\left(\eta_2(n_1,n_2)-\eta_1(n_1+i,n_2+j)\right)\\
\nonumber&+m_1(n_1+i,n_2+j)\left(\eta_1(n_1+1,n_2)-\eta_2(n_1+i,n_2+j)\right)\Big)\\
\nonumber&\quad\times\frac{1}{\sqrt{m_1(n_1+i,n_2+j)}\sqrt{m_2(n_1+i,n_2+j)}\sqrt{m_1(n_1+1,n_2)}\sqrt{m_2(n_1,n_2)}}\\ 
\nonumber&\quad\times \frac{1}{m_1(n_1+1,n_2)m_2(n_1,n_2)+m_1(n_1+i,n_2+j)m_2(n_1+i,n_2+j)}\Bigg)\\
\nonumber&\quad\times\alpha_{i-1,j}^{l_1,l_2}f_2(n_1+i,n_2+j)\Bigg|^2\\
\nonumber&+\sum_n\Bigg|\Lambda^{\gamma'}(n_1,n_2)\sum_{(l_1,l_2)\in S}\sum_{(i,j)\in I^{0}_{l_1,l_2}}\left(il_1-jl_2\right)\Bigg(\sqrt{m_1(n_1+i,n_2+j)}\sqrt{m_2(n_1+i,n_2+j)}\\
\nonumber&\quad\times\left(\varepsilon(n_1+i,n_2+j,p_1,p_2)-
\varepsilon((n_1,n_2,p_1),(n_1,n_2+1,p_2))\right)\\
\nonumber&\quad\times\frac{1}{\sqrt{m_1(n_1+i,n_2+j)}\sqrt{m_2(n_1+i,n_2+j)}\sqrt{m_1(n_1,n_2+1)}\sqrt{m_2(n_1,n_2)}}\\ 
\nonumber&+\Ec(n_1+i,n_2+j,p_1,p_2)
\Big(m_1(n_1,n_2+1)\left(\eta_2(n_1,n_2)-\eta_1(n_1+i,n_2+j)\right)\\
\nonumber&+m_1(n_1+i,n_2+j)\left(\eta_1(n_1,n_2+1)-\eta_2(n_1+i,n_2+j)\right)\Big)\\
\nonumber&\quad\times\frac{1}{\sqrt{m_1(n_1+i,n_2+j)}\sqrt{m_2(n_1+i,n_2+j)}\sqrt{m_1(n_1,n_2+1)}\sqrt{m_2(n_1,n_2)}}\\ 
\nonumber&\quad\times \frac{1}{m_1(n_1,n_2+1)m_2(n_1,n_2)+m_1(n_1+i,n_2+j)m_2(n_1+i,n_2+j)}\Bigg)\\
\nonumber&\quad\times\alpha_{i,j-1}^{l_1,l_2}f_2(n_1+i,n_2+j)\Bigg|^2\\
\nonumber&+\sum_n\Bigg|\Lambda^{\gamma'}(n_1,n_2)\sum_{(l_1,l_2)\in S}\sum_{(i,j)\in I^{0}_{l_1,l_2}}\left((n_1+i)l_1-(n_2+j)l_2\right)\\
\label{metric18}&\Bigg(\left(\frac{\Ec(n_1+i,n_2+j,p_1,p_2)}{\sqrt{m_1(n_1+i,n_2+j)}\sqrt{m_2(n_1+i,n_2+j)}}
-\frac{\Ec(n_1,n_2,p_1,p_2)}{\sqrt{m_1(n_1,n_2)}\sqrt{m_2(n_1,n_2)}}\right)\alpha_{i,j}^{l_1,l_2}\\
\nonumber&+\left(\frac{\Ec(n_1+i,n_2+j,p_1,p_2)}{\sqrt{m_1(n_1+i,n_2+j)}\sqrt{m_2(n_1+i,n_2+j)}}
-\frac{\Ec((n_1,n_2,p_1),(n_1+1,n_2,p_2))}{\sqrt{m_1(n_1+1,n_2)}\sqrt{m_2(n_1,n_2)}}\right)\alpha_{i-1,j}^{l_1,l_2}\\
\nonumber&+\left(\frac{\Ec(n_1+i,n_2+j,p_1,p_2)}{\sqrt{m_1(n_1+i,n_2+j)}\sqrt{m_2(n_1+i,n_2+j)}}
-\frac{\Ec((n_1,n_2,p_1),(n_1,n_2+1,p_2))}{\sqrt{m_1(n_1,n_2+1)}\sqrt{m_2(n_1,n_2)}}\right)\alpha_{i,j-1}^{l_1,l_2}\Bigg)\\
\nonumber&\quad\times f_2(n_1+i,n_2+j)\Bigg|^2\\
\nonumber\leq& c_{\gamma'}\|f_2\|^2
\end{align}
We treat \eqref{metric18} as follows:
\begin{align}
\nonumber&\sum_n\Bigg|\Lambda^{\gamma'}(n_1,n_2)\sum_{(l_1,l_2)\in S}\sum_{(i,j)\in I^{0}_{l_1,l_2}}\sum_{h=0}^{|i|-1}\left(n_1l_1-n_2l_2\right)\\
\nonumber&\quad\times\Bigg(\sqrt{m_1(n_1+sgn(i)(h+1),n_2)}\sqrt{m_2(n_1+sgn(i)(h+1),n_2)}\\
\label{metric21}&\quad\times\Big(\varepsilon(n_1+sgn(i)(h+1),n_2,p_1,p_2)-
\varepsilon(n_1+sgn(i)h,n_2,p_1,p_2)\Big)\\
\nonumber&\quad\times\frac{1}{\sqrt{m_1(n_1+sgn(i)(h+1),n_2)}\sqrt{m_2(n_1+sgn(i)(h+1),n_2)}}\\ 
\nonumber&\quad\times\frac{1}{\sqrt{m_1(n_1+sgn(i)h,n_2)}\sqrt{m_2(n_1+sgn(i)h,n_2)}}\\
\label{metric22}&+\Ec(n_1+sgn(i)(h+1),n_2,p_1,p_2)\Big(m_1(n_1+sgn(i)h,n_2)\Big(\eta_2(n_1+sgn(i)h+,n_2)\\
\nonumber&\quad\times -\eta_1(n_1+sgn(i)(h+1),n_2)\Big)\\
\label{metric23}&+m_1(n_1+sgn(i)(h+1),n_2)\left(\eta_1(n_1+sgn(i)h+1,n_2)-\eta_2(n_1+sgn(i)(h+1),n_2)\right)\Big)\\
\nonumber&\quad\times\frac{1}{\sqrt{m_1(n_1+sgn(i)(h+1),n_2)}\sqrt{m_2(n_1+sgn(i)(h+1),n_2)}}\\ 
\nonumber&\quad\times\frac{1}{\sqrt{m_1(n_1+sgn(i)h,n_2)}\sqrt{m_2(n_1+sgn(i)h,n_2)}}\\
\nonumber&\quad\times \Big(m_1(n_1+sgn(i)h,n_2)m_2(n_1+sgn(i)h,n_2)\\
\nonumber&+m_1(n_1+sgn(i)(h+1),n_2)m_2(n_1+sgn(i)(h+1),n_2)\Big)^{(-1)}\Bigg)\\
\nonumber&\quad\times\alpha_{i,j}^{l_1,l_2}f_2(n_1+i,n_2+j)\Bigg|^2\end{align}
and we estimate \eqref{metric10} like in the proof of Proposition \ref{H'}.
Here, we have used $(H_{3,2})$ for \eqref{metric1} and \eqref{metric4}, $(H_{3,1})$ for \eqref{metric3} and \eqref{metric6}, $(H_{4,1})$ for \eqref{metric2} and \eqref{metric5}, $(H_{8,1})$ and $(H_{8,2})$ for \eqref{metric7}, $(H_{5,2})$ for \eqref{metric8}, $(H_{3,2})$ for \eqref{metric21} and $(H_{3,1})$ for \eqref{metric22} and \eqref{metric23}. The other terms have the same treatment.\\
Then, there exists a constant $c_{\gamma'}>0$ such that
\begin{align*}
\left\|\Lambda^{\gamma'}(n_1,n_2)\left(M_{\{1,0,1,0\}}+M_{\{1,1,1,0\}}+M_{\{1,0,1,1\}}+M_{\{-1,-\min(l_1,l_2),-1,0\}}\right)f_2(n_1,n_2)\right\|^2\leq c_{\gamma'}\|f_2\|^2.\end{align*}
To conclude, we obtain \eqref{eq1} and \eqref{eq2}.
\qed
\subsection{Proof of the main result}\label{P.result}
The main result of this section is Theorem \ref{t_LAP}.
We begin by establishing the Mourre estimate in the case of perturbation. Since the ambient space is now $\ell^2(\widetilde{\Vc},m;\mathbb{C})$, we extend the operators acting on $\ell^2(\widetilde{\Vc},m;\mathbb{C})$ into this space. First, let us begin with a remark.
\begin{remark}\label{xxxx}
Let $A_{m,\Ec}:=T_{1\rightarrow m}A_HT^{-1}_{1\rightarrow m}, \forall z\in\mathbb{C}\setminus\mathbb{R},\ T_{1\rightarrow m}(A_H-z)^{-1}T^{-1}_{1\rightarrow m}=(A_{m,\Ec}-z)^{-1}$. By functional calculus, this implies $T_{1\rightarrow m}e^{\rmi tA_H}T^{-1}_{1\rightarrow m}=e^{\rmi tA_{m,\Ec}}$. Moreover if $S$ bounded in $\ell^2(\widetilde{\Vc},1;\mathbb{C})$, then $S\in \Cc^{\alpha}(A_H)\Leftrightarrow T_{1\rightarrow m}ST^{-1}_{1\rightarrow m}\in \Cc^{\alpha}(A_{m,\Ec})$, where $\alpha\in \{$1$;$2$;$0,1$;$1,1$\}$ and for $\alpha=1$, we also have $[T_{1\rightarrow m}ST^{-1}_{1\rightarrow m},\rmi A_{m,\Ec}]_{\circ}=T_{1\rightarrow m}[S,\rmi A_H]_{\circ}T^{-1}_{1\rightarrow m}.$
\end{remark}
\noindent Next, since $V(Q)$ is an operator of multiplication, so $V(Q):=T_{1\rightarrow m}V(Q)T^{-1}_{1\rightarrow m}$. Consequently, we have $$T_{1\rightarrow m}(\widetilde{\Delta}+V(Q))T_{1\rightarrow m}^{-1}=T_{1\rightarrow m}\widetilde{\Delta}T^{-1}_{1\rightarrow m}+T_{1\rightarrow m}V(Q)T^{-1}_{1\rightarrow m}=\Delta_{m,\Ec}+V(Q):=H_{m,\Ec}.$$ 
\begin{theorem}\label{mou}
Let $V:\widetilde{\Vc}\rightarrow \mathbb{R}^2$. We assume that $(H_0),\ (H_1),\ (H_2)$, $(H_3)\hbox{ and }(H_4) $ hold true. Then $\widetilde{\Delta}+V(Q)\in \Cc^{1,1}(A_H)$. Furthermore, for all compact interval $\Ic\subset[-1,1]\backslash\kappa(H_{m,\Ec})$, there are $c>0$ and a compact operator $\widetilde{K}$ such that:
\begin{align}\nonumber
&E_\Ic(\widetilde{\Delta}+V(Q))\left[\widetilde{\Delta}+V(Q), \rmi A_H\right]_\circ E_\Ic(\widetilde{\Delta}+V(Q))\\
&\label{Mour}\hspace*{+3.8cm}\geq c E_\Ic(\widetilde{\Delta}+V(Q)) +\widetilde{K}.
\end{align}
Equivalently, we obtain:
\begin{align}
&E_\Ic(H_{m,\Ec})\left[H_{m,\Ec}, \rmi A_{m,\Ec}\right]_\circ E_\Ic(H_{m,\Ec})\label{Mou}\geq c E_\Ic(H_{m,\Ec}) +\widetilde{K_{m,\Ec}},
\end{align}
where $\widetilde{K_{m,\Ec}}:=T_{1\rightarrow m}\widetilde{K}T^{-1}_{1\rightarrow m}.$
\end{theorem}
\proof
By Proposition \ref{H}, Proposition \ref{H'} and Proposition \ref{C2}, we have $\widetilde{\Delta}+V(Q)\in \Cc^{1,1}(A_H)$. By hypothesis $V(Q)$ is compact and by Proposition \ref{compact}, we have $\left(\Delta_H-\widetilde{\Delta}\right)$ is a compact operator. Then, by using Proposition \ref{Mourre} and by \cite[Theorem 7.2.9]{ABG}, we obtain \eqref{Mour}. Using the transformation unitary $T_{1\rightarrow m}$, Remark \ref{xxxx}  proves \eqref{Mou}.\qed\\
\textbf{Proof of Theorem \ref{t_LAP}}:
Proposition \ref{com} provides point $1.$ and Theorem \ref{mou} gives the points $2.$
To prove point $4.$, it suffices to consider $s>\frac{1}{2}$. Applying \cite[Proposition 7.5.6]{ABG}, we obtain:
$$\lim_{\rho\rightarrow 0^{+}}\sup_{\lambda\in[a,b]}\|\langle A_{m,\Ec}\rangle^{-s}(H_{m,\Ec}-\lambda-\rmi\rho)^{-1}\langle A_{m,\Ec}\rangle^{-s}\| \mbox{ is finite.}$$
Furthermore, in the norm topology of bounded operators, the boundary values of the
resolvent:
\[ [a,b] \ni\lambda\mapsto\lim_{\rho\to0^{\pm}}\langle A_{m,\Ec}\rangle^{-s}(H_{m,\Ec}-\lambda-\rmi\rho)^{-1}\langle A_{m,\Ec}\rangle^{-s} \mbox{ exists and is continuous},\]
where $[a,b]$ is included in $\mathbb{R}\setminus\left(\kappa\cup\sigma_{\rm p}(H_{m,\Ec})\right)$. In particular, this proves Point 3.
By Remark \ref{Lambdavarepsilon}, there exists $c>0$ such that: $$\|\langle H_{m,\Ec}\rangle^{s}f\|\leq c\|\Lambda^{s}(Q)f\|,$$
for all $f\in\Dc(\Lambda^{s}(Q)).$ We conclude that: $$\lim_{\rho\rightarrow 0^{+}}\sup_{\lambda\in[a,b]}\| \Lambda^{-s}(Q)(H_{m,\Ec}-\lambda-\rmi\rho)^{-1}\Lambda^{-s}(Q)\| \mbox{ is finite.}$$
 The point $5.$ follows directly as a consequence of $4.$\qed
 \section{Appendix}
The appendix presents the proof of Lemma \ref{remark}, using structured tables. First we recall \eqref{1}, \eqref{2}, \eqref{3} and \eqref{4}.
The highlighted cells correspond to the pairs $(i,j)$ such that $\alpha^{l_1,l_2}_{i,j}=0$.
\item[1.] The following tables provide points $1.,\ 2.$ and $3.$. 
The cells highlighted in same color other than pink, contain the pairs $(h,k)$ such that $\alpha_{h,k}^{l_1,l_2}= 0$ and satisfying one of the following properties:
\begin{align}\label{PROVE1}
   &\text{If }(-h,-k-1)\in I^5_{l_1,l_2}, \text{ then }(h-1,k)\in I^1_{l_1,l_2}\text{ or }(h,k+1)\in I^2_{l_1,l_2}\\
  &\nonumber\text{ or }(-h-\min(l_1,l_2),-k)\in I^{4-\min(l_1,l_2)}_{l_1,l_2}.\end{align}
  \begin{align}\label{PROVE2}&\text{If }(-h-1+\min(l_1,l_2),-k)\in I^{4-\min(l_1,l_2)}_{l_1,l_2}, \text{ then }
(h-1,k)\in I_{l_1,l_2}^1 \text{ or }(h,k-1)\in I_{l_1,l_2}^2\\
 &\nonumber\text{ or }
 (-h-\min(l_1,l_2),-k)\in I_{l_1,l_2}^{3+\min(l_1,l_2)}.
\end{align}
To sum up, \eqref{PROVE1} establishes Point $1.$, \eqref{PROVE2} proves Point $2.$ and Point $3.$ follows directly from  \eqref{PROVE1} and \eqref{PROVE2}.\\
Note that the colors of one table are independent of the colors an other table.
\setlength{\arrayrulewidth}{0.75pt}
\captionof{table}{$R_{0,1}$}
\label{tab6}

In this case, we have $\min(l_1,l_2)=\min(0,1)=0$, then  $I^{7+\min(l_1,l_2)}=I^7$.
For example, here if we consider the lime cells, we have $(h,k)=(-1,0)$ therefore $(-h-1+\min(0,1),-k)=(0,0)\in I^4_{0,1}$,
$(h-1,k)=(-2,0)\in I^1_{0,1}$, $(h,k-1)=(-1,-1)\in I^2_{0,1}$ and $(-h-\min(0,1),-k)=(1,0)\in I^3_{0,1}$. For the yellow cells $(h,k)=(2,-1)$, we have $(-h,-k-1)=(-2,0)\in I^5_{0,1}$ and $(-h-\min(0,1),-k)=(-2,1)\in I^3_{0,1}$. For the blue cells $(h,k)=(2-3)$, we have $(-h,-k-1)=(-2,2)\in I^5_{0,1}$ and $(h-1,k)\in I^1_{0,1}$.\\
From this table we can easily see that:
$$I^1_{0,1}=\left\{(-2,0),(1,-3),(0,2),(1,1),(1,-2),(0,-1),(1,0)\right\},$$
$$I^2_{0,1}=\left\{(0,2),(1,1),(-2,2),(1,-2),(-1,2),(-1,-1)\right\},$$
$$I^3_{0,1}=\left\{(0,-3),(-2,0),(1,-3),(-2,-1),(-2,1),(-1,-2),(-1,1),(2,0)\right\},$$
$$I^4_{0,1}=\left\{(0,0),(0,-3),(1,-3),(-2,1)\right\},$$
$$I^5_{0,1}=\left\{(-2,0),(-2,-1),(-2,1),(-2,2)\right\},$$
$$I^6_{0,1}=\left\{(0,2),(-2,2),(-1,2)\right\},$$
$$I^7_{0,1}=\left\{(-2,1)\right\},$$
$$I^8_{0,1}=I^9_{0,1}=\emptyset.$$\\\\
\newpage
\begin{tabular}{|C{1.4cm}|C{1.4cm}|C{1.4cm}|C{1.4cm}|C{1.4cm}|C{1.4cm}|}
\hlineB{2} \textbf{(i,j)}&\textbf{(i+1,j)}&\textbf{(i,j+1)}&\textbf{(-i,-j)}&\textbf{(-i-1,-j)}&\textbf{(-i,-j-1)}\\
\hlineB{2} (0,0)&(1,0)&(0,1)&(0,0)&\cellcolor{lime}(-1,0)&(0,-1)\\
\hline(0,-3)&(1,-3)&(0,-2)&\cellcolor{pink}(0,3)&\cellcolor{purple}(-1,3)&(0,2)\\
\hline(-2,0)&\cellcolor{lime}(-1,0)&(-2,1)&\cellcolor{orange}(2,0)&(1,0)&\cellcolor{yellow}(2,-1)\\
\hline(1,-3)&\cellcolor{blue}(2,-3)&(1,-2)&\cellcolor{purple}(-1,3)&\cellcolor{teal}(-2,3)&(-1,2)\\
\hline(-2,-1)&(-1,-1)&(-2,0)&\cellcolor{pink}(2,1)&(1,1)&\cellcolor{orange}(2,0)\\
\hline(0,2)&\cellcolor{pink}(1,2)&\cellcolor{pink}(0,3)&(0,-2)&(-1,-2)&(0,-3)\\
\hline(1,1)&\cellcolor{pink}(2,1)&\cellcolor{pink}(1,2)&(-1,-1)&(-2,-1)&(-1,-2)\\
\hline(-2,1)&(-1,1)&(-2,2)&\cellcolor{yellow}(2,-1)&\cellcolor{magenta}(1,-1)&\cellcolor{gray}(2,-2)\\
\hline(-2,2)&(-1,2)&\cellcolor{teal}(-2,3)&\cellcolor{gray}(2,-2)&(1,-2)&\cellcolor{blue}(2,-3)\\
\hline(-1,-2)&(0,-2)&(-1,-1)&\cellcolor{pink}(1,2)&(0,2)&(1,1)\\
\hline(1,-2)&\cellcolor{gray}(2,-2)&\cellcolor{magenta}(1,-1)&(-1,2)&(-2,2)&(-1,1)\\
\hline(-1,2)&(0,2)&\cellcolor{purple}(-1,3)&(1,-2)&(0,-2)&(1,-3)\\
\hline(0,-2)&(1,-2)&(0,-1)&(0,2)&(-1,2)&(0,1)\\
\hline(-1,-1)&(0,-1)&\cellcolor{lime}(-1,0)&(1,1)&(0,1)&(1,0)\\
\hline(-1,1)&(0,1)&(-1,2)&\cellcolor{magenta}(1,-1)&(0,-1)&(1,-2)\\
\hline(0,1)&(1,1)&(0,2)&(0,-1)&(-1,-1)&(0,-2)\\
\hline(0,-1)&\cellcolor{magenta}(1,-1)&(0,0)&(0,1)&(-1,1)&(0,0)\\
\hline(1,0)&\cellcolor{orange}(2,0)&(1,1)&\cellcolor{lime}(-1,0)&(-2,0)&(-1,-1)\\
\hline
\end{tabular}\\
  
\captionof{table}{$R_{1,0}$}
\label{tab8}

Here, we have $\min(l_1,l_2)=\min(1,0)=0$, hence $I^{7+\min(l_1,l_2)}=I^7$. From this table we can easy see that:
$$I^1_{1,0}=\left\{(2,0),(1,1),(2,-2),(-2,1),(2,-1),(-1,-1)\right\},$$
$$I^2_{1,0}=\left\{(0,-2),(-3,1),(2,0),(1,1),(-2,1),(0,1),(-1,0)\right\},$$
$$I^3_{1,0}=\left\{(-3,0),(0,-2),(-3,1),(-1,-2),(1,-2),(2,-2),(-2,1),(0,1),(1,-1)\right\},$$
$$I^4_{1,0}=\left\{(0,-2),(-1,-2),(1,-2),(2,-2)\right\},$$
$$I^5_{1,0}=\left\{(0,0),(-3,0),(-3,1),(1,-2)\right\},$$
$$I^6_{1,0}=\left\{(0,1),(-3,1),(1,1),(-2,1)\right\},$$
$$I^7_{1,0}=\left\{(1,-2)\right\},$$
$$I^8_{1,0}=\left\{(0,-2)\right\}$$ $$I^9_{1,0}=\emptyset.$$
\newpage
\setlength{\arrayrulewidth}{0.75pt}
\begin{tabular}{|C{1.4cm}|C{1.4cm}|C{1.4cm}|C{1.4cm}|C{1.4cm}|C{1.4cm}|}
\hlineB{2} \textbf{(i,j)}&\textbf{(i+1,j)}&\textbf{(i,j+1)}&\textbf{(-i,-j)}&\textbf{(-i-1,-j)}&\textbf{(-i,-j-1)}\\
\hlineB{2} (0,0)&(1,0)&(0,1)&(0,0)&(-1,0)& \cellcolor{brown}(0,-1)\\
\hline (-3,0)&(-2,0)&(-3,1)& \cellcolor{pink}(3,0)&(2,0)& \cellcolor{purple}(3,-1)\\
\hline (0,-2)&(1,-2)& \cellcolor{brown}(0,-1)& \cellcolor{yellow}(0,2)& \cellcolor{magenta}(-1,2)&(0,1)\\
\hline (-3,1)&(-2,1)& \cellcolor{cyan}(-3,2)& \cellcolor{purple}(3,-1)&(2,-1)& \cellcolor{blue}(3,-2)\\
\hline (-1,-2)&(0,-2)&(-1,-1)& \cellcolor{pink}(1,2)& \cellcolor{yellow}(0,2)&(1,1)\\
\hline (2,0)& \cellcolor{pink}(3,0)& \cellcolor{pink}(2,1)&(-2,0)&(-3,0)&(-2,-1)\\
\hline (1,1)& \cellcolor{pink}(2,1)& \cellcolor{pink}(1,2)&(-1,-1)&(-2,-1)&(-1,-2)\\
\hline (1,-2)&(2,-2)&(1,-1)& \cellcolor{magenta}(-1,2)& \cellcolor{lime}(-2,2)& \cellcolor{red}(-1,1)\\
\hline (2,-2)& \cellcolor{blue}(3,-2)&(2,-1)& \cellcolor{lime}(-2,2)& \cellcolor{cyan}(-3,2)&(-2,1)\\
\hline (-2,-1)&(-1,-1)&(-2,0)& \cellcolor{pink}(2,1)&(1,1)&(2,0)\\
\hline (-2,1)& \cellcolor{red}(-1,1)& \cellcolor{lime}(-2,2)&(2,-1)&(1,-1)&(2,-2)\\
\hline (0,1)&(1,1)& \cellcolor{yellow}(0,2)& \cellcolor{brown}(0,-1)&(-1,-1)&(0,-2)\\
\hline (2,-1)& \cellcolor{purple}(3,-1)&(2,0)&(-2,1)&(-3,1)&(-2,0)\\
\hline (-2,0)&(-1,0)&(-2,1)&(2,0)&(1,0)&(2,-1)\\
\hline (-1,-1)& \cellcolor{brown}(0,-1)&(-1,0)&(1,1)&(0,1)&(1,0)\\
\hline (1,-1)&(2,-1)&(1,0)& \cellcolor{red}(-1,1)&(-2,1)&(-1,0)\\
\hline (1,0)&(2,0)&(1,1)&(-1,0)&(-2,0)&(-1,-1)\\
\hline (-1,0)&(0,0)& \cellcolor{red}(-1,1)&(1,0)&(0,0)&(1,-1)\\
\hline
\end{tabular}

\captionof{table}{$R_{0,0}$}
\label{tab5}

Here, we have $\min(l_1,l_2)=\min(0,0)=0$, hence $I^{7+\min(l_1,l_2)}=I^7$. From this table we can easy see that:
$$I^1_{0,0}=\left\{(2,0),(1,1),(2,-2),(2,-1),(0,2),(2,-3)\right\},$$
$$I^2_{0,0}=\left\{(2,0),(1,1),(0,2),(-1,2),(-2,2),(-3,2)\right\},$$
$$I^3_{0,0}=\left\{(-3,0),(-3,1),(-1,-2),(-2,-1),(0,-3),(1,-3),(2,-3),(-3,2)\right\},$$
$$I^4_{0,0}=\left\{(0,-3),(1,-3),(2,-3)\right\},$$
$$I^5_{0,0}=\left\{(-3,0),(-3,1),(-3,2)\right\},$$
$$I^6_{0,0}=\left\{(0,2),(-1,2),(-2,2),(3,-2)\right\},$$
$$I^7_{0,0}=I^8_{0,0}=I^9_{0,0}=\emptyset.$$
\newpage
\begin{tabular}{|C{1.4cm}|C{1.4cm}|C{1.4cm}|C{1.4cm}|C{1.4cm}|C{1.4cm}|}
\hlineB{2} \textbf{(i,j)}&\textbf{(i+1,j)}&\textbf{(i,j+1)}&\textbf{(-i,-j)}&\textbf{(-i-1,-j)}&\textbf{(-i,-j-1)}\\
\hlineB{2} (0,0)&(1,0)&(0,1)&(0,0)&(-1,0)&(0,-1)\\
\hline (-3,0)&(-2,0)&(-3,1)& \cellcolor{green}(3,0)&(2,0)& \cellcolor{red}(3,-1)\\
\hline (0,-2)&(1,-2)&(0,-2)&(0,2)&(-1,2)&(0,1)\\
\hline (-3,1)&(-2,1)&(-3,2)& \cellcolor{red}(3,-1)&(2,-1)& \cellcolor{blue}(3,-2)\\
\hline (-1,-2)&(0,-2)&(-1,-1)& \cellcolor{brown}(1,2)&(0,2)&(1,1)\\
\hline (2,0)& \cellcolor{green}(3,0)& \cellcolor{orange}(2,1)&(-2,0)&(-3,0)&(-2,-1)\\
\hline (1,1)& \cellcolor{orange}(2,1)& \cellcolor{brown}(1,2)&(-1,-1)&(-2,-1)&(-1,-2)\\
\hline (1,-2)&(2,-2)&(1,-1)&(-1,2)&(-2,2)&(-1,1)\\
\hline (2,-2)& \cellcolor{blue}(3,-2)&(2,-1)&(-2,2)&(-3,2)&(-2,1)\\
\hline (-2,-1)&(-1,-1)&(-2,0)& \cellcolor{orange}(2,1)&(1,1)&(2,0)\\
\hline (-2,1)&(-1,1)&(-2,2)&(2,-1)&(1,-1)&(2,-2)\\
\hline (0,1)&(1,1)&(0,2)&(0,-1)&(-1,-1)&(0,-2)\\
\hline (2,-1)& \cellcolor{red}(3,-1)&(2,0)&(-2,1)&(-3,1)&(-2,0)\\
\hline (-2,0)&(-1,0)&(-2,1)&(2,0)&(1,0)&(2,-1)\\
\hline (-1,-1)&(0,-1)&(-1,0)&(1,1)&(0,1)&(1,0)\\
\hline (1,-1)&(2,-1)&(1,0)&(-1,1)&(-2,1)&(-1,0)\\
\hline (1,0)&(2,0)&(1,1)&(-1,0)&(-2,0)&(-1,-1)\\
\hline (0,2)& \cellcolor{brown}(1,2)& \cellcolor{purple}(0,3)&(0,-2)&(-1,-2)&(0,-3)\\
\hline (0,-3)&(1,-3)&(0,-2)& \cellcolor{purple}(0,3)& \cellcolor{gray}(-1,3)&(0,2)\\
\hline (-1,1)&(0,1)&(-1,2)&(1,-1)&(0,-1)&(1,-2)\\
\hline (-1,2)&(0,2)& \cellcolor{gray}(-1,3)&(1,-2)&(0,-2)&(1,-3)\\
\hline (1,-3)&(2,-3)&(1,-2)& \cellcolor{gray}(-1,3)& \cellcolor{magenta}(-2,3)&(-1,2)\\
\hline (-2,2)&(-1 , 2)& \cellcolor{magenta}(-2 ,3)&( 2, -2)&( 1, -2)&( 2 ,-3)\\
\hline (2,-3)& \cellcolor{yellow}(3,-3)&(2,-2)& \cellcolor{magenta}(-2,3)& \cellcolor{lime}(-3,3)&(-2,2)\\
\hline (-3,2)&(-2,2)& \cellcolor{lime}(-3,3)& \cellcolor{blue}(3,-2)&(2,-2)& \cellcolor{yellow}(3,-3)\\
\hline (0,-1)&(1,-1)&(0, 0)&(0, 1)&(-1 , 1)&(0 ,0)\\
\hline (-1,0)&(0,0)&(-1,1)&(1,0)&(0,0)&(1,-1)\\
\hline
\end{tabular}

\captionof{table}{$R_{1,1}$}
\label{tab9}

In this table, we have $\min(l_1,l_2)=\min(1,1)=1$, so $I^{7+\min(l_1,l_2)}=I^7$. From this table we can easy see that:
$$I^1_{1,1}=\left\{(-1,0),(0,1),(-1,2),(2,-2),(1,0),(2,-3),(2,-1)\right\},$$
$$I^2_{1,1}=\left\{(0,1),(-1,2),(-3,2),(1,0),(2,-1),(-2,2),(0,-1)\right\},$$
$$I^3_{1,1}=\left\{(-3,1),(-3,2),(0,-2),(2,-3),(1,-3),(2,0)\right\},$$
$$I^4_{1,1}=\left\{(-1,0),(0,1),(2,-3),(1,-3)\right\},$$
$$I^5_{1,1}=\left\{(-3,1),(-3,2),(1,0),(0,-1)\right\},$$
$$I^6_{1,1}=\left\{(-1,2),(-3,2),(-2,2)\right\},$$
$$I^8_{1,1}=\left\{(0,1)\right\},$$
$$I^7_{1,1}=I^9_{1,1}=\emptyset.$$
\newpage
\begin{tabular}{|C{1.4cm}|C{1.4cm}|C{1.4cm}|C{1.4cm}|C{1.4cm}|C{1.4cm}|}
\hlineB{2} \textbf{(i,j)}&\textbf{(i+1,j)}&\textbf{(i,j+1)}&\textbf{(-i,-j)}&\textbf{(-i-1,-j)}&\textbf{(-i,-j-1)}\\
\hlineB{2} (-1,0)& \cellcolor{pink}(0,0)&(-1,1)&(1,0)& \cellcolor{pink}(0,0)&(1,-1)\\
\hline (0,1)& \cellcolor{pink}(1,1)& \cellcolor{red}(0,2)&(0,-1)& \cellcolor{orange}(-1,-1)&(0,-2)\\
\hline (-1,2)& \cellcolor{red}(0,2)& \cellcolor{blue}(-1,3)&(1,-2)&(0,-2)&(1,-3)\\
\hline (-3,1)&(-2,1)&(-3,2)& \cellcolor{gray}(3,-1)&(2,-1)& \cellcolor{green}(3,-2)\\
\hline (-3,2)&(-2,2)& \cellcolor{pink}(-3,3)& \cellcolor{green}(3,-2)&(2,-2)& \cellcolor{brown}(3,-3)\\
\hline (2,-2)& \cellcolor{green}(3,-2)&(2,-1)&(-2,2)&(-3,2)&(-2,1)\\
\hline (0,-2)&(1,-2)&(0,-1)& \cellcolor{red}(0,2)&(-1,2)&(0,1)\\
\hline (1,0)& \cellcolor{magenta}(2,0)& \cellcolor{pink}(1,1)&(-1,0)&(-2,0)& \cellcolor{orange}(-1,-1)\\
\hline (2,-3)& \cellcolor{brown}(3,-3)&(2,-2)& \cellcolor{purple}(-2,3)& \cellcolor{pink}(-3,3)&(-2,2)\\
\hline (1,-3)&(2,-3)&(1,-2)& \cellcolor{blue}(-1,3)& \cellcolor{purple}(-2,3)&(-1,2)\\
\hline (2,-1)& \cellcolor{gray}(3,-1)& \cellcolor{magenta}(2,0)&(-2,1)&(-3,1)&(-2,0)\\
\hline (2,0)&(-1,0)&(-2,1)& \cellcolor{magenta}(2,0)&(1,0)&(2,-1)\\
\hline (-2,2)&(-1,2)& \cellcolor{purple}(-2,3)&(2,-2)&(1,-2)&(2,-3)\\
\hline (-1,1)&(0,1)&(-1,2)&(1,-1)&(0,-1)&(1,-2)\\
\hline (-2,1)&(-1,1)&(-2,2)&(2,-1)&(1,-1)&(2,-2)\\
\hline (1,-1)&(2,-1)&(1,0)&(-1,1)&(-2,1)&(-1,0)\\
\hline (1,-2)&(2,-2)&(1,-1)&(-1,2)&(-2,2)&(-1,1)\\
\hline (0,-1)&( 1,-1)& \cellcolor{pink}(0,0)&(0,1)&(-1,1)& \cellcolor{pink}(0,0)\\
\hline
\end{tabular}\\\\
\item[2.]
The following four tables provide point $1.$ of the lemma. For all $(l_1,l_2)\in \left\{(0,0), (0,1), (1,0), (1,1)\right\}$, we define  $$S_{1,0}^{l_1,l_2}:=\alpha _{i,j}^{l_1,l_2}+\alpha _{i-1,j}^{l_1,l_2}+\alpha _{i,j-1}^{l_1,l_2}$$ and $$S_{2,0}^{l_1,l_2}:=\alpha _{-i,-j}^{l_1,l_2}+\alpha _{-i-1,-j}^{l_1,l_2}+\alpha _{-i,-j-1}^{l_1,l_2}$$
\newpage
\setlength{\arrayrulewidth}{0.75pt}
\captionof{table}{Property of $R_{0,0}$ in $I^0_{0,0}$ - computing $S_{1,0}^{0,0}$ and $S_{2,0}^{0,0}$}
The highlighted cells are not included in $I^0_{0,0}$, as they contain the pairs $(h,k)$ such that $\alpha_{h,k}^{0,0}=0.$ Thanks to Table 3, we have:\\\\
\label{tab1}
\begin{tabular}{|C{1.4cm}|C{1.4cm}|C{1.4cm}|C{1.4cm}|C{1.4cm}|C{1.4cm}|C{.8cm}|C{.8cm}|}
\hlineB{2} \textbf{(i,j)}&\textbf{(i-1,j)}&\textbf{(i,j-1)}&\textbf{(-i,-j)}&\textbf{(-i-1,-j)}&\textbf{(-i,-j-1)}&$\boldsymbol{S_{1,0}^{0,0}}$&$\boldsymbol{S_{2,0}^{0,0}}$\\
\hlineB{2} (0, 0)&(-1, 0)&( 0, -1)&(0, 0)&(-1, 0)&( 0, -1)&12&12\\
\hline (-3, 0)& \cellcolor{pink}(-4, 0)& \cellcolor{pink}(-3, -1)& \cellcolor{pink}(3, 0)&(2, 0)& \cellcolor{pink}( 3, -1)&5&5\\
\hline ( 0, -2)&(-1, -2)&( 0, -3)&(0, 2)&(-1, 2)&(0, 1)&33&33\\
\hline (-3, 1)& \cellcolor{pink}(-4, 1)&(-3, 0)& \cellcolor{pink}( 3, -1)&( 2, -1)& \cellcolor{pink}( 3, -2)&12&12\\
\hline (-1, -2)& \cellcolor{pink}(-2, -2)& \cellcolor{pink}(-1, -3)& \cellcolor{pink}(1, 2)&(0, 2)&(1, 1)&9&9\\
\hline (2, 0)&(1, 0)&( 2, -1)&(-2, 0)&(-3, 0)&(-2, -1)&33&33\\
\hline (1, 1)&(0, 1)&(1, 0)&(-1, -1)&(-2, -1)&(-1, -2)&36&36\\
\hline ( 1, -2)&( 0, -2)&( 1, -3)&(-1, 2)&(-2, 2)&(-1, 1)&51&51\\
\hline ( 2, -2)&( 1, -2)&( 2, -3)&(-2, 2)&(-3, 2)&(-2, 1)&48&48\\
\hline (-2, -1)& \cellcolor{pink}(-3, -1)& \cellcolor{pink}(-2, -2)& \cellcolor{pink}(2, 1)&(1, 1)&(2, 0)&9&9\\
\hline (-2, 1)&(-3, 1)&(-2, 0)&( 2, -1)&( 1, -1)&( 2, -2)&51&51\\
\hline (0, 1)&(-1, 1)&(0, 0)&( 0, -1)&(-1, -1)&( 0, -2)&42&42\\
\hline ( 2, -1)&( 1, -1)&( 2, -2)&(-2, 1)&(-3, 1)&(-2, 0)&51&51\\
\hline (-2, 0)&(-3, 0)&(-2, -1)&(2, 0)&(1, 0)&( 2, -1)&33&33\\
\hline (-1, -1)&(-2, -1)&(-1, -2)&(1, 1)&(0, 1)&(1, 0)&36&36\\
\hline ( 1, -1)&( 0, -1)&( 1, -2)&(-1, 1)&(-2, 1)&(-1, 0)&54&54\\
\hline (1, 0)&(0, 0)&( 1, -1)&(-1, 0)&(-2, 0)&(-1, -1)&42&42\\
\hline (0, 2)&(-1, 2)&(0, 1)&( 0, -2)&(-1, -2)&( 0, -3)&33&33\\
\hline ( 0, -3)& \cellcolor{pink}(-1, -3)& \cellcolor{pink}( 0, -4)& \cellcolor{pink}(0, 3)& \cellcolor{pink}(-1, 3)&(0, 2)&5&5\\
\hline (-1, 1)&(-2, 1)&(-1, 0)&( 1, -1)&( 0, -1)&( 1, -2)&54&54\\
\hline (-1, 2)&(-2, 2)&(-1, 1)&( 1, -2)&( 0, -2)&( 1, -3)&51&51\\
\hline ( 1, -3)&( 0, -3)& \cellcolor{pink}( 1, -4)& \cellcolor{pink}(-1, 3)& \cellcolor{pink}(-2, 3)&(-1, 2)&12&12\\
\hline (-2, 2)&(-3, 2)&(-2, 1)&( 2, -2)&( 1, -2)&( 2, -3)&48&48\\
\hline ( 2, -3)&( 1, -3)& \cellcolor{pink}( 2, -4)& \cellcolor{pink}(-2, 3)& \cellcolor{pink}(-3, 3)&(-2, 2)&15&15\\
\hline (-3, 2)& \cellcolor{pink}(-4, 2)&(-3, 1)& \cellcolor{pink}( 3, -2)&( 2, -2)& \cellcolor{pink}( 3, -3)&15&15\\
\hline ( 0, -1)&(-1, -1)&( 0, -2)&(0, 1)&(-1, 1)&(0, 0)&42&42\\
\hline (-1, 0)&(-2, 0)&(-1, -1)&(1, 0)&(0, 0)&( 1, -1)&42&42\\
\hline \end{tabular}

\newpage

\captionof{table}{Property of $R_{0,1}$ in $I_{0,1}^0$ - computing $ S_{1,0}^{0,1}$ and $S_{2,0}^{0,1}$}
The highlighted cells are not included in $I^0_{0,1}$, because they contain the pairs $(h,k)$ such that $\alpha_{h,k}^{0,1}=0.$ By Table 1, we obtain:\\\\
\label{tab2}

\begingroup
\renewcommand{\arraystretch}{1} 
\begin{tabular}{|C{1.4cm}|C{1.4cm}|C{1.4cm}|C{1.4cm}|C{1.4cm}|C{1.4cm}|C{.8cm}|C{.8cm}|}
\hlineB{2}  \textbf{(i,j)}&\textbf{(i-1,j)}&\textbf{(i,j-1)}&\textbf{(-i,-j)}&\textbf{(-i-1,-j)}&\textbf{(-i,-j-1)}&$ \boldsymbol{S_{1,0}^{0,1}}$ & $\boldsymbol{S_{2,0}^{0,1}}$ \\
\hlineB{2} (0, 0)& (-1, 0)&( 0, -1)&(0, 0)& (-1, 0)&( 0, -1)&0&0\\
\hline ( 0, -3)& \cellcolor{pink}(-1, -3)& \cellcolor{pink}( 0, -4)& \cellcolor{pink}(0, 3)& \cellcolor{pink}(-1, 3)&(0, 2)&1&-1\\
\hline (-2, 0)& \cellcolor{pink}(-3, 0)&(-2, -1)& \cellcolor{pink}(2, 0)&(1, 0)& \cellcolor{pink}( 2, -1)&2&-2\\
\hline ( 1, -3)&( 0, -3)& \cellcolor{pink}( 1, -4)& \cellcolor{pink}(-1, 3)& \cellcolor{pink}(-2, 3)&(-1, 2)&2&-2\\
\hline (-2, -1)& \cellcolor{pink}(-3, -1)& \cellcolor{pink}(-2, -2)& \cellcolor{pink}(2, 1)&(1, 1)& \cellcolor{pink}(2, 0)&1&-1\\
\hline (0, 2)&(-1, 2)&(0, 1)&( 0, -2)&(-1, -2)&( 0, -3)&-8&8\\
\hline (1, 1)&(0, 1)&(1, 0)&(-1, -1)&(-2, -1)&(-1, -2)&-8&8\\
\hline (-2, 1)& \cellcolor{pink}(-3, 1)&(-2, 0)& \cellcolor{pink}( 2, -1)& \cellcolor{pink}( 1, -1)& \cellcolor{pink}( 2, -2)&0&0\\
\hline (-2, 2)& \cellcolor{pink}(-3, 2)&(-2, 1)& \cellcolor{pink}( 2, -2)&( 1, -2)& \cellcolor{pink}( 2, -3)&-2&2\\
\hline (-1, -2)& \cellcolor{pink}(-2, -2)& \cellcolor{pink}(-1, -3)& \cellcolor{pink}(1, 2)&(0, 2)&(1, 1)&2&-2\\
\hline ( 1, -2)&( 0, -2)&( 1, -3)&(-1, 2)&(-2, 2)&(-1, 1)&8&-8\\
\hline (-1, 2)&(-2, 2)&(-1, 1)&( 1, -2)&( 0, -2)&( 1, -3)&-8&8\\
\hline (1, 0)&(0, 0)& \cellcolor{pink}( 1, -1)& \cellcolor{pink}(-1, 0)&(-2, 0)&(-1, -1)&-6&6\\
\hline ( 0, -2)&(-1, -2)&( 0, -3)&(0, 2)&(-1, 2)&(0, 1)&8&-8\\
\hline (-1, -1)&(-2, -1)&(-1, -2)&(1, 1)&(0, 1)&(1, 0)&8&-8\\
\hline (-1, 1)&(-2, 1)& \cellcolor{pink}(-1, 0)& \cellcolor{pink}( 1, -1)&( 0, -1)&( 1, -2)&-6&6\\
\hline (0, 1)&(-1, 1)&(0, 0)&( 0, -1)&(-1, -1)&( 0, -2)&-14&14\\
\hline ( 0, -1)&(-1, -1)&( 0, -2)&(0, 1)&(-1, 1)&(0, 0)&14&-14\\
\hline \end{tabular}

\captionof{table}{Property of $R_{1,0}$ in $I^0_{1,0}$ - computing $S_{1,0}^{1,0}$ and $S_{2,0}^{1,0}$ }
The highlighted cells are not included in $I^0_{1,0}$, as they contain the pairs $(h,k)$ such that $\alpha_{h,k}^{1,0}=0.$ Using Table 2, we get:\\\\
\label{tab2}
\begin{tabular}{|C{1.4cm}|C{1.4cm}|C{1.4cm}|C{1.4cm}|C{1.4cm}|C{1.4cm}|C{.8cm}|C{.8cm}|}
\hlineB{2} \textbf{(i,j)}&\textbf{(i-1,j)}&\textbf{(i,j-1)}&\textbf{(-i,-j)}&\textbf{(-i-1,-j)}&\textbf{(-i,-j-1)}&$\boldsymbol{S_{1,0}^{1,0}}$&$\boldsymbol{S_{2,0}^{1,0}}$\\
\hlineB{2} (0,0)&(-1,0)&\cellcolor{pink}(0,-1)&(0,0)&(-1,0)&\cellcolor{pink}(0,-1)&0&0\\
\hline(-3,,0)&\cellcolor{pink}(-4,,0)&\cellcolor{pink}(-3,-1)&\cellcolor{pink}(3,0)&(2,0)&\cellcolor{pink}(3,-1)&1&-1\\
\hline(,0,-2)&(-1,-2)&\cellcolor{pink}(0,-3)&\cellcolor{pink}(0,2)&\cellcolor{pink}(-1,,2)&(0,1)&2&-2\\
\hline(-3,,1)&\cellcolor{pink}(-4,1)&(-3,0)&\cellcolor{pink}(3,-1)&(2,-1)&\cellcolor{pink}(3,-2)&2&-2\\
\hline(-1,-2)&\cellcolor{pink}(-2,-2)&\cellcolor{pink}(-1,-3)&\cellcolor{pink}(1,2)&\cellcolor{pink}(0,2)&(1,1)&1&-1\\
\hline(2,0)&(1,0)&(,2,-1)&(-2,,0)&(-3,,0)&(-2,-1)&-8&8\\
\hline(1,1)&(0,1)&(1,0)&(-1,-1)&(-2,-1)&(-1,-2)&-8&8\\
\hline(1,-2)&(0,-2)&\cellcolor{pink}(1,-3)&\cellcolor{pink}(-1,2)&\cellcolor{pink}(-2,2)&\cellcolor{pink}(-1,1)&0&0\\
\hline(2,-2)&(1,-2)&\cellcolor{pink}(2,-3)&\cellcolor{pink}(-2,2)&\cellcolor{pink}(-3,2)&(-2,1)&-2&2\\
\hline(-2,-1)&\cellcolor{pink}(-3,-1)&\cellcolor{pink}(-2,-2)&\cellcolor{pink}(2,1)&(1,1)&(2,0)&2&-2\\
\hline(-2,,1)&(-3,1)&(-2,0)&(2,-1)&(1,-1)&(2,-2)&8&-8\\
\hline(0,1)&\cellcolor{pink}(-1,,1)&(0,0)&\cellcolor{pink}(0,-1)&(-1,-1)&(0,-2)&-6&6\\
\hline(2,-1)&(,1,-1)&(,2,-2)&(-2,1)&(-3,1)&(-2,0)&-8&8\\
\hline(-2,0)&(-3,,0)&(-2,-1)&(2,0)&(1,0)&(2,-1)&8&-8\\
\hline(-1,-1)&(-2,-1)&(-1,-2)&(1,1)&(0,1)&(1,0)&8&-8\\
\hline(1,-1)&\cellcolor{pink}(,0,-1)&(1,-2)&\cellcolor{pink}(-1,1)&(-2,1)&(-1,0)&-6&6\\
\hline(1,0)&(0,0)&(1,-1)&(-1,0)&(-2,0)&(-1,-1)&-14&14\\
\hline(-1,0)&(-2,0)&(-1,-1)&(1,0)&(0,0)&(1,-1)&14&-14\\
\hline
\end{tabular}
\endgroup

\newpage

\captionof{table}{Property of $R_{1,1}$ in $I^0_{1,1}$ - computing $S_{1,0}^{1,1}$ and $S_{2,0}^{1,1}$ }
The highlighted cells are not included in $I^0_{1,1}$, as they contain the pairs $(h,k)$ such that $\alpha_{h,k}^{1,1}=0.$ From Table 4, we have: \\\\
\label{tab3}

\begin{tabular}{|C{1.4cm}|C{1.4cm}|C{1.4cm}|C{1.4cm}|C{1.4cm}|C{1.4cm}|C{.8cm}|C{.8cm}|}
\hlineB{2}  \textbf{(i,j)}&\textbf{(i-1,j)}&\textbf{(i,j-1)}&\textbf{(-i,-j)}&\textbf{(-i-1,-j)}&\textbf{(-i,-j-1)}&$\boldsymbol{S_{1,0}^{1,1}}$&$\boldsymbol{S_{2,0}^{1,1}}$\\
\hlineB{2} (-1  0)&(-2  0)&\cellcolor{pink}(-1 -1)&(1 0)&\cellcolor{pink}(0 0)&( 1 -1)&6&-6\\
\hline (0 ,1)&(-1 , 1)&\cellcolor{pink}(0, 0)&( 0 ,-1)&\cellcolor{pink}(-1 ,-1)&( 0 ,-2)&6&-6\\
\hline (-1,  2)&(-2  ,2)&(-1 , 1)&( 1, -2)&( 0, -2)&( 1, -3)&8&-8\\
\hline (-3,  1)& \cellcolor{pink}(-4 , 1)& \cellcolor{pink}(-3,  0)& \cellcolor{pink}( 3, -1)&( 2 ,-1)& \cellcolor{pink}( 3, -2)&1&-1\\
\hline (-3 , 2)& \cellcolor{pink}(-4 , 2)&(-3  ,1)& \cellcolor{pink}( 3 ,-2)&( 2, -2)& \cellcolor{pink}( 3, -3)&2&-2\\
\hline ( 2, -2)&( 1 ,-2)&( 2 ,-3)&(-2  ,2)&(-3 , 2)&(-2 , 1)&-8&8\\
\hline ( 0 ,-2)& \cellcolor{pink}(-1, -2)& \cellcolor{pink}( 0, -3)& \cellcolor{pink}(0 ,2)&(-1  ,2)&(0 ,1)&-2&2\\
\hline (1, 0)& \cellcolor{pink}(0 ,0)&( 1, -1)&(-1 , 0)&(-2 , 0)&\cellcolor{pink}(-1, -1)&-6&6\\
\hline ( 2 ,-3)&( 1, -3)& \cellcolor{pink}( 2, -4)& \cellcolor{pink}(-2,  3)& \cellcolor{pink}(-3 , 3)&(-2 , 2)&-2&2\\
\hline ( 1, -3)& \cellcolor{pink}( 0 ,-3)& \cellcolor{pink}( 1, -4)& \cellcolor{pink}(-1 , 3)& \cellcolor{pink}(-2 , 3)&(-1 , 2)&-1&1\\
\hline ( 2 ,-1)&( 1 ,-1)&( 2 ,-2)&(-2 , 1)&(-3 , 1)&(-2 , 0)&-8&8\\
\hline (-2,  0)& \cellcolor{pink}(-3 , 0)& \cellcolor{pink}(-2 ,-1)& \cellcolor{pink}(2, 0)&(1, 0)&( 2 ,-1)&2&-2\\
\hline (-2,  2)&(-3 , 2)&(-2 , 1)&( 2, -2)&( 1, -2)&( 2, -3)&8&-8\\
\hline (-1 , 1)&(-2 , 1)&(-1,  0)&( 1 ,-1)&( 0, -1)&( 1 ,-2)&14&-14\\
\hline (-2 , 1)&(-3 , 1)&(-2 , 0)&( 2 ,-1)&( 1, -1)&( 2 ,-2)&8&-8\\
\hline ( 1 ,-1)&( 0 ,-1)&( 1 ,-2)&(-1 , 1)&(-2 , 1)&(-1 , 0)&-14&14\\
\hline ( 1 ,-2)&( 0 ,-2)&( 1, -3)&(-1 , 2)&(-2,  2)&(-1,  1)&-8&8\\
\hline ( 0 ,-1)&\cellcolor{pink}(-1 ,-1)&( 0 ,-2)&(0 ,1)&(-1 , 1)& \cellcolor{pink}(0 ,0)&-6&6\\
\hline
\end{tabular}\\\\\\
\item[3.] Point $2.$ of the Lemma is shown in the following four tables, where
 $$S_{1,1}^{l_1,l_2}:=\alpha _{i,j}^{l_1,l_2}+\alpha _{i+1,j}^{l_1,l_2}+\alpha _{i+1,j-1}^{l_1,l_2}$$ and $$S_{2,1}^{l_1,l_2}:=\alpha _{-i-1,-j}^{l_1,l_2}+\alpha _{-i-2,-j}^{l_1,l_2}+\alpha _{-i-1,-j-1}^{l_1,l_2},$$
for all $(l_1,l_2)\in \left\{(0,0), (0,1), (1,0), (1,1)\right\}.$
\captionof{table}{Property of $R_{0,0}$ in $I^1_{0,0}$ - computing $S_{1,1}^{0,0}$ and $S_{2,1}^{0,0}$ }
The highlighted cells are not included in $I^1_{0,0}$, as they contain the pairs $(h,k)$ such that $\alpha_{h,k}^{0,0}=0.$ By Table 3, we own:\\\\
\label{tab4}
\begin{tabular}{|C{1.4cm}|C{1.4cm}|C{1.7cm}|C{1.4cm}|C{1.4cm}|C{1.7cm}|C{.8cm}|C{.8cm}|}
\hlineB{2} \textbf{(i,j)}&\textbf{(i+1,j)}&\textbf{(i+1,j-1)}&\textbf{(-i-1,-j)}&\textbf{(-i-2,-j)}&\textbf{(-i-1,-j-1)}&$\boldsymbol{S_{1,1}^{0,0}}$&$\boldsymbol{S_{2,1}^{0,0}}$\\
\hlineB{2} (2 ,0)& \cellcolor{pink}(3, 0)& \cellcolor{pink}( 3 ,-1)&(-3  ,0)& \cellcolor{pink}(-4  ,0)& \cellcolor{pink}(-3 ,-1)&5&5\\
\hline (1 ,1)& \cellcolor{pink}(2, 1)&(2, 0)&(-2 ,-1)& \cellcolor{pink}(-3, -1)& \cellcolor{pink}(-2 ,-2)&9&9\\
\hline ( 2 ,-2)& \cellcolor{pink}( 3, -2)& \cellcolor{pink}( 3 ,-3)&(-3 , 2)& \cellcolor{pink}(-4 , 2)&(-3 , 1)&15&15\\
\hline ( 2 ,-1)& \cellcolor{pink}( 3, -1)& \cellcolor{pink}(3 ,-2)&(-3 , 1)& \cellcolor{pink}(-4  ,1)&(-3 , 0)&12&12\\
\hline (0 ,2)& \cellcolor{pink}(1, 2)&(1 ,1)&(-1, -2)& \cellcolor{pink}(-2, -2)& \cellcolor{pink}(-1 ,-3)&9&9\\
\hline ( 2 ,-3)& \cellcolor{pink}( 3 ,-3)& \cellcolor{pink}(3 ,-4)& \cellcolor{pink}(-3 , 3)& \cellcolor{pink}(-4  ,3)&(-3 , 2)&8&8\\
\hline
\end{tabular}
\newpage
\captionof{table}{Property of $R_{1,0}$ in $I^1_{1,0}$ - computing  $S_{1,1}^{0,1}$ and $S_{2,1}^{0,1}$ }
The highlighted cells are not included in $I^1_{1,0}$, because they contain the pairs $(h,k)$ such that $\alpha_{h,k}^{1,0}=0.$ From Table 2, we obtain:\\\\
\label{tab3}
\begin{tabular}{|C{1.4cm}|C{1.4cm}|C{1.7cm}|C{1.4cm}|C{1.4cm}|C{1.7cm}|C{.8cm}|C{.8cm}|}
\hlineB{2} \textbf{(i,j)}&\textbf{(i+1,j)}&\textbf{(i+1,j-1)}&\textbf{(-i-1,-j)}&\textbf{(-i-2,-j)}&\textbf{(-i-1,-j-1)}& $\boldsymbol{S_{1,1}^{0,1}}$&$\boldsymbol{S_{2,1}^{0,1}}$ \\
\hlineB{2}
$( -2, 0 )$  &\cellcolor{pink} $( -1, 0 )$  & $( -1, -1 )$ & $( 1, 0 )$  & $( 0, 0 )$  & \cellcolor{pink}$( 1, -1 )$ & $6$  & $-6$  \\
\hline
$( 1, -3 )$  & \cellcolor{pink}$( 2, -3 )$  & \cellcolor{pink}$( 2, -4 )$ & \cellcolor{pink}$( -2, 3 )$ & \cellcolor{pink}$( -3, 3 )$ & $( -2, 2 )$ & $1$  & $-1$  \\
\hline
$( 0, 2 )$  & \cellcolor{pink}$( 1, 2 )$  & $( 1, 1 )$ &$( -1, -2 )$ & \cellcolor{red!20}$( -2, -2 )$ & \cellcolor{red!20}$( -1, -3 )$ & $-2$ & $2$  \\
\hline
$( 1, 1 )$  & \cellcolor{pink}$( 2, 1 )$  & \cellcolor{pink}$( 2, 0 )$ & $( -2, -1 )$ & \cellcolor{pink}$( -3, -1 )$ & \cellcolor{pink}$( -2, -2 )$ & $-1$ & $1$  \\
\hline
$( 1, -2 )$ & \cellcolor{pink}$( 2, -2 )$ & \cellcolor{pink}$( 2, -3 )$ & $( -2, 2 )$ & \cellcolor{pink}$( -3, 2 )$ & $( -2, -3 )$ & $2$  & $-2$  \\
\hline
$( 0, -1 )$ & \cellcolor{pink}$( 1, -1 )$ & $( 1, -2 )$ & $( -1, 1 )$ & $( -2, 1 )$ & \cellcolor{pink}$( -1, 0 )$ & $6$  & $-6$  \\
\hline
$( 1, 0 )$  & \cellcolor{pink}$( 2, 0 )$  & \cellcolor{pink}$( 2, -1 )$ & $( -2, 0 )$ & \cellcolor{pink}$( -3, 0 )$ & $( -2, -1 )$ & $-2$ & $2$  \\
\hline
\end{tabular}

\captionof{table}{Property of $R_{1,1}$ in $I^1_{0,1}$ - computing $S_{1,1}^{1,0}$ and $S_{2,1}^{1,0}$ }
The highlighted are not included in $I^1_{0,1}$, as they contain the pairs $(h,k)$ such that $\alpha_{h,k}^{0,1}=0.$ Thanks to Table 4, we obtain:\\\\
\label{tab5}
\begin{tabular}{|C{1.4cm}|C{1.4cm}|C{1.7cm}|C{1.4cm}|C{1.4cm}|C{1.7cm}|C{.8cm}|C{.8cm}|}
\hlineB{2} \textbf{(i,j)}&\textbf{(i+1,j)}&\textbf{(i+1,j-1)}&\textbf{(-i-1,-j)}&\textbf{(-i-2,-j)}&\textbf{(-i-1,-j-1)}&$\boldsymbol{S_{1,1}^{1,0}}$&$\boldsymbol{S_{2,1}^{1,0}}$\\
\hlineB{2} ( 2, -3)& \cellcolor{pink}( 3, -3)& \cellcolor{pink}( 3, -4)& \cellcolor{pink}(-3 , 3)& \cellcolor{pink}(-4 , 3)&(-3 , 2)&-1&1\\
\hline ( -1, 0)& \cellcolor{pink}(0,0)& (0,-1)& \cellcolor{pink}(0,0)& (-1,0)&(0,-1)&0&0\\
\hline (0,1)& \cellcolor{pink}(1,1)& (1,0)& \cellcolor{pink}(-1,-1)& (-2,-1)&(1,-2)&0&0\\
\hline (-1,2)& \cellcolor{pink}(0,2)& (0,1)& (0,-2)&\cellcolor{pink} (-1,-2)&\cellcolor{pink}(0,-3)&2&-2\\
\hline (2,-2)& \cellcolor{pink}(3,-2)& \cellcolor{pink}(3,-3)& (-3,2)& \cellcolor{pink}(-4,2)&(-3,1)&-2&2\\
\hline (1,0)& \cellcolor{pink}(2,0)& (2,-1)& (-2,0)&\cellcolor{pink}(-3,0)&\cellcolor{pink}(-2,-1)&-2&2\\
\hline (2,-3)& \cellcolor{pink}(3,-3)& \cellcolor{pink}(3,-4)& \cellcolor{pink}(-3,3)& \cellcolor{pink}(-4,3)&(-3,2)&-1&1\\
\hline
\end{tabular}

\captionof{table}{Property of $R_{0,1}$ in $I^1_{1,1}$ - computing $S_{1,1}^{1,1}$ and $S_{2,1}^{1,1}$}
The highlighted cells are not included in $I^1_{1,1}$, as they contain the pairs $(h,k)$ such that $\alpha_{h,k}^{1,1}=0.$ Using Table 1, we get:\\\\
\label{tab3}
\begin{tabular}{|C{1.4cm}|C{1.4cm}|C{1.7cm}|C{1.4cm}|C{1.4cm}|C{1.7cm}|C{.8cm}|C{.8cm}|}
\hlineB{2} \textbf{(i,j)}&\textbf{(i+1,j)}&\textbf{(i+1,j-1)}&\textbf{(-i-1,-j)}&\textbf{(-i-2,-j)}&\textbf{(-i-1,-j-1)}& $\boldsymbol{S_{1,1}^{1,1}}$&$\boldsymbol{S_{2,1}^{1,1}}$ \\
\hlineB{2}
$( 0, -2 )$  & \cellcolor{pink}$( 0, -1 )$  & $( -1, -1 )$ & $( 0, 1 )$  & $( 0, 0 )$  & \cellcolor{pink}$( -1, 1 )$ & $6$  & $-6$  \\
\hline
$( -3, 1 )$  & \cellcolor{pink}$( -3, 2 )$  & \cellcolor{pink}$( -4, 2 )$ & \cellcolor{pink}$( 3, -2 )$ & \cellcolor{pink}$( 3, -3 )$ & $( 2, -2 )$ & $1$  & $-1$  \\
\hline
$( 2, 0 )$  & \cellcolor{pink}$( 2, 1 )$  & $( 1, 1 )$ & $( -2, -1 )$ & \cellcolor{pink}$( -2, -2 )$ & \cellcolor{pink}$( -3, -1 )$ & $-2$ & $2$  \\
\hline
$( 1, 1 )$  & \cellcolor{pink}$( 1, 2 )$  & \cellcolor{pink}$( 0, 2 )$ & $( -1, -2 )$ & \cellcolor{pink}$( -1, -3 )$ & \cellcolor{pink}$( -2, -2 )$ & $-1$ & $1$  \\
\hline
$( -2, 1 )$ & \cellcolor{pink}$( -2, 2 )$ & \cellcolor{pink}$( -3, 2 )$ & $( 2, -2 )$ & \cellcolor{pink}$( 2, -3 )$ & $( 1, -2 )$ & $2$  & $-2$  \\
\hline
$( 0, 1 )$ & \cellcolor{pink}$( 0, 2 )$ & \cellcolor{pink}$( -1, 2 )$ & $( 0, -2 )$ &\cellcolor{pink} $( 0, -3 )$ & $( -1, -2 )$ & $-2$  & $2$  \\
\hline
$( -1, 0 )$  & \cellcolor{pink}$( -1, 1 )$  & $( -2, 1 )$ & $( 1, 1 )$ & $( 1, -2 )$ & \cellcolor{pink}$( 0, -1 )$ & $6$ & $-6$  \\
\hline
\end{tabular}
\\\\\\

\item[4.] The following tables provide the proof of point $3.$. Let
$$S_{1,2}^{l_1,l_2}:=\alpha _{i,j}^{l_1,l_2}+\alpha _{i,j+1}^{l_1,l_2}+\alpha _{i-1,j+1}^{l_1,l_2}$$ and $$S^{l_1,l_2}_{2,2}:=\alpha _{-i,-j-1}^{l_1,l_2}+\alpha _{-i-1,-j-1}^{l_1,l_2}+\alpha _{-i,-j-2}^{l_1,l_2},$$
for all $(l_1,l_2)\in \left\{(0,0), (0,1), (1,0), (1,1)\right\}.$
\newpage
\captionof{table}{Property of $R_{0,0}$ in $I^2_{0,0}$ - computing $S_{1,2}^{0,0}$ and $S^{0,0}_{2,2}$ }
The highlighted cells are not included in $I^2_{0,0}$, as they contain the pairs $(h,k)$ such that $\alpha_{h,k}^{0,0}=0.$ By Table 3, we have:\\\\
\label{tab5}
\begin{tabular}{|C{1.4cm}|C{1.4cm}|C{1.7cm}|C{1.4cm}|C{1.4cm}|C{1.7cm}|C{.8cm}|C{.8cm}|}
\hlineB{2} \textbf{(i,j)}&\textbf{(i,j+1)}&\textbf{(i-1,j+1)}&\textbf{(-i,-j-1)}&\textbf{(-i,-j-2)}&\textbf{(-i-1,-j-1)}&$\boldsymbol{S_{1,2}^{0,0}}$&$\boldsymbol{S^{0,0}_{2,2}}$\\
\hlineB{2} ( 0 ,-2)& \cellcolor{pink}( 0 ,-1)&(-1 ,-1)&(0 ,1)&(0 ,0)&(-1 , 1)&42&42\\
\hline (2, 0)& \cellcolor{pink}(2, 1)&(1 ,1)&(-2 ,-1)& \cellcolor{pink}(-2 ,-2)& \cellcolor{pink}(-3, -1)&9&9\\
\hline (1, 1)& \cellcolor{pink}(1, 2)&(0, 2)&(-1 ,-2)& \cellcolor{pink}(-1, -3)& \cellcolor{pink}(-2, -2)&9&9\\
\hline (-1 ,-1)& \cellcolor{pink}(-1 , 0)&(-2 , 0)&(1, 0)&(1 ,-1)&(0 ,0)&42&42\\
\hline (0 ,2)& \cellcolor{pink}(0 ,3)& \cellcolor{pink}(-1 , 3)&( 0 ,-3)& \cellcolor{pink}( 0 ,-4)& \cellcolor{pink}(-1 ,-3)&5&5\\
\hline (-1 , 2)& \cellcolor{pink}(-1 , 3)& \cellcolor{pink}(-2  ,3)&( 1 ,-3)& \cellcolor{pink}(1, -4)&( 0 ,-3)&12&12\\
\hline (-2 , 2)& \cellcolor{pink}(-2 , 3)& \cellcolor{pink}(-3 , 3)&( 2 ,-3)& \cellcolor{pink}( 2 ,-4)&(1 ,-3)&15&15\\
\hline (-3 , 2)& \cellcolor{pink}(-3 , 3)& \cellcolor{pink}(-4 , 3)& \cellcolor{pink}( 3 ,-3)& \cellcolor{pink}( 3 ,-4)&( 2 ,-3)&8&8\\
\hline
\end{tabular}
\captionof{table}{Property of $R_{1,0}$ in $I^2_{1,0}$ - computing $S_{1,2}^{1,0}$ and $S^{1,0}_{2,2}$ }
The highlighted cells are not included in $I^2_{1,0}$, as they contain the pairs $(h,k)$ such that $\alpha_{h,k}^{1,0}=0.$ From Table 2, we get:\\\\
\label{tab9}
\begin{tabular}{|C{1.4cm}|C{1.4cm}|C{1.7cm}|C{1.4cm}|C{1.4cm}|C{1.7cm}|C{.8cm}|C{.8cm}|}
\hlineB{2} \textbf{(i,j)}&\textbf{(i,j+1)}&\textbf{(i-1,j+1)}&\textbf{(-i,-j-1)}&\textbf{(-i,-j-2)}&\textbf{(-i-1,-j-1)} & $\boldsymbol{S_{1,2}^{1,0}}$&$\boldsymbol{S^{1,0}_{2,2}}$ \\
\hlineB{2}
(0,-2) & \cellcolor{pink}(0,-1) & (-1,-1) & (0,1) & (0,0) & \cellcolor{pink}(-1,1) & 6 & -6 \\
\hline
(-3,1) &\cellcolor{pink} (-3,2) & \cellcolor{pink}(-4,2) & \cellcolor{pink}(3,-2) & \cellcolor{pink}(3,-3) & (2,-2) & 1 & -1 \\
\hline
(2,0) &\cellcolor{pink} (2,1) & (1,1) & (-2,-1) & \cellcolor{pink}(-2,-2) & \cellcolor{pink}(-3,-1) & -2 & 2 \\
\hline
(1,1) & \cellcolor{pink}(1,2) & \cellcolor{pink}(0,2) & (-1,-2) & \cellcolor{pink}(-1,-3) & \cellcolor{pink}(-2,-2) & -1 & 1 \\
\hline
(-2,1) & \cellcolor{pink}(-2,2) & \cellcolor{pink}(-3,2) & (2,-2) & \cellcolor{pink}(2,-3) & (1,-2) & 2 & -2 \\
\hline
(0,1) & \cellcolor{pink}(0,2) & \cellcolor{pink}(-1,2) & (0,-2) & \cellcolor{pink}(0,-3) & (-1,-2) & -2 & 2 \\
\hline
(-1,0) & \cellcolor{pink}(-1,1) & (-2,1) & (1,1) & (1,-2) & \cellcolor{pink}(0,-1) & 6 & -6 \\
\hline
\end{tabular}

\captionof{table}{Property of $R_{0,1}$ in $I^2_{0,1}$ - computing $S_{1,2}^{0,1}$ and $S^{0,1}_{2,2}$ }
The highlighted cells are not included in $I^2_{0,1}$, as they contain the pairs $(h,k)$ such that $\alpha_{h,k}^{0,1}=0.$ Thanks to Table 1, we obtain:\\\\
\label{tab9}
\begin{tabular}{|C{1.4cm}|C{1.4cm}|C{1.7cm}|C{1.4cm}|C{1.4cm}|C{1.7cm}|C{1.4cm}|C{1.4cm}|}
\hlineB{2}
 \textbf{(i,j)}&\textbf{(i,j+1)}&\textbf{(i-1,j+1)}&\textbf{(-i,-j-1)}&\textbf{(-i,-j-2)}&\textbf{(-i-1,-j-1)} & $\boldsymbol{S_{1,2}^{0,1}}$&$\boldsymbol{S^{0,1}_{2,2}}$ \\
\hlineB{2}
(-2,2) & \cellcolor{pink}(-2,3) & \cellcolor{pink}(-3,3) & \cellcolor{pink}(2,-3) & \cellcolor{pink}(2,-4) & (1,-3) & -1 & 1 \\
\hline
(-1,2) & \cellcolor{pink}(-1,3) & \cellcolor{pink}(-2,3) & (1,-3) & \cellcolor{pink}(1,-4) & (0,-3) & -2 & 2 \\
\hline
(-1,-1) & \cellcolor{pink}(-1,0) & (-2,0) &(1,0) & \cellcolor{pink}(1,-1) & (0,0) & 1 & -1 \\
\hline
(0,2) & \cellcolor{pink}(0,3) & \cellcolor{pink}(-1,3) & (0,-3) & \cellcolor{pink}(0,-4) & \cellcolor{pink}(-1,-3) & 6 & -6 \\
\hline
(1,1) & \cellcolor{pink}(1,2) & (0,2) & (-1,-2) & \cellcolor{pink}(-1,-3) & \cellcolor{pink}(-2,-2) & -2 & 2 \\
\hline
(1,-2) & \cellcolor{pink}(1,-1) & \cellcolor{pink}(0,-1) & (-1,1) & \cellcolor{pink}(-1,0) & (-2,1) & 6 & -6 \\
\hline
\end{tabular}

\captionof{table}{Property of $R_{1,1}$ in $I^2_{1,1}$ - computing $S_{1,2}^{1,1}$ and $S^{1,1}_{2,2}$}
The highlighted cells are not included in $I^2_{1,1}$, as they contain the pairs $(h,k)$ such that $\alpha_{h,k}^{1,1}=0.$ From Table 4, we have:\\\\
\begin{tabular}{|C{1.4cm}|C{1.4cm}|C{1.7cm}|C{1.4cm}|C{1.4cm}|C{1.7cm}|C{1.4cm}|C{1.4cm}|}
\hlineB{2}
\textbf{(i,j)}&\textbf{(i,j+1)}&\textbf{(i-1,j+1)}&\textbf{(-i,-j-1)}&\textbf{(-i,-j-2)}&\textbf{(-i-1,-j-1)} & $\boldsymbol{S_{1,2}^{1,1}}$&$\boldsymbol{S^{1,1}_{2,2}}$ \\
\hlineB{2}
(0,1)   & \cellcolor{pink}(0,2)   & (-1,2)  & (0,-2)   & \cellcolor{pink}(0,-3)   & \cellcolor{pink}(-1,-2) & 2 & -2 \\
\hline
(-1,2)  & \cellcolor{pink}(-1,3)  & \cellcolor{pink}(-2,3)  & (1,-3)   & \cellcolor{pink}(1,-4)   & \cellcolor{pink}(0,-3)  & 1 & -1 \\
\hline
(-3,2)  & \cellcolor{pink}(-3,3)  & \cellcolor{pink}(-4,3)  & \cellcolor{pink}(3,-3)   & \cellcolor{pink}(3,-4)   & (2,-3)  & 1 & -1 \\
\hline
(1,0)   & \cellcolor{pink}(1,1)   & (0,1)   & \cellcolor{pink}(-1,-1) & \cellcolor{pink}(-1,-2) & \cellcolor{pink}(-2,-1) & 0 & 0  \\
\hline
(2,-1)  & \cellcolor{pink}(2,0)   & (1,0)   & (-2,0)  & \cellcolor{pink}(-2,-1) & \cellcolor{pink}(-3,0)  & -2 & 2  \\
 \hline
(-2,2)  & \cellcolor{pink}(-2,3)  & \cellcolor{pink}(-3,3)  & (2,-3)   & \cellcolor{pink}(2,-4)   & (1,-3)   & 2 & -2 \\
 \hline
(0,-1)  &\cellcolor{pink} (0,0)   & (-1,0)  & \cellcolor{pink}(0,0)   & (0,-1)   & (-1,0)  & 0 & 0  \\
 \hline
\end{tabular}
\\\\\\\
\item[5.] Point $4.$ is demonstrated in the following tables. Let
$$S_{1,4}^{l_1,l_2}:=\alpha _{i,j}^{l_1,l_2}+\alpha _{i+1,j}^{l_1,l_2}+\alpha _{i-1,j+1}^{l_1,l_2}$$ and $$S_{2,4}^{l_1,l_2}:=\alpha _{-i-1,-j}^{l_1,l_2}+\alpha _{-i-2,-j}^{l_1,l_2}+\alpha _{-i-1,-j-1}^{l_1,l_2},$$
where  $(l_1,l_2)\in \left\{(0,0), (0,1), (1,0), (1,1)\right\}.$
\captionof{table}{Property of $R_{0,0}$ in $I^4_{0,0}$ - computing $S_{1,4}^{0,0}$ and $S_{2,4}^{0,0}$}
The highlighted cells are not included in $I^4_{0,0}$, as they contain the pairs $(h,k)$ such that $\alpha_{h,k}^{0,0}=0.$ By Table 3, we own:\\\\
\label{tab5}
\begin{tabular}{|C{1.4cm}|C{1.4cm}|C{1.7cm}|C{1.4cm}|C{1.4cm}|C{1.7cm}|C{1.4cm}|C{1.4cm}|}
\hlineB{2} \textbf{(i,j)} & \textbf{(i+1,j)} & \textbf{(i+1,j-1)} & \textbf{(-i-1,-j)} & \textbf{(-i-2,-j)} & \textbf{(-i-1,-j-1)} & $\boldsymbol{S_{1,4}^{0,0}}$ & $\boldsymbol{S_{2,4}^{0,0}}$ \\
\hlineB{2}
(0,-3)  & \cellcolor{pink}(1, -3 ) & \cellcolor{pink}(1, -4 ) & \cellcolor{pink}( -1, 3 ) & \cellcolor{pink}( -2, 3 ) & ( -1, 2 ) & 12 & 12  \\
\hline
(1,-3) & ( 2,-3 ) & \cellcolor{pink}( 2, -4 ) & ( -2, -3 ) & \cellcolor{pink}( -3,3 ) & ( -2,2)& -2  & 2 \\
\hline
( 2, -3 ) & \cellcolor{pink}( 3, -3 ) & \cellcolor{pink}( 3, -4 ) & \cellcolor{pink}( -3, 3 ) & \cellcolor{pink}( -4, 3 ) & ( -3, 2 ) & -1  & 1  \\
\hline
( 2, -1 )  & \cellcolor{pink}( 3, -1 )  & \cellcolor{pink}( 3, -2 ) & \cellcolor{pink}( -3, 1 ) & \cellcolor{pink}( -4, 1 ) & \cellcolor{pink}( -3, 0 ) & -1 & -1  \\
\hline
\end{tabular}

\setlength{\arrayrulewidth}{0.75pt}
\captionof{table}{Property of $R_{1,0}$ in $I^4_{1,0}$ - computing $S_{1,4}^{1,0}$ and $S_{2,4}^{1,0}$}
The highlighted cells are not included in $I^4_{1,0}$, as they contain the pairs $(h,k)$ such that $\alpha_{h,k}^{1,0}=0.$ From Table 2, we get:\\\\
\begin{tabular}{|C{1.4cm}|C{1.4cm}|C{1.7cm}|C{1.4cm}|C{1.4cm}|C{1.7cm}|C{1.4cm}|C{1.4cm}|}
\hlineB{2}
\textbf{(i,j)} & \textbf{(i+1,j)} & \textbf{(i+1,j-1)} & \textbf{(-i-1,-j)} & \textbf{(-i-2,-j)} & \textbf{(-i-1,-j-1)} & $\boldsymbol{S_{1,4}^{1,0}}$ & $\boldsymbol{S_{2,4}^{1,0}}$ \\ \hlineB{2}
(0,-2)   &(1,-2)   & \cellcolor{pink}(1,-3)   & \cellcolor{pink}(-1,2)   & \cellcolor{pink}(-2,2)   & \cellcolor{pink}(-1,1)  & 0 & 0 \\ \hline
(-1,-2)  & (0,-2)   & \cellcolor{pink}(0,-3)   & \cellcolor{pink}(0,2)   & \cellcolor{pink}(-1,2)   & (0,1)   & 2 & -2 \\ \hline
(1,-2)   & \cellcolor{pink}(2,-2)   & \cellcolor{pink}(2,-3)   & \cellcolor{pink}(-2,2)   & \cellcolor{pink}(-3,2)   & (-2,1)  & -2 & 2 \\ \hline
(2,-2)   & \cellcolor{pink}(3,-2)   & \cellcolor{pink}(3,-3)   & \cellcolor{pink}(-3,2)   & \cellcolor{pink}(-4,2)   & (-3,1)  & -1 & 1 \\ \hline
\end{tabular}

\captionof{table}{Property of $R_{0,1}$ in $I^4_{0,1}$ - computing $S_{1,4}^{0,1}$ and $S_{2,4}^{0,1}$ }
The highlighted cells are not included in $I^4_{0,1}$, as they contain the pairs $(h,k)$ such that $\alpha_{h,k}^{0,1}=0.$ Using Table 1, we obtain:\\\\
\begin{tabular}{|C{1.4cm}|C{1.4cm}|C{1.7cm}|C{1.4cm}|C{1.4cm}|C{1.7cm}|C{1.4cm}|C{1.4cm}|}
\hlineB{2}
\textbf{(i,j)} & \textbf{(i+1,j)} & \textbf{(i+1,j-1)} & \textbf{(-i-1,-j)} & \textbf{(-i-2,-j)} & \textbf{(-i-1,-j-1)} & $\boldsymbol{S_{1,4}^{0,1}}$ & $\boldsymbol{S_{2,4}^{0,1}}$ \\ \hlineB{2}
(0,0) &  (1,0) & \cellcolor{pink} (1,-1) & \cellcolor{pink} (-1,0) & (-2,0) &  (-1,-1) & -14 & 14 \\ \hline
(0,-3) &  (1,-3) & \cellcolor{pink} (1,-4) & \cellcolor{pink} (-1,3) & \cellcolor{pink} (-2,3) &  (-1,2) & 2 & -2 \\ \hline
(1,-3) & \cellcolor{pink} (2,-3) & \cellcolor{pink} (2,-4) & \cellcolor{pink} (-2,3) & \cellcolor{pink} (-3,3) &  (-2,2) & -1 & 1\\ \hline
(-2,1) &  (-1,1) & \cellcolor{pink} (-1,0) & \cellcolor{pink} (1,-1) & (0,-1) &  (1,-2) & 6 & -6 \\ \hline
\end{tabular}

\newpage
\setlength{\arrayrulewidth}{0.75pt}
\captionof{table}{Property of $R_{1,1}$ in $I^4_{1,1}$ - computing $S_{1,4}^{1,1}$ and $S_{2,4}^{1,1}$}
The highlighted cells are not included in $I^4_{1,1}$, as they contain the pairs $(h,k)$ such that $\alpha_{h,k}^{1,1}=0.$ From Table 4, we have:\\\\
\begin{tabular}{|C{1.4cm}|C{1.4cm}|C{1.7cm}|C{1.4cm}|C{1.4cm}|C{1.7cm}|C{1.4cm}|C{1.4cm}|}
\hlineB{2}
\textbf{(i,j)} & \textbf{(i+1,j)} & \textbf{(i+1,j-1)} & \textbf{(-i-1,-j)} & \textbf{(-i-2,-j)} & \textbf{(-i-1,-j-1)} & $\boldsymbol{S_{1,4}^{1,1}}$ & $\boldsymbol{S_{2,4}^{1,1}}$ \\ \hlineB{2}
(-1,0) & \cellcolor{pink} (0,0) &  (0,-1) & \cellcolor{pink} (0,0) & (-1,0) &  (0,-1) & 0 & 0 \\ \hline
(0,1) & \cellcolor{pink} (1,1) &  (1,0) & \cellcolor{pink} (-1,-1) & \cellcolor{pink} (-2,-1) & \cellcolor{pink} (-1,-2) & 0 & 0 \\ \hline
(2,-3) & \cellcolor{pink} (3,-3) & \cellcolor{pink} (3,-4) & \cellcolor{pink} (-3,3) & \cellcolor{pink} (-4,3) &  (-3,2) & -1 & 1 \\ \hline
(1,-3) & \cellcolor{pink} (2,-3) & \cellcolor{pink} (2,-4) & \cellcolor{pink} (-2,3) & \cellcolor{pink} (-3,3) &  (-2,2) & -2 & 2 \\ \hline
\end{tabular}
\\

\end{document}